\documentclass[runningheads]{llncsMod}
\usepackage{amsmath,amssymb}
\usepackage{xspace}
\usepackage[dvipsnames,table]{xcolor}
\usepackage{wrapfig,graphicx}
\DeclareGraphicsExtensions{.pdf,.png}
\graphicspath{{figs/}{plots/}}
\usepackage{subcaption}
\usepackage{alltt}
\usepackage{tikz}
\usepackage{bm}
\usepackage[inline]{enumitem}

\newenvironment{ttdescription}
  {\begin{description}[font=\ttfamily]}
  {\end{description}}
\usepackage{booktabs}
\usepackage{tabularx}
\usepackage{multirow}
\usepackage{array}
\usepackage{siunitx}
\newcommand{\phZ}{\phantom{1}}

\usepackage{mathtools}
\DeclarePairedDelimiter\set{\lbrace}{\rbrace}
\DeclarePairedDelimiter\abs{\lvert}{\rvert}
\DeclarePairedDelimiter\ceil{\lceil}{\rceil}
\DeclarePairedDelimiter\floor{\lfloor}{\rfloor}
\DeclarePairedDelimiter\croc{\langle}{\rangle}

\def\cC{\ensuremath{\mathcal{C}}}

\def\cT{\ensuremath{\mathcal{T}}}

\DeclareMathOperator{\prob}{Pr}
\DeclareMathOperator{\dRF}{d_{\textup{RF}}}
\DeclareMathOperator{\ERF}{\textup{ERF}}
\def\ESS{\ensuremath{N_\textup{Eff}}}
\def\logP{\texttt{logP}\xspace}
\def\medianRF{\texttt{pseudoMedianRF}\xspace}
\def\minRF{\texttt{pseudoMinRF}\xspace}
\def\mapRF{\texttt{mapRF}\xspace}
\def\expRF{\texttt{expRF}\xspace}
\def\approximateESS{\texttt{approximateESS}\xspace}
\def\frechess{\texttt{fréchetIPSE}\xspace}
\def\edcf{\texttt{EDCF}\xspace}
\def\edcfml{\edcf-\texttt{ML}\xspace}
\def\edcfmcmc{\edcf-\texttt{B}\xspace}
\def\cladeInd{\texttt{cladeInd}\xspace} 

\usepackage{url}
\usepackage{cite}
\definecolor{linkblue}{rgb}{0.198,0.198,0.5392}
\usepackage[colorlinks=true,
    linkcolor=linkblue,
    anchorcolor=linkblue,
    citecolor=linkblue,
    filecolor=linkblue,
    menucolor=linkblue,
    urlcolor=linkblue,
    bookmarks=true,
    bookmarksopen=true,
    bookmarksopenlevel=2,
    bookmarksnumbered=true,
    hyperindex=true,
    plainpages=false,
    pdfpagelabels=true
]{hyperref}
\usepackage{bookmark}
\usepackage[capitalise,nameinlink]{cleveref}
\newcommand{\etal}{{et~al.}}

\renewcommand{\orcidID}[1]{\href{https://orcid.org/#1}{\includegraphics[scale=.03]{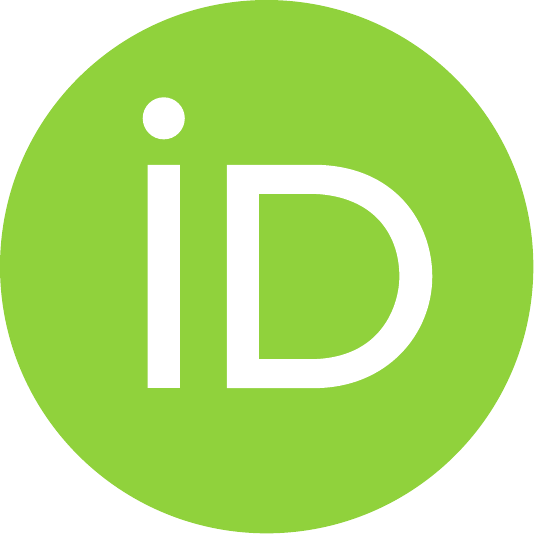}}}
\usepackage{siunitx}
\usepackage{placeins}  

\newcommand{\signatureGrid}[1]{
\begin{figure}[htb]
		\centering
	\includegraphics[trim={0 0 0 1.1cm},clip,width=\linewidth]{actSignature/#1-ACT25-run1-signatures}
	\caption{For \texttt{\uppercase{#1}}, scaled autocorrelation signatures of \logP{}1 and \expRF{}1 traces
	for the three chain simulation methods with a target ACT of 25
	and of the posterior (density) and $\kappa$ (transition/transversion rate ratio) for the corresponding MCMC~chain.}
	\label{fig:signature:#1}
\end{figure}
}

\newcommand{\robustnessAppendix}[1]{
	\begin{figure}[htb]
		\centering
		\includegraphics[trim={0 0.3cm 0 1cm},clip,width=0.9\linewidth,page=#1]{robustness/robustness-10k-run1-DS_full}
		\caption{Robustness of ESS estimators under different chain fragmentations for DS#1 with 10k trees.}
		\label{sfig:robustness:ds#1}
	\end{figure}
}

\newcommand{\mdsFigureYule}[1]{
	\begin{figure}[htb]
		\centering

		\includegraphics[width=.7\textwidth,height=.25\textheight,keepaspectratio]
		{plots/MDS/Yule50/Yule-Rep#1-bonsai_tsne-RNNI.pdf}

		\caption{MDS and clustering for Yule simulation #1.}
		\label{fig:mds:yule#1}
	\end{figure}
}

\makeatletter
\newcommand{\mdsFigure}[1]{%
	\@ifnextchar[{%
		\mdsFigure@i{#1}%
	}{%
		\mdsFigure@ii{#1}{}{}%
	}%
}

\def\mdsFigure@i#1[#2]{%
	\@ifnextchar[{%
		\mdsFigure@iii{#1}{#2}%
	}{%
		\mdsFigure@ii{#1}{#2}{}%
	}%
}

\def\mdsFigure@iii#1#2[#3]{%
	\mdsFigure@ii{#1}{#2}{#3}%
}

\def\mdsFigure@ii#1#2#3{%
	\begin{figure}[htb]
		\centering

		\includegraphics[width=.7\textwidth,height=.25\textheight,keepaspectratio]
		{plots/MDS/DS#1-bonsai_tsne-RNNI.pdf}

		\ifx#2\empty\else
		\par\medskip
		\includegraphics[width=.7\textwidth,height=.25\textheight,keepaspectratio]
		{plots/MDS/DS#1-bonsai_c#2.pdf}
		\fi

		\ifx#3\empty\else
		\par\medskip
		\includegraphics[width=.7\textwidth,height=.25\textheight,keepaspectratio]
		{plots/MDS/DS#1-bonsai_c#3.pdf}
		\fi

		\caption{MDS and clustering for dataset DS#1.}
		\label{fig:mds:ds#1}
	\end{figure}
}
\makeatother

\newcommand{\jumpDistance}[2]{
	\begin{figure}[p]
		\centering
		\includegraphics[width=.9\textwidth, height=.9\textheight, keepaspectratio]{DSs_trace/trace-ds#1-c#2-run1}
		\caption{Tree trace plots for different ESS statistics for DS{#1}, colored according to a {#2}-clustering.}
		\label{sfig:jumpdistance:ds#1}
	\end{figure}
}

\newcommand{\stabilitySingle}[1]{
\begin{figure}[htb]
	\centering
	\includegraphics[trim={0 0.3cm 0 1.1cm},clip,width=\linewidth,page=#1]{plots/stability/stability_full.pdf}
	\caption{Tree ESS estimates for the oversampled DS{#1} under different thinning intervals as per our stability experiment.}
	\label{sfig:stab:ds#1}
\end{figure}
}

\newcommand{\fullAccuracyRel}[1]{
\begin{figure}[htb!]
	\centering
	\includegraphics[clip,trim={0 0.5cm 0 1cm},width=0.9\linewidth,page=#1]{plots/accuracy/RNNI/RNNI-ACT5-final-rel_full.pdf}
	\caption{Relative estimated tree ESS by different estimators for DS#1 with respect to lower bound on the ESS of the simulated RNNI chains with target ACT 5.}
	\label{sfig:accuracy:rel:DS#1}
\end{figure}
}

\title{On estimating the effective sample size of phylogenetic trees in an autocorrelated chain}
\titlerunning{On estimating the ESS of phylogenetic trees}
\author{Jonathan~Klawitter\inst{1}$^{,\star}$\orcidID{0000-0001-8917-5269}
\and Lars~Berling\inst{2,4}\orcidID{0000-0002-1303-0669}
\and Jordan~Douglas\inst{1,3}\orcidID{0000-0003-0371-9961}
\and\\ Dong~Xie\inst{1}\orcidID{0000-0003-0878-3380}
\and Alexei~J.~Drummond\inst{1}\orcidID{0000-0003-4454-2576}}
\authorrunning{J. Klawitter et al.}
\institute{University of Auckland, Aotearoa New Zealand
\and Simon Fraser University, Canada
\and Australian National University, Australia
\and University of Canterbury, Aotearoa New Zealand \\
$^\star$ \href{mailto:jo[dot]klawitter[at]gmail[dot]com}{jo.klawitter[at]gmail.com}}

\usepackage{soul}

\begin{document}

\maketitle

\pdfbookmark[1]{Abstract}{Abstract}
\begin{abstract}
Estimating the effective sample size (ESS) is fundamental in Bayesian phylogenetic inference to account for autocorrelation in MCMC samples.
While methods for continuous parameters are well established, the discrete and high-dimensional nature of treespace poses substantial challenges.
In this paper, we introduce a new approach based on Conditional Clade Distributions (CCDs) to emulate MCMC chains with known ESS bounds.
We compare existing tree ESS estimators with novel approaches that utilize CCDs as tractable tree distributions, as well as a new probabilistic estimator based on differences in clade frequencies between independent chains.
Using our emulated chains from both simulated and real datasets, we assess estimators for their ability to recover the true ESS (accuracy), their insensitivity to the removal of autocorrelated samples from the chain (stability), and their tolerance to combining multiple chains together (robustness).
We further examine how multimodality and poor inter-mode chain mixing can compromise the reliability of tree ESS estimates.
These findings indicate that our new CCD-based estimators perform comparably to existing approaches with respect to accuracy, stability, and robustness.
In contrast, the probabilistic estimator and two previously recommended methods incur prohibitive computational costs for long chains.
Together, these results establish a new framework for evaluating tree ESS estimators using CCD-based emulated MCMC chains and demonstrate their performance and current limitations in practical phylogenetic analyses.
The expanded suite of tree ESS estimators is available in the user-friendly TreeStat2 software package for BEAST 2.
\end{abstract}

\section{Introduction} 
Bayesian inference with a Markov chain Monte Carlo (MCMC) algorithm is one of the main inference paradigms in phylogenetics~\cite{ronquist12mrbayes,hoehna16revbayes,suchard2018bayesian,bouckaert2019beast}.
MCMC algorithms explore the parameter space by proposing changes to the current state (such as tree topology modifications or parameter value adjustments) and accepting or rejecting these proposals based on their likelihood and prior probability.
The objective is to construct a Markov chain that effectively converges to the posterior distribution of phylogenetic trees along with associated parameters, such as substitution and clock rates.
This posterior distribution spans the entire space of possible trees (the \emph{treespace}) and their parameters.
In practice, however, we can only approximate this posterior through a finite sequence of samples collected during the MCMC run.

A key challenge of MCMC is autocorrelation: consecutive samples are typically dependent, which reduces the \emph{effective sample size (ESS)} and lowers the precision of posterior estimates~\cite{gelman95bayesian,rambaut2018tracer,vats2021revisiting}.
This dependence is quantified by the \emph{autocorrelation time (ACT)}, which measures the average number of iterations needed to obtain effectively independent samples.
For a chain of length~$N$, the ESS ($\ESS$) is inversely related to the ACT ($\tau$) by the equation  $\ESS = N / \tau$.
Thus, a higher ACT directly translates to a lower ESS, requiring more iterations to achieve reliable posterior estimates.
This challenge is particularly acute in phylogenetics, where MCMC must navigate the complex, high-dimensional, and non-Euclidean phylogenetic treespace using local tree rearrangement operations~\cite{allen01,bordewich05,collienne2021computing}.

Understanding the ESS is crucial for properly accounting for autocorrelation in downstream analyses and assessing the quality of posterior estimates.
The ESS quantifies how many independent samples would provide the same statistical precision as the autocorrelated MCMC samples.
A low ESS relative to the total number of draws indicates that the chain is moving slowly through the parameter space, signalling that more samples are needed for accurate estimates.
See Magee \etal~\cite{magee23} for an in-depth study of the relation between the ESS of phylogenetic trees and the Monte Carlo error.
Monitoring ESS is therefore essential for assessing MCMC efficiency and determining whether sufficient effectively independent samples have been collected.
Because MCMC explores parameter space with varying efficiency across parameters, the ESS must be calculated separately for each parameter.
In  phylogenetics, a common rule of thumb suggests a minimum ESS of 200 for all important parameters, including the posterior density and likelihood.
Magee \etal~\cite{magee23} and Fabreti and Höhna~\cite{fabreti22} advocate for higher thresholds of 500 and 625, respectively.

Reliable ESS estimation provides several critical benefits for phylogenetic MCMC analyses.
First, a low ESS often indicates that an MCMC chain should be run for additional iterations.
While a large ESS does not guarantee convergence, it serves as a valuable diagnostic for assessing convergence when used alongside other lines of evidence, as discussed by Berling, Bouckaert, and Gavryushkin~\cite{berling24automated}.
Reliable ESS estimation is thus an essential \textit{in situ} or \textit{post hoc} step for interpreting and ensuring the robustness of MCMC results.
Second, ESS and ACT values quantify sampling efficiency, enabling researchers to compare different MCMC algorithms, evaluate the effectiveness of proposal mechanisms, and optimize tuning parameters.
Third, the ESS is fundamental for computing Monte Carlo standard errors, which quantify the precision of posterior estimates due to finite sampling~\cite{fabreti22,magee23}.
Finally, in Bayesian model averaging~\cite{wasserman2000bayesian} and model selection approaches such as AIC and BIC, the ESS can be used to adjust effective sample sizes or serve as weights to prevent overestimation of model support~\cite{berger14}.

We tackle the challenging problem of estimating the ESS for samples of phylogenetic tree topologies (i.e., without branch lengths).
The fundamental difficulty stems from the discrete, high-dimensional nature of treespace, which lacks the metric properties of Euclidean spaces and therefore cannot accommodate existing ESS estimation methods designed for continuous parameters.
Compounding this effect, treespace is difficult to traverse during MCMC due to its rugged posterior distribution landscape~\cite{gao2026biological}, and wide variation in the acceptance rates of tree topology proposals among different landscapes~\cite{harrington2021properties}.
Moreover, it is common practice to downsample the tree sets beyond the initial MCMC sampling interval for experimental convenience (i.e., for smaller file sizes and faster processing).
Together, these challenges constitute a significant problem in phylogenetic analyses: while ESS is routinely treated as a procedural checkpoint, estimating it properly for tree topologies presents unique methodological obstacles compared to continuous parameters.

\paragraph{Related Work.}
To estimate the ESS for samples of phylogenetic tree topologies, previous approaches have mapped the sampled trees to traces of real-valued parameters, enabling the application of existing ESS estimators developed for continuous variables~\cite{gelman95bayesian}.
Lanfear, Hua, and Warren~\cite{lanfear16estimating} introduced the tree ESS estimator \texttt{pseudoESS}, which computes the distance from each sampled tree to a set of randomly selected reference (focal) trees from the sample, derives ESS estimates from these distance traces, and reports the median value along with summary statistics.
They proposed plotting these distances as tree topology traces and in addition also propose jump distances which visualize tree distances between pairs of trees at increasing sampling intervals.
In a related vein, the RNNI (Ranked Nearest Neighbor Interchange) distance~\cite{collienne2021computing} to a fixed tree estimate (UPGMA and neighbour joining) is used to address gene tree topology convergence in the StarBeast3 multispecies coalescent program \cite{douglas2022starbeast3}.
Fabreti and Höhna~\cite{fabreti22} took a different approach, converting (unrooted) trees into binary traces that indicate the presence or absence of specific splits (or clades for rooted trees).
Magee \etal~\cite{magee23} adapted the initial positive sequence estimation method---a popular ESS estimator for Euclidean variables---for use with trees by approximating the covariance structure through a tree metric.
They named their method \texttt{frechetCorrelationESS} to reflect its use of the Fréchet mean and variance. Lanfear \etal~\cite{lanfear16estimating} also proposed a similar approach, \approximateESS, which employs a more heuristic approach to estimating ACT and ESS.
In a comparative study, Magee \etal~\cite{magee23} evaluated the accuracy of these and additional ESS estimators using simulated chains.
Their evaluation was based on comparing estimated values to known Monte Carlo sampling errors, using simulated chains that replicate key characteristics of real phylogenetic posterior tree samples.
Beyond such benchmarks, the practical need for reliable tree ESS estimates and tree convergence methodology is emphasized in Berling~\etal~\cite{berling24automated}, where methods were developed and implemented to show that solely assessing continuous parameter ESS and convergence is inadequate.
Thus, it is important in practice to diagnose both continuous parameters and tree topologies, a necessity that was later corroborated by Brusselmans \etal~\cite{brusselmans2024importance}.

\paragraph{Contribution.}
In this paper, we present new methods for estimating the effective sample size of phylogenetic tree topologies, a refined evaluation framework based on simulated chains with known ESS bounds, and a case study examining the effect of multimodality on tree ESS estimation.
Our comparison shows that the new estimators perform comparably to existing ones but do not outperform them, and we recommend \cladeInd and \medianRF for practical use.

We contribute two distinct methodological approaches.
First, we extend the class of tree-to-trace mappings by leveraging \emph{Conditional Clade Distributions (CCD)}~\cite{hoehna12guided,larget13estimation,berling2025accurate}.
We map each sampled tree to (i) its posterior log probability, (ii) its Robinson-Foulds (RF) distance to the CCD MAP tree (point estimate)~\cite{berling2025accurate}, and (iii) its expected RF distance to the full posterior tree distribution, computed using a new efficient algorithm that avoids focal tree selection.
Second, we propose a probabilistic estimator based on observed clade frequencies across multiple independent MCMC chains, which operates independently of distance metrics, autocorrelation time, and tuning parameters.
While this approach is theoretically appealing, it ultimately proves less accurate and more computationally demanding than existing approaches.

Additionally, we refine the evaluation methodology using simulated MCMC chains with known lower bounds on the ESS.
Inspired by Fabreti and Höhna~\cite{fabreti22} and Magee \etal~\cite{magee23}, we generate i.i.d.\ tree samples from CCDs fitted to real posterior distributions~\cite{berling2025accurate}, which allows us to sample trees beyond those observed in the original MCMC sample.
We then introduce controlled perturbations, ranging from simple repetition to structured transitions in RNNI treespace~\cite{collienne2021computing}, to emulate MCMC autocorrelation patterns.
The expanded suite of tree ESS estimators is available through the user-friendly \texttt{TreeStat2} software package for BEAST 2 \cite{bouckaert2019beast}; with further software contribution detailed under Availability (Section~\ref{sec:availability}).

\section{Methods}
\label{sec:methods}
We begin by briefly recalling CCDs and introducing a new algorithm for computing the expected RF distance.
We then present both existing and new tree ESS estimators, followed by the evaluation framework used to test them.
Finally, we outline the real datasets used in our experiments, both for direct analysis and for simulating chains, and explain our approach for assessing their modality.
We assume throughout that all phylogenetic trees are rooted and binary (fully resolved).

\subsection{Conditional Clade Distributions} 
\label{sec:methods:ccd}
\begin{figure}[tbh]
  \centering
  \includegraphics{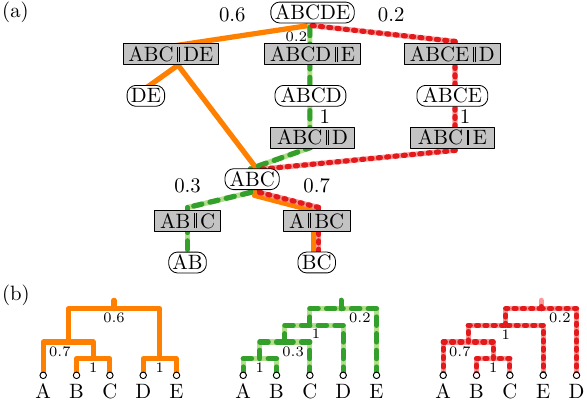}
  \caption{A CCD graph and three trees it contains: (a)~Truncated CCD graph (cherry splits and leaf clades omitted) annotated with clade split probabilities.
    Only the clades ABCDE and ABC split in multiple ways. (b)~Three of the six trees represented by the CCD graph with probabilities $0.42$, $0.06$, and $0.14$, respectively.}
  \label{fig:ccdExample}
\end{figure}

A \emph{conditional clade distribution (CCD)} is a distribution of phylogenetic trees (topologies) on the same taxa that parametrises treespace.
A CCD is based on a \emph{CCD graph}, which is bipartite graph on clades and clade splits.
The CCD graph contains all trees that can be combined by these elements.
The probability of picking a child clade split of a clade, called the \emph{conditional clade probability}, yields probabilities on trees (and for other elements)~\cite{larget13estimation}.
Depending on the exact CCD$[i]$ model~\cite{zhang18generalizing,jun23,berling2025accurate}, independence of clades (CCD0) or clade splits (CCD1) or pairs of clade splits (CCD2) is assumed.
See Berling \etal~\cite{berling2025accurate} for more details on CCDs and \cref{fig:ccdExample} for an example.

When a CCD is constructed based on an MCMC sample, both to form the CCD graph and populate its parameters, a CCD essentially smooths the Monte Carlo probability of the (in practice often sparse) sample to a larger treespace.
This allows it to provide a better estimate of the posterior distribution for non-trivial instances~\cite{berling2025accurate}.
Originally developed by Höhna and Drummond~\cite{hoehna12guided} and Larget~\cite{larget13estimation}, CCDs have found application in species tree–gene tree reconciliation~\cite{szollosi13efficient}, the efficient computation of the phylogenetic entropy~\cite{lewis16estimating}, detecting rogue taxa~\cite{klawitter24rogue}, and computing credible sets~\cite{klawitter25}.
Furthermore, similar tree distributions and parameter population methods have been investigated~\cite{zhang18generalizing,zhang18variational,jun23}.

Here we use that a CCD $D$ allows us to perform the following tasks efficiently:
\begin{enumerate}[label=(\roman*),leftmargin=*]
  \item sample a tree from $D$,
  \item compute the probability $\prob(T)$ of a tree $T$ in $D$,
  \item compute the probability $\prob(C)$ of a clade $C$ appearing in any tree in $D$,~and
  \item find the tree with maximum probability in $D$, the \emph{CCD MAP tree} of $D$.
\end{enumerate}
As a summary tree, the CCD MAP tree outperforms the popular MCC tree in terms of accuracy and precision and performs at least as well as the greedy consensus tree while further guaranteeing being fully resolved~\cite{berling2025accurate,hipstr25}.

\subsubsection{Expected RF Distance.}
Recall that for two trees~$T$ and~$T'$, their \emph{Robinson-Foulds (RF) distance} $\dRF(T, T')$ equals the symmetric distance of their clade sets~$\cC(T)$ and~$\cC(T')$, that is, the number of clades in $\cC(T) \setminus \cC(T')$ plus the number of clades in $\cC(T') \setminus \cC(T)$. For fully resolved trees, these two terms are equal and we thus use only one as RF distance.

Let $D$ be a CCD and let $T$ be a phylogenetic tree on the same taxon set as $D$. We want to define an RF distance between $T$ and $D$. To this end, we compute, for each tree $T' \in D$, the RF distance of $T$ and $T'$ and weigh it by the probability of~$T'$. We call this the \emph{expected RF distance} of $T$ to $D$, denoted by $\ERF(T, D)$:
\begin{equation} \label{eq:erf:trees}
	\ERF(T, D) = \mathbb E_{T' \in D}[\dRF(T, T')] = \sum_{T' \in D} \prob(T') \dRF(T, T')
\end{equation}

The expected RF distance $\ERF(T, D)$ can be computed efficiently with the CCD graph, noting that both computational runtimes scale with the number of observed clades. Let $C$ be a clade in $D$. Note that if $C \in \cC(T)$, then $C$ does not contribute to $\ERF(T, D)$ regardless of what tree it is part of. Otherwise--i.e., if $C \not\in \cC(T)$--note that $C$ contributes with $\prob(T')$ to $\ERF(T, D)$ for every tree $T' \in D$ that $C$ appears in (i.e., $C \in\cC(T')$). Fortunately, these contributions of~$C$ sum up exactly to $\prob(C)$. Hence, we can compute $\ERF(T, D)$ without iterating over all trees in $D$ by instead iterating over all clades in $D$:
\begin{equation} \label{eq:erf:clades}
	\ERF(T, D) = \sum_{\mathclap{\substack{C \in \cC(D)\\C \not\in \cC(T)}}} \prob(C)
\end{equation}
We can compute $\prob(C)$ collectively for all $C$ and the sum in~\cref{eq:erf:clades} both in linear time in the size of $D$~\cite{berling2025accurate}.

For example, the first tree in \cref{fig:ccdExample}(b) has an expected RF distance of 0.7 to the CCD in \cref{fig:ccdExample}(a):
clades ABCD and ABCE each contribute 0.2, clade AB contributes 0.3.
The second tree has an expected RF distance of 1.5 to the CCD as clade DE contributes 0.6, clade ABCE contributes 0.2, and clade BC contributes 0.7.

\subsection{Tree ESS Estimators} 
\label{sec:methods:ess}
The sequence of values of a specific parameter in the sample is known as the \emph{trace}.
Estimating the ESS and/or the ACT of a trace is a challenging statistical problem
and so several different approaches have been developed,
each addressing the impact of autocorrelation differently.
Fabreti and Höhna~\cite{fabreti22} compared the three commonly used ESS estimation implementations used in phylogenetics:
\texttt{Tracer} (an initial positive sequence estimator~\cite{straatsma86,geyer11,rambaut2018tracer}; see also Magee \etal~\cite{magee23}),
\texttt{coda} (a spectral density estimator~\cite{heidelberger81spectral,hamilton94,coda}),
and \texttt{MCMCSE} (a batch means estimator~\cite{flegal10,MCMCSE21}).
In summary, they found that the initial positive sequence estimator of \texttt{Tracer} behaved most robust for small and large ACT,
with consistent slight underestimation though no overestimation, which both other methods showed. We thus rely on \texttt{Tracer}'s method.

We start with the ESS estimators that map the trace of trees to a trace of values in an Euclidean space.
We then discuss how Magee \etal~\cite{magee23} adapted the initial positive sequence estimator, before describing our new probabilistic approach.

\subsubsection{Mapping Trees.}
Let $\cT = \croc{T_1, \ldots, T_N}$ be a trace of trees.
Our first set of estimators map each tree $T_j$, $j \in \set{1, \ldots, N}$, to a value (set of values),
for which we can then use \texttt{Tracer} to obtain an ESS estimate.
For those mappings utilizing a CCD, we tested both the CCD0 and CCD1 model;
for $i \in \set{0,1}$, let~$D_i$ be the CCD[$i$] on $\cT$.
We observed that CCD1 was generally a more reliable method for ESS estimation, and therefore our interpretation of CCD0-based estimators can be found in the Supplementary Material.

In non-trivial phylogenetic problems, most or even all trees in an MCMC chain are distinct
and hence the Monte Carlo probability is not reliable (as each different topology gets probability $1/N$).
However, CCDs offer a more accurate and refined probability of a tree $T_j$~\cite{lewis16estimating,berling2025accurate}.
Our first estimator \texttt{logP$i$}, $i \in \set{0,1}$, utilizes the probability $\prob_i(T_j)$ given by $D_i$
with a log transformation, since the variance can be span many orders of magnitude:
\begin{ttdescription}
	\item[\logP{}$i$] Map $T_j$ to $-\log\prob_i(T_j)$.
\end{ttdescription}
Note that this estimator does not require a choice of tree distance or focal trees.

The next set of estimators relate $T_j$ to the posterior distribution based on a tree distance, in particular the RF distance.
Lanfear \etal~\cite{lanfear16estimating} also considered the path distance~\cite{steel93distributions} and we tested the RNNI distance.
Since our focus is on the ESS of (unranked) tree topologies, we restrict the analysis to the RF distance (noting that the tree metric can matter in other settings~\cite{lanfear16estimating}).

Lanfear \etal~\cite{lanfear16estimating} suggested with their method \texttt{pseudoESS} to map each $T_j$ to its distance to a \emph{focal} tree,
repeating this several times, say $m = 100$, and then reporting the median estimated ESS. Since we use the RF distance, we call this method \medianRF.
We also introduce an estimator that uses the CCD$i$ MAP tree as focal tree.
Instead of having to choose focal trees, we suggest using the relation to the whole distribution by using the expected RF distance.
\begin{ttdescription}
	\item[\medianRF] Map $T_j$ to a vector 
	of RF distances to 100 (or less, if $N < 100$) random focal trees of $\cT$,
	estimate the ESS for each focal tree, and report the median.
	\item[\mapRF{}$i$] Map $T_j$ to $d_\text{RF}(T_j, T)$ with $T$ being the MAP tree of $D_i$.
	\item[\expRF{}$i$] Map $T_j$ to $\ERF(T_j, D_i)$, the expected RF distance of $T_j$ to $D_i$ (with the algorithm in \cref{sec:methods:ccd}).
\end{ttdescription}
There exist several other estimators that use distance-based mappings, though none showed an overall better performance than \medianRF (see Magee \etal~\cite[SM]{magee23}).
However, for completeness, we also evaluate \minRF in the Supplementary Material.

Another estimator using mappings is by Fabreti and Höhna~\cite{fabreti22},
who suggested to use binary vectors of clade presence and absence:
\begin{ttdescription}
	\item[\cladeInd] For each clade $C$ in $\cT$, create a trace with entry $j$ set to $\mathbf{1}_{C \in T_j}$,
	i.e., 1 if $C$ is in $T_j$ and 0 otherwise. Estimate the ESS for each trace and report the mean.
\end{ttdescription}

\subsubsection{Adaptation to Trees.}\label{sec:tree:adaptation}
Magee \etal~\cite{magee23} used the RF distance with Fréchet generalizations of mean and variance. They applied these to adapt both the initial positive sequence and the batch means estimators, and also proposed further ad hoc approaches. Here we focus only on the former.
\begin{ttdescription}
	\item[\frechess] Adapt the initial positive sequence estimator (IPSE) to trees by estimating covariance and correlations via Fréchet means and variances using squared RF distances.
\end{ttdescription}

In their experiments, Magee \etal~\cite[SM]{magee23} also included another new estimator \texttt{foldedRankmedoidESS}
and the ad hoc estimator \approximateESS by Lanfear \etal~\cite{lanfear16estimating}.
Based on their experiments, they recommend the use of \medianRF and \frechess.
It is worth noting that Lanfear et al.'s \approximateESS and \texttt{foldedRankmedoidESS} performed nearly as good.
Nonetheless, we included only \frechess in our list of evaluated estimators.

A downside of \frechess is that it computes all pairwise distances of trees in the chain. (\texttt{foldedRankmedoidESS} and \approximateESS share this downside.)
In practice, this can take a considerable amount of time and, if the full distance matrix is stored, also space\footnotemark.
\footnotetext{Concerning the space usage of \frechess,
we improved the implementation by Magee \etal~by not storing the whole distance matrix.
Instead we directly compute all values derived from the distance matrix either directly or offsets between consecutive values (for different lags).}%

This is a problem because chain convergence is commonly performed while the MCMC chain is still running, either manually by the user or in automated chain convergence tests~\cite{berling24automated}, which may lead to hundreds of requests to compute an estimate (e.g.,in a multilocus analysis).
Therefore, any given ESS measure should ideally take no more than a few milliseconds to compute, or a few seconds maximum.
Since \frechess is substantially more expensive per evaluation than existing measures, we explored whether random subsamples of trees could suffice to estimate the Fréchet variance and mean; however, in some instances this caused the underlying algorithm not to converge, and we did not pursue this speedup further.
Hence, in our main experiments we impose a five-minute time limit on each \frechess computation, and we report runtimes for different chain sizes and estimators in \autoref{sec:runtime}.

\subsubsection{Expected Difference of Clade Frequencies.}
When assessing convergence across multiple independent MCMC runs in phylogenetics, a key challenge is determining whether observed differences in clade frequencies between chains reflect genuine non-convergence or merely sampling variation. Fabreti and Höhna~\cite{fabreti22} addressed this through their \emph{Expected Difference of Split Frequencies (EDSF)} method. They demonstrated that the expected difference in split frequencies between two converged chains depends critically on both the true split frequency and the effective sample size. By modeling split presence/absence as binomial draws, they derived the theoretical distribution of frequency differences and used the 95th percentile as a convergence threshold. Crucially, their approach assumes that a reliable ESS estimate is already available from other methods, which is then used to determine appropriate thresholds for each split.

Building on this foundation, we propose a complementary approach that inverts the logic: rather than assuming a known ESS to test convergence, we directly estimate the ESS from observed clade frequencies across multiple independent chains. Our method requires no prior ESS estimates, no distance metrics, and no explicit autocorrelation time calculations. Instead, we model the relationship between raw MCMC samples and effectively independent samples through a probabilistic framework. We call this approach \emph{Expected Difference of Clade Frequencies (EDCF)}.
We present this method in full as the derivation is non-trivial and the negative result is informative; we discuss the method's limitations in \cref{sec:discussion}.

\noindent\textit{Model.} Suppose we have $m$ independent MCMC chains of phylogenetic trees where chain $i$ has $\ell_i$ samples. Let $K_i$ denote the number of times a clade was observed in chain $i$. We model the sampling process hierarchically: there exists an unknown parameter $R$ with $0 < R \leq 1$ representing the proportion of samples that are effectively independent. The \textit{effective} number of observations of a clade in chain $i$ is $\floor{ RK_i }$, and we assume these effective counts are binomially distributed:

\begin{align}
    \floor{ RK_i } & \sim \text{Binomial}(n = \floor{ R\ell_i }, p = C),  \\
    \text{where } C		  & \sim \text{Beta}(\alpha = a, \beta = b),
\end{align}
Here, $C$ represents the true posterior probability of the clade, which is itself modeled with a Beta prior. This hierarchical structure allows us to integrate over our uncertainty about true clade probabilities while estimating the autocorrelation factor $R$.

The probability model gives the following:
\begin{align}
    \prob\big(\floor{ RK_i } \big| C \big)
    	&= {\floor{ R\ell_i } \choose \floor{ RK_i }} C^{\floor{ RK_i }} (1-C)^{\floor{ R\ell_i } - \floor{ RK_i }} \\
  	\prob(C)
    	&= \frac{C^{a-1} (1-C)^{b-1}}{B(a,b)}
\end{align}

\begin{figure}[tbh]
  \centering
  \begin{subfigure}{0.34\textwidth}
  \includegraphics[trim={0 0 12.8cm 0.1cm},clip,height=3.6cm]{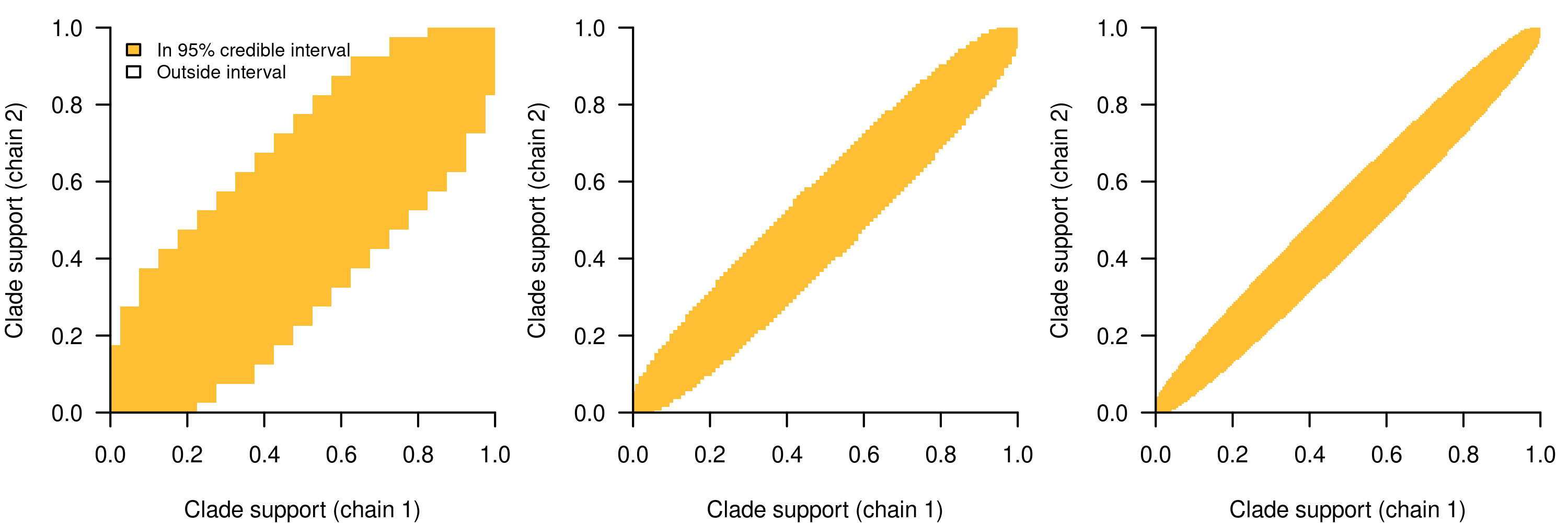}
  \caption{ESS $= 20$}
  \end{subfigure}
  \begin{subfigure}{0.31\textwidth}
  \includegraphics[trim={6.8cm 0 6.4cm 0.1cm},clip,height=3.6cm]{figs/EDCF.png}
  \caption{ESS $= 100$}
  \end{subfigure}
  \begin{subfigure}{0.31\textwidth}
  \includegraphics[trim={13.2cm 0 0 0.1cm},clip,height=3.6cm]{figs/EDCF.png}
  \caption{ESS $= 200$}
  \end{subfigure}
  \caption{Each panel shows the 95\% credible intervals in clade support probabilities under the EDCF strategy, given a fixed ESS. Larger ESSes give smaller sampling errors across the $m=2$ independent chains. This sampling error can be used to estimate the ESS using both maximum likelihood and Bayesian inference approaches.}
  \label{fig:EDCF}
\end{figure}

This strategy is depicted in \cref{fig:EDCF} for a fixed ESS.

\noindent\textit{Likelihood for Two Chains.} Consider the case of two independent chains ($m = 2$). By integrating out clade probability $C$ and accounting for the combinatorial relationship between a count $K_i$ and a rounded fraction of that count $\floor{ RK_i}$, we obtain the joint probability of $K_1$ and $K_2$:
\begin{align}
\prob\big(K_1, K_2 \big| R \big)
	&= \int\limits_0^1 \frac{\prob\big(\floor{ RK_1 } \big| C \big)}{\mathcal{N}(K_1,R)}
					\; \frac{\prob\big(\floor{ RK_2 } \big| C \big)}{\mathcal{N}(K_2,R)}
					   \prob( C ) \; dC \nonumber \\
    &= {\floor{ R\ell_1 } \choose \floor{ RK_1 }}{\floor{ R\ell_2 } \choose \floor{ RK_2 }}
    \frac{B(f_1, f_2 )}{\mathcal{N}(K_1,R) \; \mathcal{N}(K_2,R) \; B(a,b)}, \\
    \text{where }
    f_1 &= \floor{ RK_1 } + \floor{ RK_2 } + a\text{,}\nonumber\\
    f_2 &= \floor{ R L_1 } + \floor{ R L_2 } - \floor{ R K_1 } - \floor{ R K_2 } + b\text{,}\nonumber\\
    \text{and } B(x,y) &= \int\limits_0^1 q^{x-1} (1-q)^{y-1} \; dq = \frac{\Gamma(x) \; \Gamma(y)}{\Gamma(x+y)}\text{ is the beta function.}\nonumber
\end{align}

The combinatorial term $\mathcal{N}(k,r)$ accounts for the discrete nature of the floor operation, counting the total number of integers $z$ such that $\floor{ zr } = \floor{ kr }$, for $0 < r \leq 1$.
That is, find the number of values in $z \in \mathbb{Z}$ such that $x \leq zr < x + 1$, where $x = \floor{ kr }$.
This count can be directly solved as:
\begin{align}
    \mathcal{N}(k,r) = \floor[\Big]{ \frac{x+1}{r} } - \ceil[\Big]{ \frac{x}{r} }  +
    \begin{cases}
        0 \text { if }  \frac{x+1}{r}  \in \mathbb{Z}\\
        1 \text { if }  \frac{x+1}{r}  \notin \mathbb{Z}
    \end{cases} \text{where } x = \floor{ kr }
\end{align}

Then, given two vectors of $n_c$ clade frequencies $(\vec K_1) = (K_{1,1}, \ldots, K_{1,n_c})$ and $(\vec K_2) = (K_{2,1}, \ldots, K_{2,n_c})$, and assuming independence between observations, we evaluate the joint likelihood. To avoid the intractably large space of unobserved clades, we condition on observing each clade at least once across both chains, i.e., $K_{(1,j)} + K_{(2,j)} > 0$:
\begin{align}
	L\big( \vec{K}_1, \vec{K}_2 \big| R \big)
		&= \prob\big( \vec{K}_1, \vec{K}_2 \big| R,  \vec{K}_1 > 0 \land \vec{K}_2 > 0 \big) \nonumber \\
		&= \prod\limits_{j=1}^{n_c} \frac{ \prob\big(K_{1,j}, K_{2,j} \big| R \big)}{{1 - \prob\big( K_{1,j}=0, K_{2,j}=0 \big| R \big)}}.
\end{align}

A key limitation in this arises from the assumption of independence between clade counts, when in reality clades share hierarchical relationships.
Additionally, the proportion $R$ describes the fraction of all samples that are independent, yet in practice this term may vary between the two chains.

\noindent\textit{ESS Estimation.} The effective sample size is computed using $\ESS = R (\ell_1 + \ell_2 + \dotso + \ell_m)$. This represents the number of effectively independent samples across all chains, interpreted as the number of independent draws from the posterior needed to achieve the same sampling precision as the $m$ autocorrelated chains. Importantly, this ESS estimate depends only on observed clade frequencies and requires no distance metrics or explicit autocorrelation calculations.

We estimate $R$ either through maximum likelihood of $L( \vec{K}_1, \vec{K}_2 | R)$ or through Bayesian inference with posterior:
\begin{align}
	\prob\big(R \big| \vec{K}_1, \vec{K}_2 \big) \propto L\big( \vec{K}_1, \vec{K}_2 \big| R \big) \prob\big( R \big)
\end{align}

For Bayesian inference, we perform MCMC using a single unidimensional proposal on $R$ -- the interval operator in BEAST 2 \cite{bouckaert2019beast}, with a step size sampled from a Bactrian kernel \cite{yang2013searching}. We set $a = b = 1$, corresponding to a uniform prior distribution of clade support probabilities. This approach estimates not just the mean $\ESS$ but its full posterior distribution, enabling conservative convergence assessment by ensuring that the 95\% credible interval of $\ESS$ exceeds recommended thresholds. The tree ESS across chains is thus estimated by running a comparatively computationally inexpensive auxillary MCMC on $R$, ensuring this auxiliary chain itself achieves an ESS over 500.

We will restrict our focus to the Bayesian MCMC estimator \edcfmcmc(\textit{B} for Bayesian), noting that additional experiments on the maximum likelihood estimator \edcfml can be found in the Supplementary Material.

\subsection{Evaluation Datasets} 
\label{sec:methods:eval}
We evaluate the accuracy and robustness of the tree ESS estimators on both simulated and empirical data sets, building on ideas from Lanfear \etal~\cite{lanfear16estimating},  Magee \etal~\cite{magee23}, and Fabreti and Höhna~\cite{fabreti22}.
For our simulated chains, we aim to establish ground truth by constructing chains with known, controllable ESS values and to ensure samples mimic real world posterior distributions.

Lanfear \etal~\cite{lanfear16estimating} generated samples of 1000 trees on 50 taxa by performing a fixed number of subtree prune \& regraft (SPR) moves~\cite{hein90,bordewich05} between subsequent samples, with the number of moves ranging from 1 to 50. Increasing the number of SPR moves reduces correlation between consecutive samples and therefore raises the ESS estimate. This relationship was recovered by their ESS estimators. However, this approach has two limitations: the true ESS remains unknown, preventing assessment of estimator accuracy, and the samples are generated through a random walk over the entire treespace rather than from a concentrated posterior distribution.

Magee \etal~\cite{magee23} improved the realism of simulated samples by using the 95\% credible set from an inferred posterior tree sample as a \emph{reference distribution}. They simulated MCMC chains within this distribution using nearest neighbour interchange (NNI) moves. They then applied each tested tree ESS estimator. For each simulated chain and estimator, they drew as many i.i.d.\ samples from the reference distribution as the estimated ESS, then compared both samples (simulated chain and i.i.d.\ sample) by estimating clade probabilities, tree probabilities, and majority rule consensus trees. By comparing the standard error of these estimates across many replicates, they evaluated whether the estimated ESS correctly predicted the precision of posterior estimates. While we build on their simulation approach, it is limited by restricting sampling to the discrete set of trees in the original MCMC sample, which itself represents only a sparse sample from the true posterior.

Fabreti and Höhna~\cite{fabreti22} developed a more general evaluation framework that also considered Metropolis-coupled MCMC with chain swapping and adaptive MCMC algorithms. For their baseline evaluation, they generated samples of length $N$ with \textit{known} ESS $\ESS$ by using $\ESS$ many i.i.d.\ samples from a posterior distribution and repeating each roughly ACT $\tau = N / \ESS$ many times.
While this repetition is quite artificial, it may mimic scenarios with concentrated posteriors or low MCMC acceptance rates.
Testing with the ESS ranging from 100 to 1000 and ACT from 1 to 1000, they found that the initial positive sequence estimator (as used by \texttt{Tracer}) is the most robust method with only slight but consistent underestimation; CODA and MCMCSE were less robust and in particular for higher ACT also overestimating the ESS. Following their recommendation and as already mentioned above, we use \texttt{Tracer} for the trace based estimators.

Building on these approaches, we now describe our evaluation datasets and simulation procedures. We first detail the real datasets and testing procedures, then explain our chain simulation methodology.

\subsubsection{Real Datasets.} 
\label{sec:data}
We use the datasets DS1--11, which are popular for benchmarking computational phylogenetic software~\cite{lakner2008efficiency,whidden2015quantifying,meyer2021adaptive,magee23}; see \cref{tbl:ds} for details. For each dataset, we ran a BEAST 2 MCMC chain with a length of 10 million iterations, using a sampling interval of 100 and discarding a pre-burnin of \num{500000} trees, resulting in a final sample of \num{100000} trees per chain. We ran two independent analyses for each dataset. The remaining MCMC setup follows the one in Berling \etal~\cite{berling24automated} and the BEAST 2 XML files can be accessed through the ASM package\footnote{GitHub ASM package \href{https://github.com/rbouckaert/asm/releases/tag/v0.0.1}{\texttt{github.com/rbouckaert/asm}}}.
For the experiments in \cref{sec:stability}, we used the same setup for DS1--11 but increased the sampling rate to oversample on purpose.

Since multimodal posteriors can challenge both MCMC mixing and ESS estimation, we assessed the modality of each tree posterior. We used a subsample of \num{1000} trees from each dataset, computed all pairwise RNNI distances, and applied multidimensional scaling (MDS). The number of clusters reported in~\cref{tbl:ds} are based on visual assessment of the resulting MDS plots, which are available in the Supplementary Material~\ref{asec:mds:cluster}.
We further report in~\cref{tbl:ds} the mean RNNI distance between consecutive draws from a CCD0 based on each dataset (sample size \num{1000}).

\begin{table}[htbp]
    \centering
    \caption[DS Table]{Datasets used for our experiments with $n$ the number of taxa,
    $H$ the phylogenetic entropy\protect\footnotemark,
    the mean RNNI distance given with standard deviation,
  and a visual assessment of multimodality using RNNI-based MDS (see~\cref{sec:multi:mds} and Supplementary Material~\ref{asec:mds:cluster}).}
    \label{tbl:ds}
    \begin{tabular}{cccS[table-format=3.2]S[table-format=2.2(2)]ccc}
      \toprule
        Dataset & Description & \phantom{hih}$n$\phantom{ihi} & {\phantom{hih}$H$\phantom{ihi}}
              & {mean RNNI dist.} & \phantom{xx}\# Modes\phantom{xx} & Reference \\
      \midrule
        \texttt{DS1} & Tetrapods &\phZ27 &  5.7 &   8.75 (4.00) & 2 & \cite{DS1-hedges90} \\
        \texttt{DS2} & Wheel animals &\phZ29 &  5.2 &   4.74 (2.52) & 4 & \cite{DS2-garey96} \\
        \texttt{DS3} & Mouse lemurs  &\phZ36 &  5.1 &   3.9  (1.91) & 2 & \cite{DS3-yang03} \\
        \texttt{DS4} & Parasitic fungi  &\phZ41 &  5.7 &   7.46 (4.16) & 2 & \cite{DS4-henk03} \\
        \texttt{DS5} & Butterflies  &\phZ50 & 26.4 &  42.33 (10.69)& 1 & \cite{lakner2008efficiency} \\
        \texttt{DS6} & Sac fungi  &\phZ50 & 12.8 &  30.19 (11.11)& 2 & \cite{DS6-zhang01} \\
        \texttt{DS7} & Mouse lemurs &\phZ59 & 10.5 &   8.41 (2.78) & 1 & \cite{DS7-yoder04} \\
        \texttt{DS8} & Sac fungi  &\phZ64 & 16.1 &  27.91 (9.78) & 1 & \cite{DS8-rossman01} \\
        \texttt{DS9} &  Grasses &\phZ67 & 37.7 &  71.17 (11.38)& 1 & \cite{DS9-ingram04} \\
        \texttt{DS10}& Sac fungi  &\phZ67 & 15.0 &  32.31 (12.93)& 2 & \cite{DS10-suh99} \\
        \texttt{DS11}& Green algae &\phZ71 & 46.7 & 123.28 (15.68)& 1 & \cite{DS11-kroken00} \\
      \bottomrule
    \end{tabular}
\end{table}
	\footnotetext{The \emph{phylogenetic entropy} of a tree distribution $D$ is defined as the sum of $-\prob(T) \log \prob(T)$ over all trees of $D$,
	here computed with a CCD1~\cite{klawitter24rogue}.}

\subsubsection{Robustness and Stability Testing on Real Datasets.} 
While the real datasets do not provide ground truth for ESS values, we used them to assess estimator robustness and stability through two complementary tests.

\noindent\textit{Robustness to partitioning.} We used chains of \num{10000} tree samples from each dataset and partitioned each into $k$ equally-sized fragments for $k \in \set{1, 2, \ldots, 10}$. We then applied each tree ESS estimator to every fragment and summed the resulting estimates. A robust estimator should yield similar total ESS values regardless of how the chain is fragmented, resulting in low variance across different values of $k$. However, as we demonstrate below, poor mixing and multimodality can substantially affect this stability.
To concisely summarise these analyses, we report tables that contain summary statistics of the additivity across different chain fragmentations.
Columns correspond to datasets and rows to ESS estimators.
Each table entry reports the mean and standard deviation of the absolute pairwise differences in the total ESS obtained from different numbers of fragments.
Specifically, for each estimator and dataset, we compute the summed ESS values $S_k$ for fragmentations into $k \in \set{1, \ldots, 10}$ equally sized subsets, and then consider all $\binom{10}{2}$ absolute differences $\abs{S_i - S_j}$ for $i < j$.
The reported mean and standard deviation are normalised by the maximum observed summed ESS for the corresponding dataset and estimator.
Here we restrict the table to a subset of estimators and complete tables can be found in the Supplementary Material~\ref{asec:summary:robustness}.
These values quantify how stable the total ESS remains under different numbers of fragments.

\noindent\textit{Stability under oversampling.} We ran MCMC analyses with intentional oversampling by reducing the sampling interval from 100 to 10 iterations while maintaining all other parameters, yielding chains of \num{1000} trees with higher autocorrelation. We estimated tree ESS on the full chain and on systematically thinned versions (retaining every 2nd, 4th, 8th, and 16th tree) to assess whether estimators remain stable under different levels of thinning.

\subsubsection{Accuracy Evaluation on Simulated Chains.} 

Building on ideas from Fabreti and Höhna~\cite{fabreti22} and Magee \etal~\cite{magee23}, we simulate chains with known bounds on ESS and ACT.
We first construct a CCD1 based on the posterior tree sample from the real dataset DS1--11 as reference distribution.
We sample $k = 1000$ independent trees from this CCD1, which establishes our maximum tested ESS value; let $T_i$ denote the $i$th sampled tree for $i \in \set{1, \ldots, k}$.
If the target ACT is $\tau = 1$, then the independent samples form the chain directly. For $\tau > 1$, we introduce autocorrelation by filling the gaps between independent samples using one of the following methods:

\begin{description}
 	\item[Simple Repetition.] Repeat each tree $T_i$ exactly $\tau$ times before proceeding to $T_{i+1}$.
	\item[Noisy Repetition.] For each $i \in \set{1, \ldots, k}$, sample $\tau_i$ from a normal distribution with mean $\tau$ and variance such that 90\% of values lie within $[0.75\tau, 1.25\tau]$, rounding to the nearest integer. Build the chain by repeating each tree $T_i$ exactly $\tau_i$ times.
	\item[Shortest RNNI Path.] As in noisy repetition, first sample $\tau_i$ for each $i \in \set{1, \ldots, k}$. To mimic MCMC movement through treespace, compute the shortest RNNI path $P_i$ from $T_i$ to $T_{i+1}$ (with indices taken modulo $k$). Fill the chain between $T_i$ and $T_{i+1}$ with $\tau_i - 1$ trees sampled uniformly at random from $P_i$.
	We call this an \emph{RNNI chain}.
\end{description}

\noindent\textit{Accuracy of estimators.} To evaluate estimators at a target ESS of $k' < k$, we truncate the simulated chain after tree $T_{k'}$.
For repetition chains, the tree ESS is exactly $k'$, proving ground truth. For RNNI chains, the ESS is at least $k'$, since the random sampling along paths may introduce some additional independence;
the number of independent samples $k'$ thus constitutes a lower bound. How tight this bound is depends on the mean RNNI distance between independent samples in the reference distribution:
small mean distances lead to more repeated trees along paths, making the true ESS closer to $k'$, particularly at small ACT values.
We report average RNNI distances for our datasets in~\cref{tbl:ds}.

\noindent\textit{Conformity of simulated and empirical chains.} To assess how well our simulated chains correspond to empirical MCMC runs, we compare the autocorrelation sequences $\rho_k$ at different lags using Tracer:
The ACT can be expressed as the sum of autocorrelation over increasing lags,
\begin{equation}
	\tau = 1 + 2\sum_{k=1}^\infty \rho_k,
\end{equation}
which forms the basis of the initial positive sequence estimator. We call the autocorrelation sequence truncated at the first negative value, following the stopping criterion of the initial positive sequence estimator, the \emph{autocorrelation signature} of that trace.

Finally, we note a computational consideration: computing shortest RNNI paths can be done in polynomial time~\cite{collienne2021computing}, whereas computing shortest NNI paths is NP-hard.
For RNNI path computation, we must specify rankings for each tree $T_i$, which we derive from common ancestor heights of clades, as our CCD model does not explicitly represent vertex heights or rankings.

\noindent\textit{Summarizing Accuracy.} Using our simulated chains, for which the true ESS (or at least a lower bound) is known, we can directly assess the accuracy of different ESS estimators.
To concisely summarize these evaluations, we employ the \emph{mean relative error (MRE)}, which quantifies the deviation of the estimated ESS values from the expected ESS (lower bound).
Formally, for each dataset and ACT value, we compute
$$ \textup{MRE} = \frac{1}{n} \sum_{i=1}^{n} \frac{(Y_i - \hat{Y}_i)}{\hat{Y}_i} $$
where $Y_i$ denotes the ESS estimated by a given method and $\hat{Y}_i$ is the corresponding expected ESS.

\subsubsection{Simulated Datasets.} 

In addition to the real datasets, we include the \texttt{Yule50} datasets from Berling \etal~\cite{berling2025accurate} that serve as a controlled baseline with well-behaved posterior distributions.
These datasets were simulated using LinguaPhylo~\cite{drummond23linguaphylo} under a Yule model with $50$ taxa. For each dataset, we ran two independent MCMC replicates in BEAST 2~\cite{bouckaert2019beast}, sampling \num{10000} trees after burn-in. The inference used a birth-death prior with a log-normal distribution (mean~3.21, s.d.\ 0.3 in log space) and a HKY+G substitution model~\cite{hasegawa1985dating}. The shape parameter for the gamma distribution of site rates was modelled using a log-normal distribution (mean ${-}$1.0, s.d.\ 0.5 in log space). The transition/transversion rate ratio ($\kappa$) also followed a log-normal distribution (mean 1.0, s.d.\ 1.25 in log space). The nucleotide base frequencies were independently simulated for each replicate from a Dirichlet distribution with a concentration parameter array of [5.0, 5.0, 5.0, 5.0]. The length of the sequence alignments was 300 sites and the mutation rate was fixed at 1.0, so that divergence ages were in units of substitutions per~site.

\section{Results} 
\label{sec:results}
We first assess our chain simulation methods by comparing their autocorrelation structure to that of real MCMC chains.
We then evaluate estimator accuracy on simulated chains with known ESS, validate stability with respect to oversampling, and test robustness under partitioning.
Finally, we examine the implications of multimodality and between-mode mixing.
We consider the following ESS estimators: the trace-based estimators \logP, \medianRF, \mapRF, \expRF, and \cladeInd, the Fréchet variance-based \frechess, and the new probabilistic method \edcfmcmc.

Throughout this section, we pay special attention to two contrasting case studies: DS3 and DS4.
These datasets were selected because the ESS estimators behave well for DS3 but poorly for DS4.
As we show later (\cref{sec:multi:mds}), both datasets appear bimodal, which offers an explanation to the differential findings of estimator behaviour.
To validate our finding being dataset related, we also evaluate accuracy and robustness on simulated Yule data, which is expected to follow a well-behaved unimodal distribution.
Summary results across all datasets are presented alongside additional plots in the Supplementary Material.

\subsection{Simulated Chains} \label{sec:simulated}
To assess how closely our simulated chains mimic the autocorrelation structure of real MCMC chains, we analyzed the autocorrelation signatures of selected~traces.

First, we selected a sub-chain from an MCMC chain by finding the smallest multiple of \num{1000} samples that would have its ESS estimate at least 200.
For the simulated chains, we use the subsample of trees that contains 200 i.i.d.\ trees from the CCD and a target ACT of 25.
As our main reference, we used the posterior trace of the MCMC chains and for all simulated chains we considered the expected RF distance and \logP{}1 traces.
The results for DS3 are shown in \cref{fig:signature} and of the others in the Supplementary Material~\ref{asec:autocorr:signature}.

Considering first the autocorrelation signatures from the real reference MCMC chains, we observe that these chains exhibit substantial variation and irregular fluctuations across lags.
Simple repetition chains produce autocorrelation signatures consisting of flat line segments with sharp breakpoints at the target ACT value.
Noisy repetition chains smooth these transitions slightly but still show clear linear segments and visible breakpoints.
RNNI chains reduce the prominence of breakpoints and introduce more variable behaviour across lags, though breakpoints remain visible.
Overall, while none of our simulation methods fully replicate real MCMC behaviour, RNNI chains represent a clear improvement, with autocorrelation signatures that more closely resemble those of real chains.

\begin{figure}[tbh]
	\centering
	\includegraphics[trim={0 0 0 1.1cm},clip,width=\linewidth]{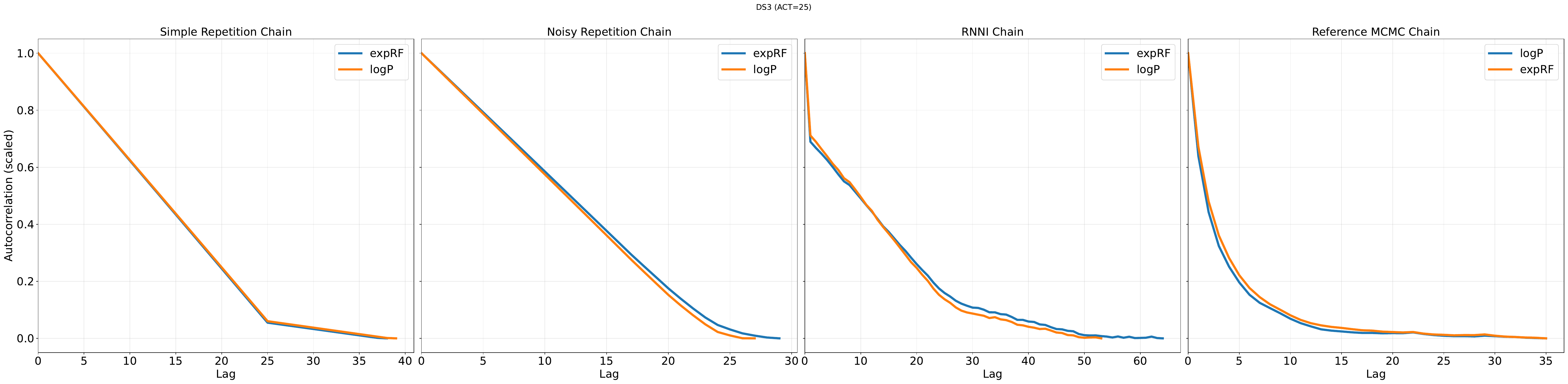}
	\caption{Scaled autocorrelation signatures for \texttt{DS3} of the \logP{}1 and \expRF{}1 traces
	for the three chain simulation methods with a target ACT of 25,
	and the posterior and $\kappa$ (transition/transversion rate ratio) traces for the corresponding MCMC~chain.}
	\label{fig:signature}
\end{figure}

\subsection{Estimator runtime}\label{sec:runtime}

Before comparing the estimators, we assess their runtimes on a set of four simulated RNNI chains from DS1.
We set the maximum number of i.i.d.\ samples to 200 and use an ACT of 5, resulting in a total chain length of approximately 1000 samples.
For \edcfmcmc, we impose a runtime limit of 60 seconds, which is consistently reached even for these relatively short chains.
All times were measured on a 3.1 GHz 6-Core Intel Core i5 iMac and are presented for comparison only, not as a benchmark.

We observe, as expected, that most estimators scale with the number of samples.
An exception is \edcfmcmc, which estimates the ESS from clade frequencies and therefore depends primarily on the number of observed clades rather than the chain length.
However, this scaling behaviour is not apparent in our experiments, as \edcfmcmc consistently reaches the imposed runtime limit and returns an ESS estimate before its MCMC chain has converged.
Overall, \edcfmcmc is the slowest estimator due to the computational cost of running an MCMC chain.
It is followed by the \frechess estimator, which was roughly \num{750}--\num{4240} times slower than the other estimators on these test data, underscoring our rationale for the imposed time limitation (\autoref{sec:tree:adaptation}).

\begin{figure}[tbh]
	\centering
	\begin{subfigure}{0.55\textwidth}
		\includegraphics[width=\linewidth]{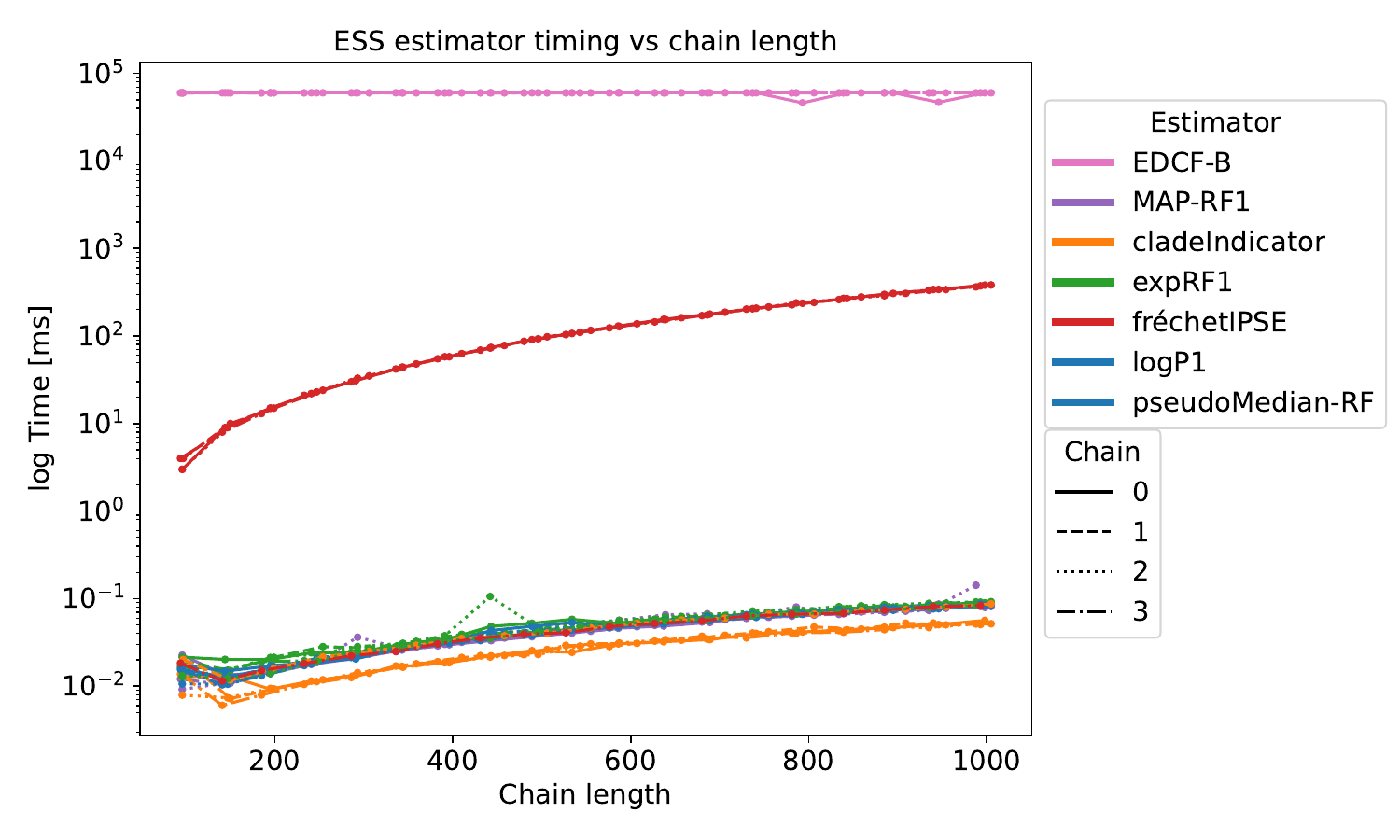}
		\caption{Runtime vs step}
		\label{fig:runtime}
	\end{subfigure}
	\hfill
	\begin{subfigure}{0.4\textwidth}
		\centering
		\resizebox{\linewidth}{!}{
			\begin{tabular}{lrrr}
\toprule
Estimator & $\approx 200$ & $\approx 500$ & $\approx 1000$ \\
\midrule
fréchetIPSE & 15.00 & 93.00 & 382.00 \\
pseudoMedian-RF & 0.02 & 0.04 & 0.08 \\
cladeIndicator & 0.01 & 0.02 & 0.06 \\
MAP-RF1 & 0.02 & 0.05 & 0.08 \\
expRF1 & 0.02 & 0.05 & 0.09 \\
logP1 & 0.02 & 0.04 & 0.08 \\
\bottomrule
\end{tabular}

		}
		\caption{Average running times (in ms) at selected chain lengths}
		\label{tab:runtime-table}
	\end{subfigure}
	\caption{Timing comparison across ESS estimators on four simulated chains over different number of samples.}
	\label{fig:runtime:comparison}
\end{figure}

Thus, while most estimators showed comparable runtimes, these experiments suggest that \edcfmcmc and \frechess are less suitable for performing hundreds to thousands of such computations, with the latter becoming increasingly costly for longer chains.
The full comparison across different chain lengths is shown in~\cref{fig:runtime:comparison}.

\subsection{Accuracy} \label{sec:accuracy} 
For each of the eleven datasets, DS1--11, we used two independent MCMC replicates, obtaining a CCD as the reference distribution from each replicate.
For each reference distribution and each target ACT (1, 2, 5, 10, 25, 50, 75, 100), we generated two simulated chains using the simple, noisy and RNNI methods.
For each dataset, each target ACT, and each chain simulation method, this yielded four simulated chains; two from each of the two independent MCMC replicates.
Because of computational limitations we were only able to compute the accuracy on simulated RNNI chains for up to an ACT value of 25.

\begin{figure}[htb]
	\centering
	\begin{subfigure}{\textwidth}
		\centering
		\includegraphics[clip,trim={0 0.5cm 0 1cm},width=0.9\linewidth,page=3]{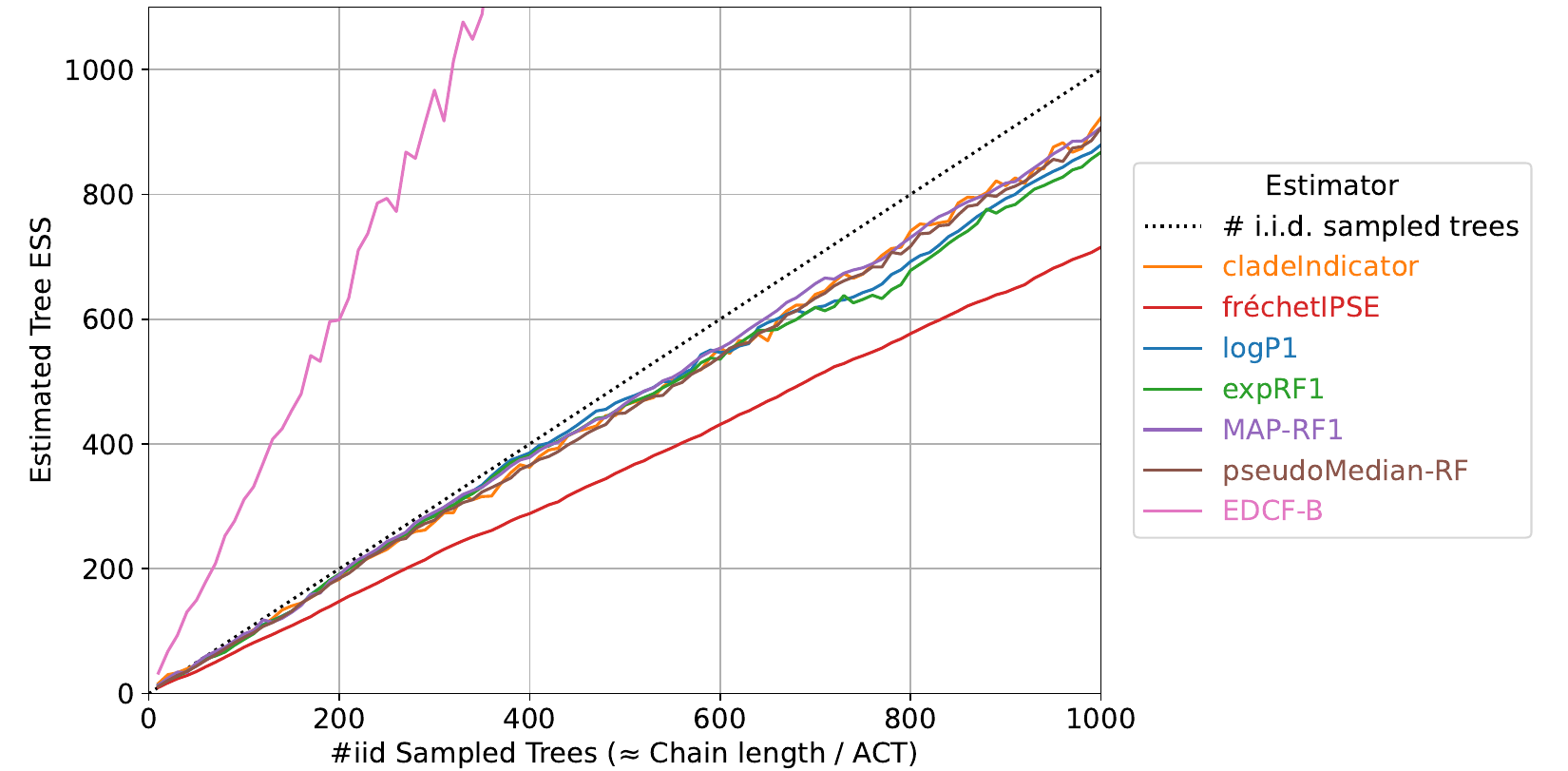}
		\caption{DS3}
	\end{subfigure}
	\begin{subfigure}{\textwidth}
		\centering
		\includegraphics[clip,trim={0 0.5cm 0 1cm},width=0.9\linewidth,page=4]{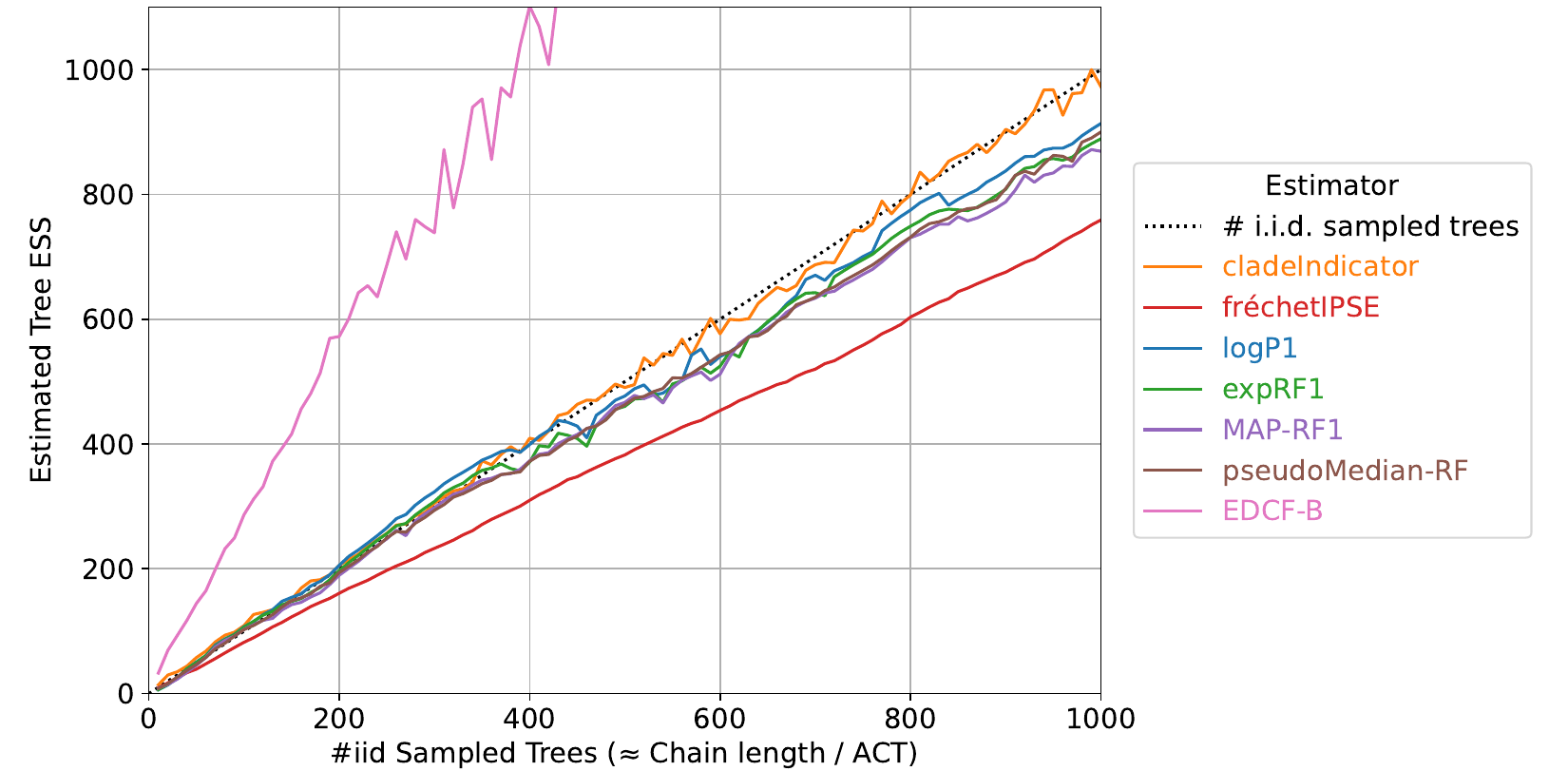}
		\caption{DS4}
	\end{subfigure}
	\caption{Estimated tree ESS for DS3 and DS4 on simulated RNNI chains with an underlying ACT of 5 across different chain lengths and different ESS estimators. Estimator accuracy is measured as the deviation from the diagonal (lower true ESS bound); summary results of this accuracy evaluation are shown in \cref{fig:accuracy:boxes:ds}.}
	\label{fig:acc_results_xy}
\end{figure}

Due to our simulation procedure, we have a known lower bound on the true ESS in these simulated chains given by the number of independent samples.
This allows us to assess the accuracy of different estimators by comparing this bound with the estimated ESS values, as shown in \cref{fig:acc_results_xy}.
We observe that all estimators except \edcfmcmc lie close to the diagonal, which represents the expected number of independently sampled trees.
For DS3, all estimators (except \edcfmcmc) underestimate the ESS, with \frechess exhibiting the strongest underestimation but with low variance.
A similar pattern is observed for the DS4 simulated chains, where the \cladeInd estimator is closest to the diagonal.

\begin{figure}[htb]
	\centering
	\begin{subfigure}{\textwidth}
		\centering
		\includegraphics[clip,trim={0 0.5cm 0 1cm},width=0.9\linewidth,page=3]{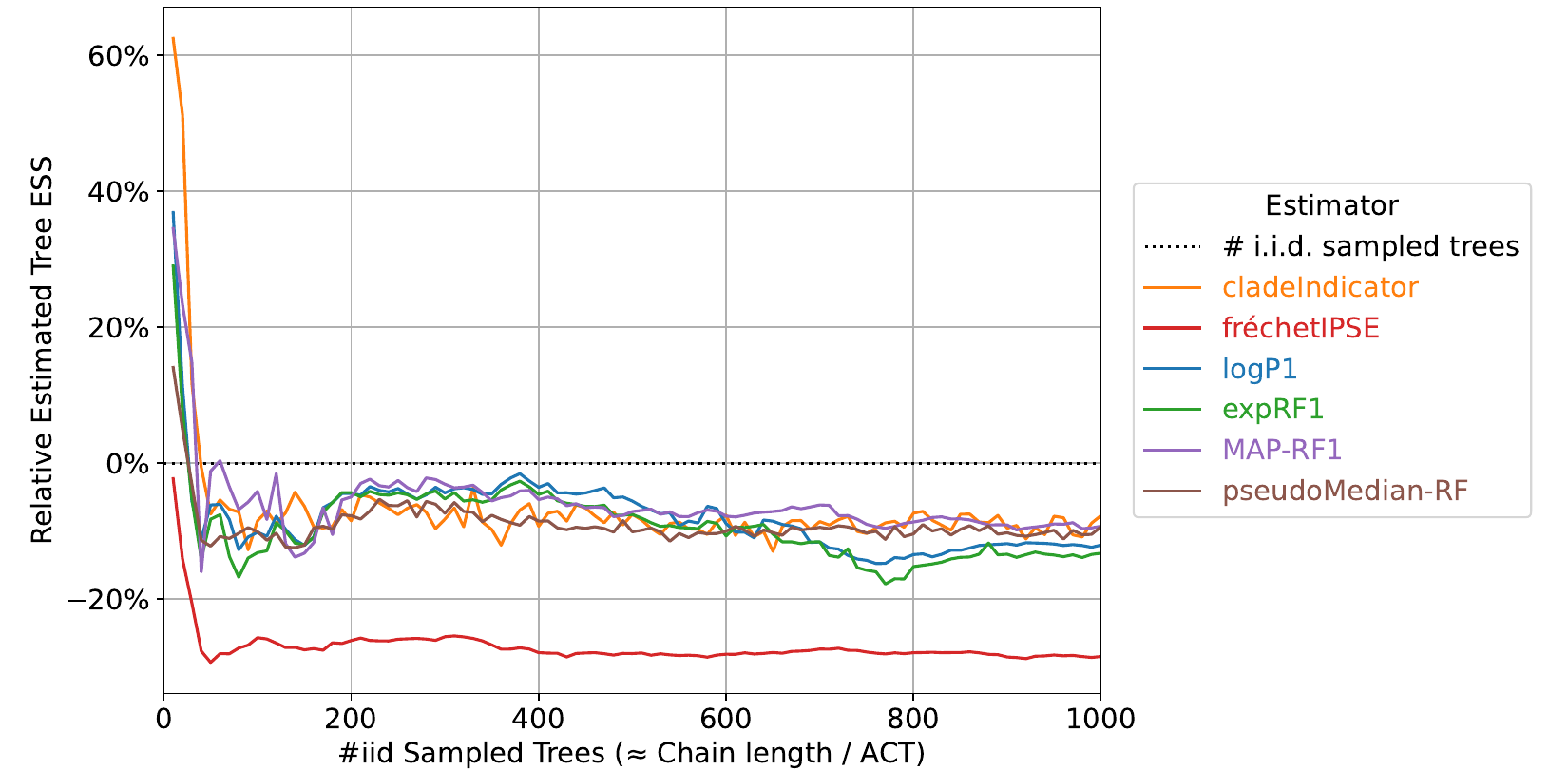}
		\caption{DS3}
	\end{subfigure}
	\begin{subfigure}{1\textwidth}
		\centering
		\includegraphics[clip,trim={0 0.5cm 0 1cm},width=0.9\linewidth,page=4]{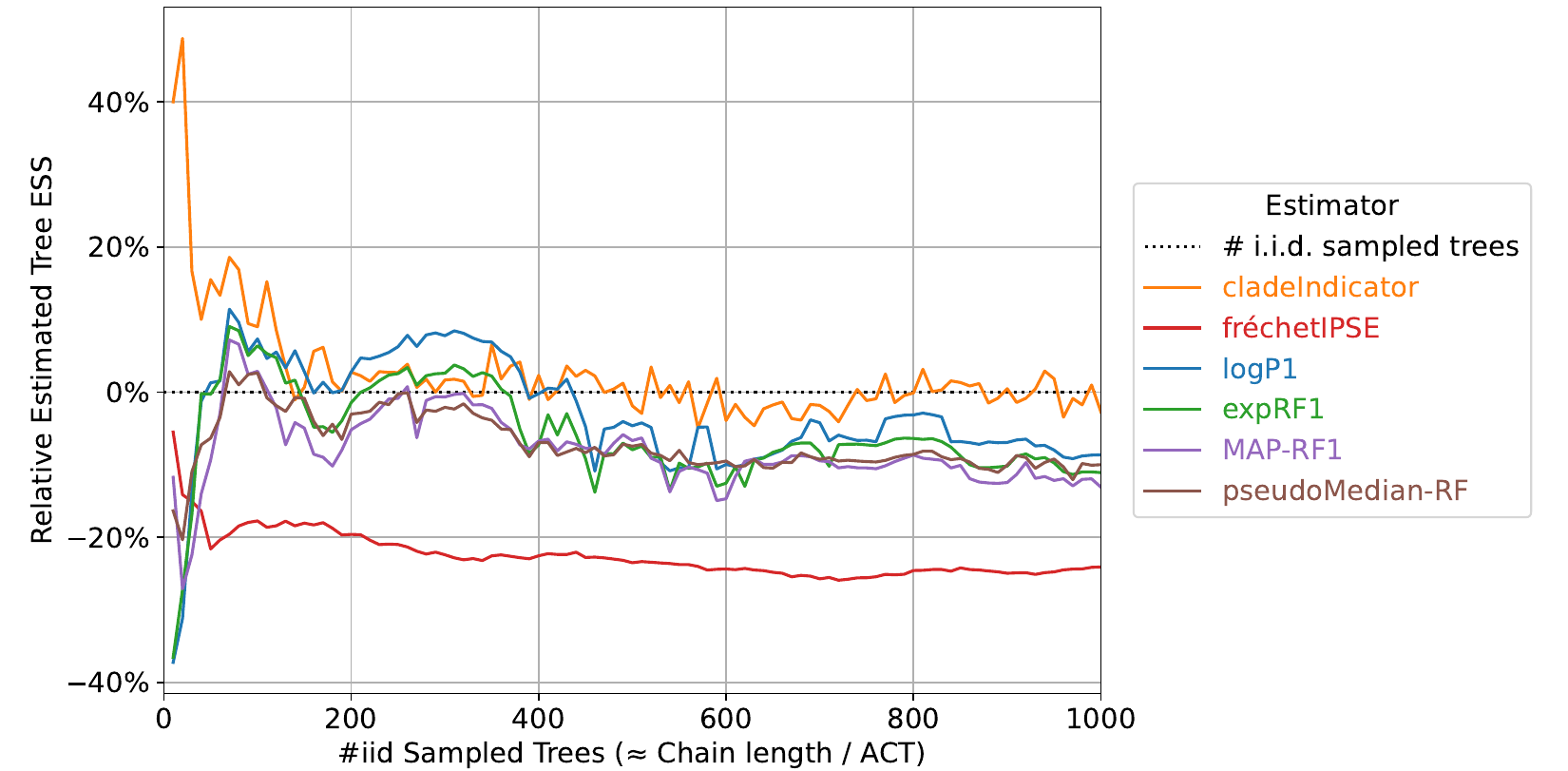}
		\caption{DS4}
	\end{subfigure}
    \caption{Relative estimated tree ESS for DS3 and DS4, showing the same data as \cref{fig:acc_results_xy} but normalised by the expected lower bound. The \edcfmcmc estimator is omitted due to its comparatively high values.}
	\label{fig:acc_results_xy:rel}
\end{figure}

In addition to these absolute estimates, we examine the relative estimates in \cref{fig:acc_results_xy:rel},
defined as the ratio between the estimated ESS and the lower bound (i.e., number of independent samples).
Here, we omit the \edcfmcmc estimator to facilitate comparison among the remaining methods.
Again, \frechess shows a consistent underestimation, while the other estimators exhibit greater variability, with no single method consistently outperforming the others in terms of proximity to the expected value.

We emphasise that the diagonal (or, in the case of the relative evaluation, the zero line) represents a lower bound on the ESS.
The RNNI simulation process may generate additional trees that effectively act as independent samples.
However, we do not expect this process to introduce a large number of new effectively independent samples when the RNNI distance is small.
In the case of the two datasets presented here, the average RNNI distance is below 10, and we therefore expect the true number of effectively independent samples to remain close to the diagonal.
Consequently, the overestimation observed for the \edcfmcmc method may only partially be attributed to a higher true ESS and more likely represents true overestimation.

\subsubsection{Summarizing Accuracy.}

We calculate the mean relative error between the estimated and expected lower-bound ESS values, as displayed in \cref{fig:acc_results_xy}.
An MRE of zero indicates that an estimator matches the expected lower bound on average, while a positive MRE signals overestimation and a negative MRE indicates underestimation.
Because the expected ESS value $\hat{Y}_i$ ranges from 10 to 1000, we use the relative error to compress the evaluation for all datasets and expected ESS values given a fixed ACT value.
Overall, we would expect a well-performing estimator to have an MRE close to zero (or slightly below, as the initial positive sequence estimator method consistently underestimates~\cite{fabreti22}),
reflecting accurate or slightly lower estimates of the ESS.

\begin{figure}[tbh]
  \centering
  \includegraphics[width=\linewidth]{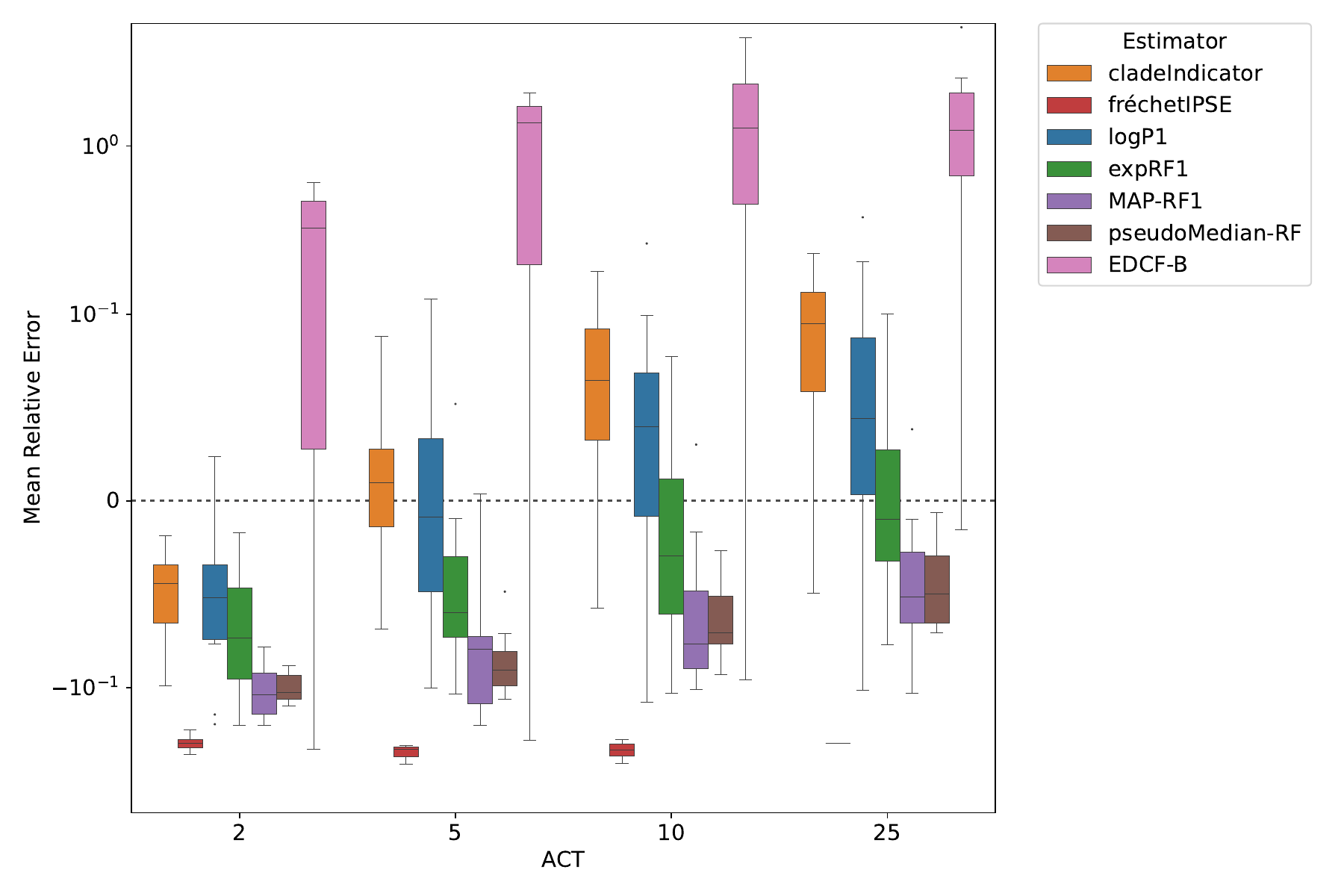}
  \caption{Summary of estimator accuracy on simulated RNNI chains for all eleven datasets. Performance is quantified using the relative mean error, taking into account the known ESS bound. Results are only reported up to an ACT of 25 due to computational limitations.}
  \label{fig:accuracy:boxes:ds}
\end{figure}

Across the eleven real datasets, we observe that the estimated ESS generally increases with larger ACT values (\cref{fig:accuracy:boxes:ds}).
Besides \edcfmcmc consistently overestimating, \cladeInd provides the next highest estimates, along with the \logP{}1 and \expRF{}1 estimators. Note that these show substantial overlap of their interquartile ranges.
The \mapRF{}1 and \medianRF estimators produce very similar estimates across all ACT values, while the \frechess consistently yields the lowest ESS estimates, albeit with a high consistency across ACT values.

\begin{figure}[tbh]
  \centering
  \includegraphics[width=\linewidth]{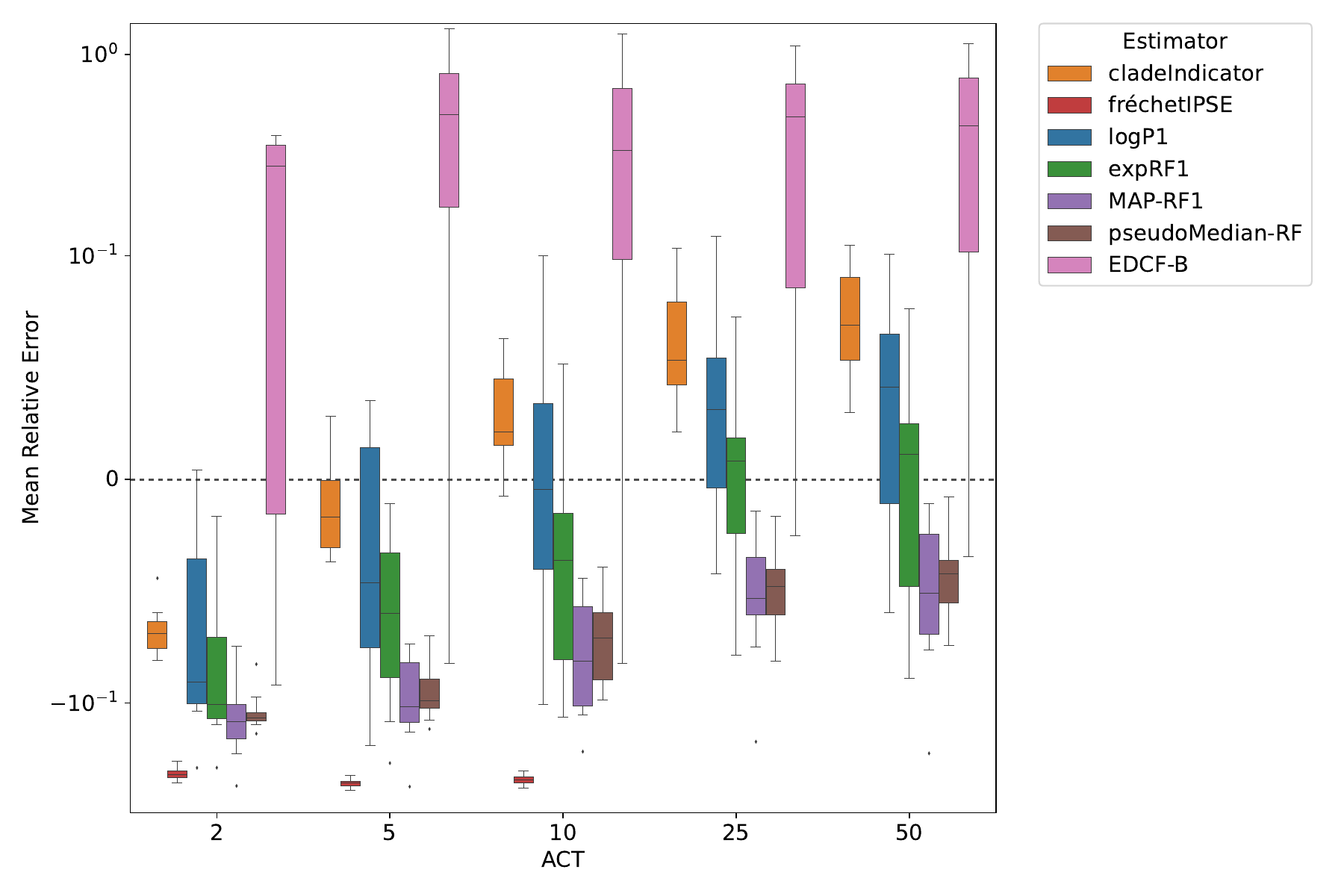}
  \caption{Summary of the estimator accuracy on the simulated RNNI chains for all 10 Yule datasets. Results are only reported up to an ACT of 50 due to computational limitations.}
  \label{fig:accuracy:boxes:yule}
\end{figure}

To establish a baseline for estimator performance, we further evaluate accuracy on simulated Yule datasets,
which we assume to represent well-behaved posterior distributions without multimodality or mixing issues (\cref{fig:accuracy:boxes:yule}).
In this setting, we use the RNNI chain simulation method with ACT values ranging from~2 to 50, constrained by computational limits.
Most estimators follow a similar trend: higher ACT values result in higher ESS estimates.
The \edcfmcmc method continues to show the highest estimate and highest variance.
Furthermore, \expRF{}1 and \logP{}1 exhibit larger variance than other trace based methods.
Overall, this comparison displays an estimator behaviour pattern almost identical to that in \cref{fig:accuracy:boxes:ds} on real datasets.
Due to its high computational cost, \frechess is only reported for ACT values between 2 and 10.

\FloatBarrier
\subsection{Stability} \label{sec:stability}

MCMC samples a chain of states from a regular, but arbitrary, interval, which are then often downsampled even further into a manageable size.
A useful ESS estimator should therefore be stable with respect to the thinning interval (up to a point): thinning an oversampled chain removes redundant, highly autocorrelated trees without discarding effectively independent ones, so estimates should stay roughly constant until thinning begins to remove genuine information.

To evaluate the stability of the different tree ESS estimators, we investigate the variance of the estimates under different thinning intervals.
If an MCMC is oversampled, its output contains large amounts of autocorrelated (i.e., non independent) trees.
Consequently, ESS estimates should remain stable across thinning intervals until thinning begins to remove effectively independent samples.

\begin{figure}[htb]
	\centering
	\begin{subfigure}{0.49\textwidth}
		\centering
		\includegraphics[trim={0 0.3cm 0 1.1cm},clip,width=\linewidth, page=3]{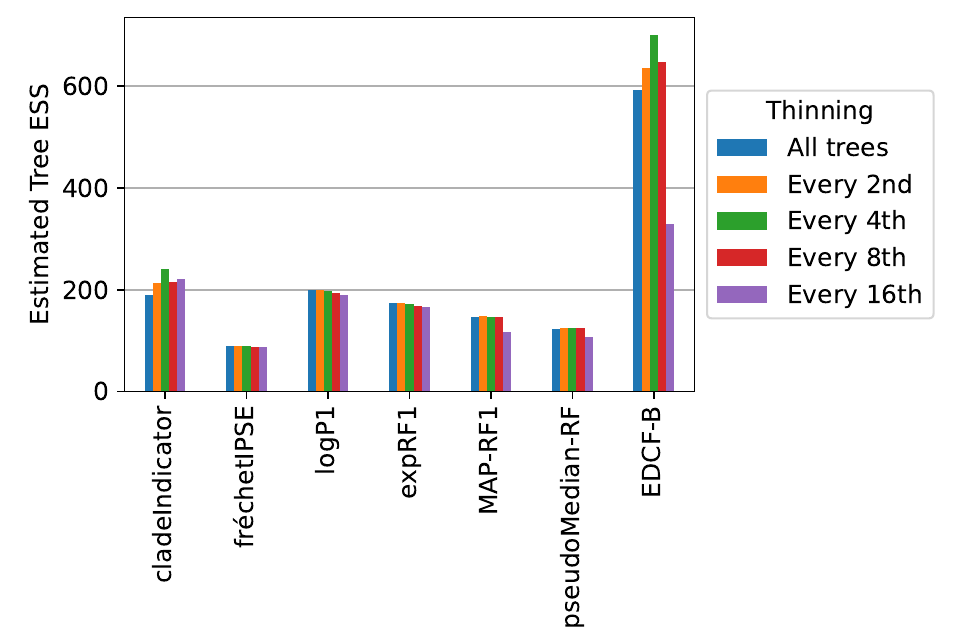}
		\caption{DS3}
	\end{subfigure}\hfill
	\begin{subfigure}{0.49\textwidth}
		\centering
		\includegraphics[trim={0 0.3cm 0 1.1cm},clip,width=\linewidth, page=4]{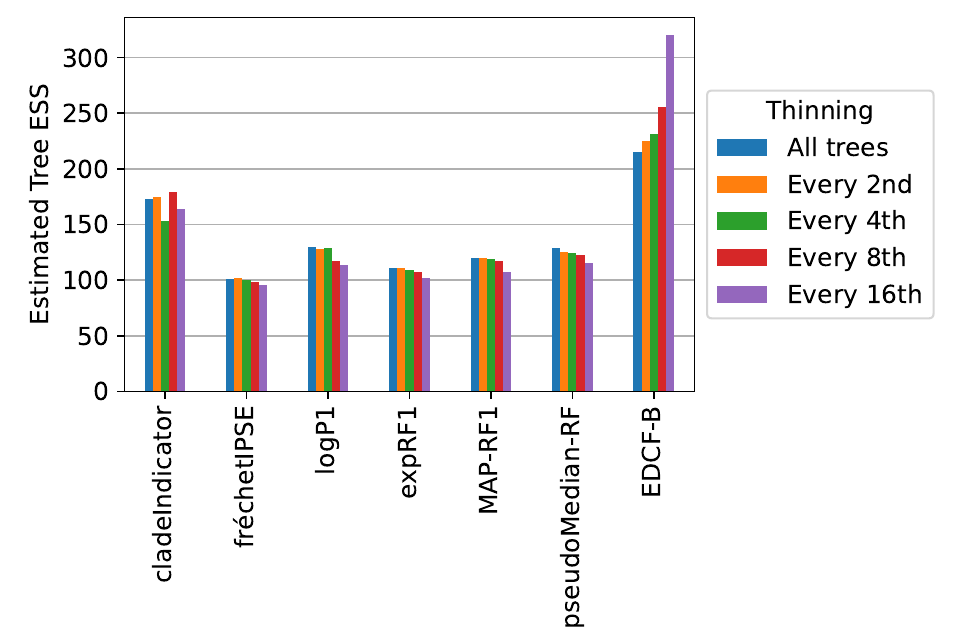}
		\caption{DS4}
	\end{subfigure}
	\caption{Tree ESS estimates for different thinning intervals of oversampled datasets as per our stability experiment.}
	\label{fig:stability}
\end{figure}

In \cref{fig:stability}, we present the raw ESS estimates for DS3 and DS4 under different thinning intervals, namely 0, 2, 4, 8, and 16
when starting with the oversampled chain (cf.\ \cref{sec:data}).
The plots for the all datasets and with more estimators shown can be found in Supplementary Material~\ref{asec:stab}.
All estimators appear to be very stable under oversampling, with the only outlier in terms of stability being the \edcf methods (only \edcfmcmc shown here).
As expected, the other estimators yield slightly lower but very similar estimates at higher thinning intervals.

\subsection{Robustness} \label{sec:robustness}

By combining multiple independent chains, or splitting up one for analysis, it is implicitly assumed that ESS estimates are additive: that is, the total ESS should not depend on how the chain is partitioned.
Here we assess the robustness of different ESS estimators with respect to additivity, by evaluating their behaviour under varying numbers of fragmented sub-chains.
If an estimator is not robust, and has an intrinsically high variance across chain fragments regardless of how~well that chain is mixing, then it will not be useful for identifying convergence issues~\cite{berling24automated}.
For a well-behaved estimator, we expect low variance in the total ESS across different numbers of fragments,
such that the sum of the ESS estimates remain approximately constant.
Substantial deviations from this behaviour may indicate issues such as poor mixing,
multimodality in the underlying distribution, or inadequate removal of burn-in; issues that are prevalent in real analyses~\cite{harrington2021properties,gao2026biological}.

For this evaluation we used chains of \num{10000} trees from each dataset and partitioned each into $k$ equally sized fragments, for $k \in \set{1, 2, \ldots, 10}$.
For each $k$, we estimate the ESS for all fragments and then combined these estimates into a single value.
In this part of the paper, we present the results for DS3 and DS4, \cref{fig:robustness:ds:three:four}, and the remaining datasets are available in the Supplementary Material~\ref{asec:robustness}.
For DS3 we observe a very stable sum of total ESS over all different fragments for all estimators.
However, for DS4 we find high variability among all estimators with a tendency for higher sums of ESS estimates with more fragments, independent of the specific estimator.

From the perspective of the estimators, these results indicate that, with the exception of \edcfmcmc, all methods behave robustly under chain fragmentation when the underlying posterior distribution is well behaved, as in the case of DS3.
In such settings, splitting the chain does not substantially affect the total ESS, suggesting that these estimators are internally consistent with respect to additivity.
Conversely, when the posterior exhibits poor mixing or multimodality (see \cref{sec:multi:mds}), as in DS4, all estimators show similar instability, with increasing fragmentation leading to inflated ESS estimates.
This suggests that the observed lack of robustness is primarily driven by properties of the underlying tree distribution rather than by deficiencies of individual estimators.
Consequently, additivity-based diagnostics are better interpreted as tools for detecting mixing and convergence issues in MCMC tree samples, rather than as a means of distinguishing between ESS estimators.

\begin{figure}[htb]
	\centering
	\begin{subfigure}{\textwidth}
		\centering
		  \includegraphics[trim={0 0.5 0 1cm},clip,width=0.9\linewidth,page=3]{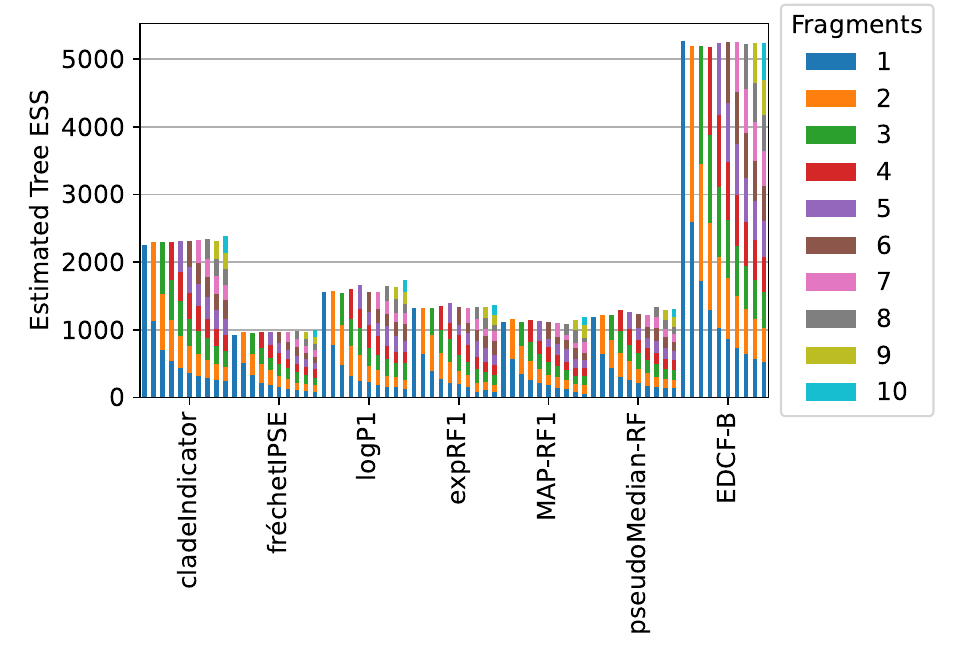}
		\caption{DS3}
	\end{subfigure}
	\begin{subfigure}{1\textwidth}
		\centering
		\includegraphics[trim={0 0.5 0 1cm},clip,width=0.9\linewidth,page=4]{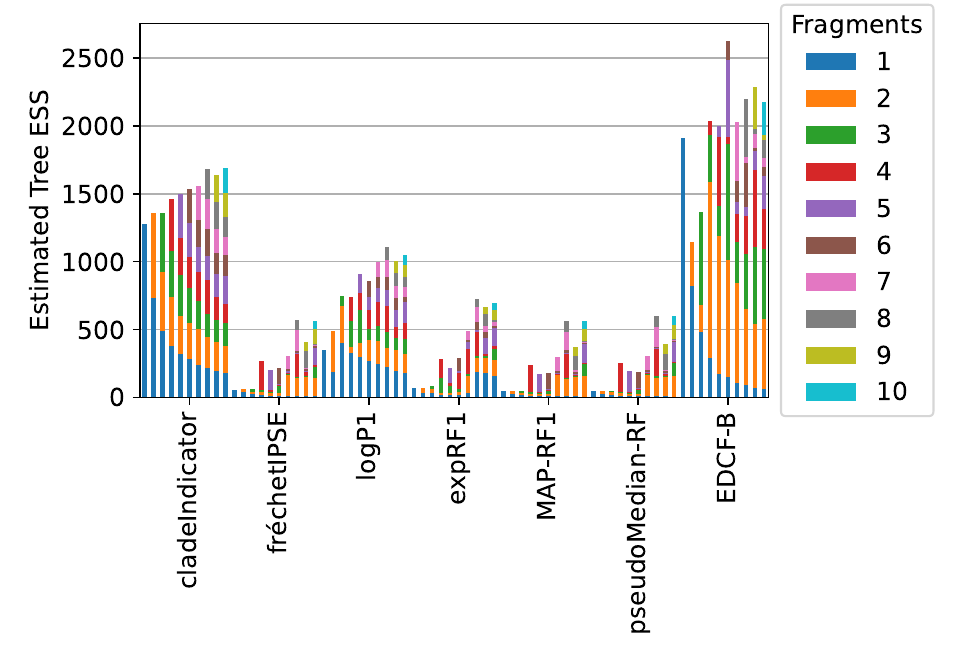}
		\caption{DS4}
	\end{subfigure}
	\caption{Robustness of ESS estimators under different chain fragmentations. The two datasets exhibit markedly different behaviour, while no estimator shows a consistently distinct trend.}
	\label{fig:robustness:ds:three:four}
\end{figure}

\clearpage
\subsubsection{Summarizing Robustness Evaluation.}

The robustness analysis for the eleven empiricial datasets is summarised in \cref{tab:robustness:summary:10k:run1}.
In the context of  the previous stacked histograms ~\cref{fig:robustness:ds:three:four}, these statistics indicate how the total heights of the bars vary for each estimator.
The table is colour-coded to indicate what we consider good behaviour in green (mean and standard deviation below $0.05$), slightly worse performance in yellow (values between $0.05$ and $0.1$), and poor behaviour in red (values above $0.1$).
It is evident from the table that datasets 2, 3, and 7 exhibit well-behaved additivity across all estimators.
In contrast, the remaining datasets mostly show values larger than $0.1$, indicating substantial variation in the summed ESS across different fragmentations.

To verify that these results are driven by the underlying datasets rather than by the estimators themselves, we present the same summary table for the 10 simulated Yule datasets in \cref{tab:yule:robustness:summary:10k:run1}.
As in the accuracy analysis, this experiment serves as a baseline, since these datasets are assumed to represent well-behaved tree distributions without multimodality or mixing issues.
Here we observe that all 10 datasets predominantly yield values below $0.05$, with only the \edcfmcmc estimator exhibiting more pronounced variability.

Overall, the robustness analysis indicates that fragmentation primarily exposes characteristics of the posterior distribution and the underlying sampling behaviour, rather than systematic differences among estimators.
These insights underscore the value of additivity-based diagnostics for identifying convergence and mixing issues in tree MCMC samples, setting the stage for a more detailed investigation in the next section.

\begin{table}[!ht]
	\centering
	\caption{Robustness of ESS estimates across the datasets DS1--11 (10,000 trees), with lower values indicating more robust performance.}
	\label{tab:robustness:summary:10k:run1}
	\resizebox{1\textwidth}{!}{
		\begin{tabular}{lccccccccccc}
			\toprule
			& 1 & 2 & 3 & 4 & 5 & 6 & 7 & 8 & 9 & 10 & 11 \\
			\midrule
			LogP1 & $\colorbox{red!25}{0.21} \pm \colorbox{red!25}{0.16}$ & $\colorbox{orange!25}{0.06} \pm \colorbox{green!25}{0.04}$ & $\colorbox{green!25}{0.04} \pm \colorbox{green!25}{0.03}$ & $\colorbox{red!25}{0.26} \pm \colorbox{red!25}{0.18}$ & $\colorbox{red!25}{0.14} \pm \colorbox{orange!25}{0.09}$ & $\colorbox{red!25}{0.18} \pm \colorbox{red!25}{0.13}$ & $\colorbox{green!25}{0.05} \pm \colorbox{green!25}{0.04}$ & $\colorbox{red!25}{0.19} \pm \colorbox{red!25}{0.12}$ & $\colorbox{red!25}{0.19} \pm \colorbox{red!25}{0.13}$ & $\colorbox{red!25}{0.15} \pm \colorbox{red!25}{0.11}$ & $\colorbox{red!25}{0.17} \pm \colorbox{red!25}{0.12}$ \\
			cladeIndicator & $\colorbox{red!25}{0.18} \pm \colorbox{red!25}{0.12}$ & $\colorbox{green!25}{0.02} \pm \colorbox{green!25}{0.02}$ & $\colorbox{green!25}{0.02} \pm \colorbox{green!25}{0.01}$ & $\colorbox{red!25}{0.10} \pm \colorbox{orange!25}{0.06}$ & $\colorbox{orange!25}{0.09} \pm \colorbox{orange!25}{0.06}$ & $\colorbox{orange!25}{0.09} \pm \colorbox{orange!25}{0.06}$ & $\colorbox{green!25}{0.01} \pm \colorbox{green!25}{0.01}$ & $\colorbox{orange!25}{0.08} \pm \colorbox{green!25}{0.05}$ & $\colorbox{orange!25}{0.08} \pm \colorbox{green!25}{0.05}$ & $\colorbox{red!25}{0.10} \pm \colorbox{orange!25}{0.06}$ & $\colorbox{green!25}{0.02} \pm \colorbox{green!25}{0.02}$ \\
			fréchetIPSE & $\colorbox{red!25}{0.27} \pm \colorbox{red!25}{0.19}$ & $\colorbox{green!25}{0.03} \pm \colorbox{green!25}{0.02}$ & $\colorbox{green!25}{0.02} \pm \colorbox{green!25}{0.02}$ & $\colorbox{red!25}{0.40} \pm \colorbox{red!25}{0.27}$ & $\colorbox{orange!25}{0.09} \pm \colorbox{orange!25}{0.06}$ & $\colorbox{red!25}{0.31} \pm \colorbox{red!25}{0.28}$ & $\colorbox{orange!25}{0.05} \pm \colorbox{green!25}{0.04}$ & $\colorbox{orange!25}{0.05} \pm \colorbox{green!25}{0.04}$ & $\colorbox{orange!25}{0.09} \pm \colorbox{orange!25}{0.06}$ & $\colorbox{red!25}{0.29} \pm \colorbox{red!25}{0.19}$ & $\colorbox{orange!25}{0.05} \pm \colorbox{green!25}{0.03}$ \\
			expRF1 & $\colorbox{red!25}{0.19} \pm \colorbox{red!25}{0.13}$ & $\colorbox{orange!25}{0.06} \pm \colorbox{green!25}{0.04}$ & $\colorbox{green!25}{0.02} \pm \colorbox{green!25}{0.01}$ & $\colorbox{red!25}{0.43} \pm \colorbox{red!25}{0.29}$ & $\colorbox{red!25}{0.16} \pm \colorbox{red!25}{0.11}$ & $\colorbox{red!25}{0.30} \pm \colorbox{red!25}{0.29}$ & $\colorbox{green!25}{0.04} \pm \colorbox{green!25}{0.03}$ & $\colorbox{red!25}{0.20} \pm \colorbox{red!25}{0.13}$ & $\colorbox{red!25}{0.23} \pm \colorbox{red!25}{0.15}$ & $\colorbox{red!25}{0.26} \pm \colorbox{red!25}{0.17}$ & $\colorbox{red!25}{0.27} \pm \colorbox{red!25}{0.20}$ \\
			MAP-RF1 & $\colorbox{red!25}{0.18} \pm \colorbox{red!25}{0.12}$ & $\colorbox{orange!25}{0.06} \pm \colorbox{green!25}{0.04}$ & $\colorbox{green!25}{0.03} \pm \colorbox{green!25}{0.02}$ & $\colorbox{red!25}{0.41} \pm \colorbox{red!25}{0.28}$ & $\colorbox{red!25}{0.12} \pm \colorbox{orange!25}{0.08}$ & $\colorbox{red!25}{0.35} \pm \colorbox{red!25}{0.31}$ & $\colorbox{green!25}{0.05} \pm \colorbox{green!25}{0.03}$ & $\colorbox{red!25}{0.12} \pm \colorbox{orange!25}{0.09}$ & $\colorbox{red!25}{0.16} \pm \colorbox{red!25}{0.11}$ & $\colorbox{red!25}{0.33} \pm \colorbox{red!25}{0.21}$ & $\colorbox{red!25}{0.18} \pm \colorbox{red!25}{0.13}$ \\
			pseudoMedian-RF & $\colorbox{red!25}{0.28} \pm \colorbox{red!25}{0.20}$ & $\colorbox{green!25}{0.03} \pm \colorbox{green!25}{0.02}$ & $\colorbox{green!25}{0.04} \pm \colorbox{green!25}{0.03}$ & $\colorbox{red!25}{0.41} \pm \colorbox{red!25}{0.28}$ & $\colorbox{orange!25}{0.08} \pm \colorbox{green!25}{0.05}$ & $\colorbox{red!25}{0.33} \pm \colorbox{red!25}{0.30}$ & $\colorbox{green!25}{0.04} \pm \colorbox{green!25}{0.03}$ & $\colorbox{orange!25}{0.07} \pm \colorbox{green!25}{0.04}$ & $\colorbox{red!25}{0.11} \pm \colorbox{orange!25}{0.07}$ & $\colorbox{red!25}{0.32} \pm \colorbox{red!25}{0.21}$ & $\colorbox{green!25}{0.03} \pm \colorbox{green!25}{0.02}$ \\
			EDCF-Bayesian & $\colorbox{red!25}{0.24} \pm \colorbox{red!25}{0.24}$ & $\colorbox{green!25}{0.03} \pm \colorbox{green!25}{0.02}$ & $\colorbox{green!25}{0.01} \pm \colorbox{green!25}{0.00}$ & $\colorbox{red!25}{0.18} \pm \colorbox{red!25}{0.14}$ & $\colorbox{orange!25}{0.07} \pm \colorbox{orange!25}{0.06}$ & $\colorbox{red!25}{0.26} \pm \colorbox{red!25}{0.24}$ & $\colorbox{red!25}{0.22} \pm \colorbox{red!25}{0.15}$ & $\colorbox{red!25}{0.15} \pm \colorbox{red!25}{0.11}$ & $\colorbox{green!25}{0.03} \pm \colorbox{green!25}{0.02}$ & $\colorbox{red!25}{0.28} \pm \colorbox{red!25}{0.18}$ & $\colorbox{orange!25}{0.05} \pm \colorbox{green!25}{0.03}$ \\
			\bottomrule
		\end{tabular}
	}
\end{table}

\begin{table}[!th]
	\centering
	\caption{Robustness of ESS estimates across 10 Yule datasets (10,000 trees), with lower values indicating more robust performance.}
	\label{tab:yule:robustness:summary:10k:run1}
	\resizebox{1\textwidth}{!}{
		\begin{tabular}{lcccccccccc}
			\toprule
			& 1 & 2 & 3 & 4 & 5 & 6 & 7 & 8 & 9 & 10 \\
			\midrule
			LogP1 & $\colorbox{green!25}{0.04} \pm \colorbox{green!25}{0.03}$ & $\colorbox{green!25}{0.02} \pm \colorbox{green!25}{0.01}$ & $\colorbox{green!25}{0.02} \pm \colorbox{green!25}{0.01}$ & $\colorbox{green!25}{0.01} \pm \colorbox{green!25}{0.01}$ & $\colorbox{green!25}{0.02} \pm \colorbox{green!25}{0.02}$ & $\colorbox{green!25}{0.04} \pm \colorbox{green!25}{0.03}$ & $\colorbox{green!25}{0.02} \pm \colorbox{green!25}{0.01}$ & $\colorbox{green!25}{0.01} \pm \colorbox{green!25}{0.01}$ & $\colorbox{green!25}{0.04} \pm \colorbox{green!25}{0.02}$ & $\colorbox{green!25}{0.01} \pm \colorbox{green!25}{0.01}$ \\
			cladeIndicator & $\colorbox{green!25}{0.00} \pm \colorbox{green!25}{0.00}$ & $\colorbox{green!25}{0.01} \pm \colorbox{green!25}{0.00}$ & $\colorbox{green!25}{0.01} \pm \colorbox{green!25}{0.01}$ & $\colorbox{green!25}{0.00} \pm \colorbox{green!25}{0.00}$ & $\colorbox{green!25}{0.00} \pm \colorbox{green!25}{0.00}$ & $\colorbox{green!25}{0.01} \pm \colorbox{green!25}{0.01}$ & $\colorbox{green!25}{0.01} \pm \colorbox{green!25}{0.00}$ & $\colorbox{green!25}{0.00} \pm \colorbox{green!25}{0.00}$ & $\colorbox{green!25}{0.00} \pm \colorbox{green!25}{0.00}$ & $\colorbox{green!25}{0.01} \pm \colorbox{green!25}{0.00}$ \\
			fréchetIPSE & $\colorbox{green!25}{0.00} \pm \colorbox{green!25}{0.00}$ & $\colorbox{green!25}{0.00} \pm \colorbox{green!25}{0.00}$ & $\colorbox{green!25}{0.00} \pm \colorbox{green!25}{0.00}$ & $\colorbox{green!25}{0.00} \pm \colorbox{green!25}{0.00}$ & $\colorbox{green!25}{0.02} \pm \colorbox{green!25}{0.01}$ & $\colorbox{green!25}{0.00} \pm \colorbox{green!25}{0.00}$ & $\colorbox{green!25}{0.00} \pm \colorbox{green!25}{0.00}$ & $\colorbox{green!25}{0.00} \pm \colorbox{green!25}{0.00}$ & $\colorbox{green!25}{0.00} \pm \colorbox{green!25}{0.00}$ & $\colorbox{green!25}{0.01} \pm \colorbox{green!25}{0.00}$ \\
			expRF1 & $\colorbox{green!25}{0.02} \pm \colorbox{green!25}{0.01}$ & $\colorbox{green!25}{0.01} \pm \colorbox{green!25}{0.01}$ & $\colorbox{green!25}{0.02} \pm \colorbox{green!25}{0.01}$ & $\colorbox{green!25}{0.01} \pm \colorbox{green!25}{0.01}$ & $\colorbox{green!25}{0.02} \pm \colorbox{green!25}{0.02}$ & $\colorbox{green!25}{0.04} \pm \colorbox{green!25}{0.04}$ & $\colorbox{green!25}{0.03} \pm \colorbox{green!25}{0.02}$ & $\colorbox{green!25}{0.01} \pm \colorbox{green!25}{0.00}$ & $\colorbox{green!25}{0.02} \pm \colorbox{green!25}{0.02}$ & $\colorbox{green!25}{0.01} \pm \colorbox{green!25}{0.01}$ \\
			MAP-RF1 & $\colorbox{green!25}{0.02} \pm \colorbox{green!25}{0.01}$ & $\colorbox{green!25}{0.01} \pm \colorbox{green!25}{0.01}$ & $\colorbox{green!25}{0.02} \pm \colorbox{green!25}{0.02}$ & $\colorbox{green!25}{0.04} \pm \colorbox{green!25}{0.02}$ & $\colorbox{green!25}{0.03} \pm \colorbox{green!25}{0.02}$ & $\colorbox{green!25}{0.02} \pm \colorbox{green!25}{0.02}$ & $\colorbox{green!25}{0.04} \pm \colorbox{green!25}{0.03}$ & $\colorbox{green!25}{0.02} \pm \colorbox{green!25}{0.02}$ & $\colorbox{green!25}{0.02} \pm \colorbox{green!25}{0.01}$ & $\colorbox{green!25}{0.02} \pm \colorbox{green!25}{0.01}$ \\
			pseudoMedian-RF & $\colorbox{green!25}{0.01} \pm \colorbox{green!25}{0.01}$ & $\colorbox{green!25}{0.01} \pm \colorbox{green!25}{0.01}$ & $\colorbox{green!25}{0.01} \pm \colorbox{green!25}{0.01}$ & $\colorbox{green!25}{0.01} \pm \colorbox{green!25}{0.01}$ & $\colorbox{green!25}{0.01} \pm \colorbox{green!25}{0.01}$ & $\colorbox{green!25}{0.01} \pm \colorbox{green!25}{0.01}$ & $\colorbox{green!25}{0.01} \pm \colorbox{green!25}{0.00}$ & $\colorbox{green!25}{0.01} \pm \colorbox{green!25}{0.01}$ & $\colorbox{green!25}{0.01} \pm \colorbox{green!25}{0.01}$ & $\colorbox{green!25}{0.01} \pm \colorbox{green!25}{0.01}$ \\
			EDCF-Bayesian & $\colorbox{green!25}{0.01} \pm \colorbox{green!25}{0.01}$ & $\colorbox{orange!25}{0.06} \pm \colorbox{orange!25}{0.06}$ & $\colorbox{green!25}{0.03} \pm \colorbox{green!25}{0.04}$ & $\colorbox{orange!25}{0.09} \pm \colorbox{red!25}{0.16}$ & $\colorbox{green!25}{0.02} \pm \colorbox{green!25}{0.02}$ & $\colorbox{green!25}{0.03} \pm \colorbox{green!25}{0.03}$ & $\colorbox{green!25}{0.00} \pm \colorbox{green!25}{0.00}$ & $\colorbox{red!25}{0.11} \pm \colorbox{red!25}{0.13}$ & $\colorbox{green!25}{0.00} \pm \colorbox{green!25}{0.00}$ & $\colorbox{red!25}{0.10} \pm \colorbox{orange!25}{0.07}$ \\
			\bottomrule
		\end{tabular}
	}
\end{table}

\subsection{Multimodality, Poor Mixing, and ESS}
\label{sec:multi:mds}

To visualise and explore multimodality in the posterior tree sets, we first sub-sampled each dataset to \num{1000} trees to reduce computational cost while preserving representative diversity.
These \num{1000} trees were sampled from the last \num{10000} trees of each chain, keeping every 10th tree.
Pairwise RNNI tree distances were then computed, and a 2-dimensional embedding was generated using t-distributed Stochastic Neighbour Embedding (t-SNE)~\cite{van2008visualizing}, which emphasises local structure in high-dimensional spaces.
To complement the visualisation, we applied spectral clustering to the affinity matrices derived from the pairwise tree distances, using a heat kernel transformation to convert distances into affinities~\cite{sklearn}.
Both t-SNE and spectral clustering were implemented using scikit-learn~\cite{sklearn}.
Visualisations combining the MDS embedding with the identified clusters were used for qualitative assessment of multimodality and mixing, with further details and individual dataset plots provided in the  Supplementary Material~\ref{asec:mds:cluster} and more trace plots are provided in the Supplementary Material~\ref{asec:traces}.

\begin{figure}[htb]
	\centering
	\begin{subfigure}{0.48\textwidth}
		\centering
		\includegraphics[width=\linewidth, trim={0 0 0 0.7cm}, clip]{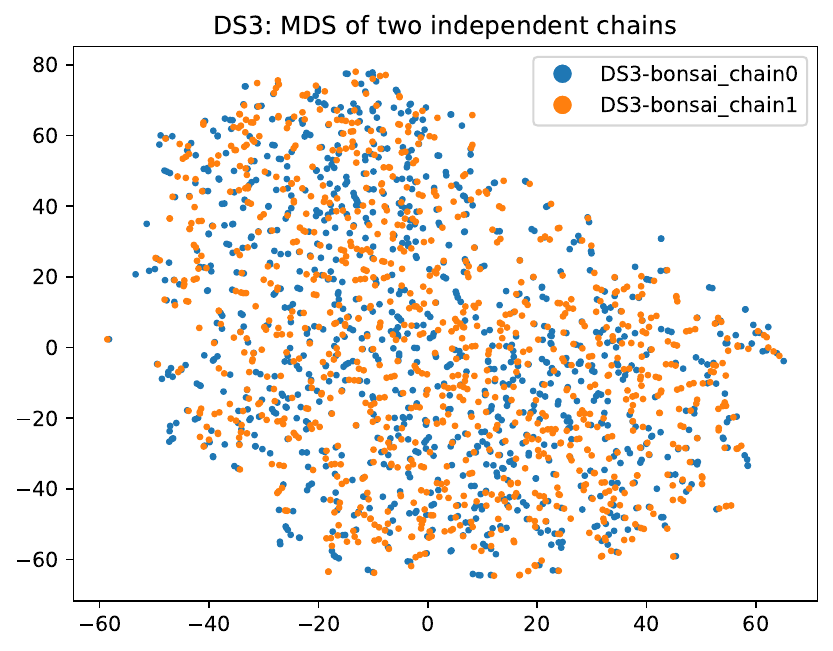}
		\caption{MDS plot with colours for two independent chains}
	\end{subfigure}\hfill
	\begin{subfigure}{0.48\textwidth}
		\centering
		\includegraphics[width=\linewidth, trim={0 0 0 0.7cm}, clip]{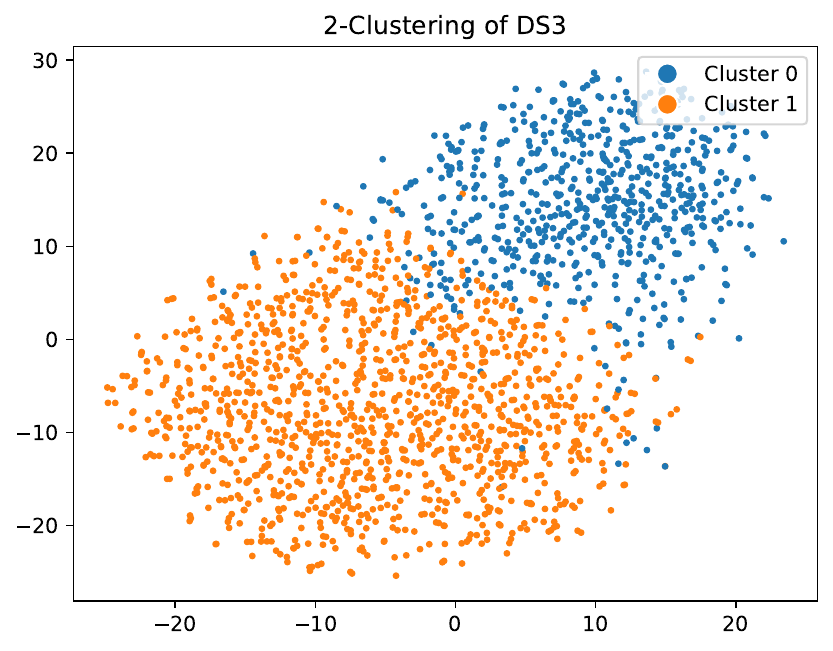}
		\caption{MDS plot with 2 clustering identified on RNNI distances.}
	\end{subfigure}\hfill
		\begin{subfigure}{0.98\textwidth}
		\centering
		\includegraphics[width=\linewidth, trim={0 13.5cm 0 76.9cm}, clip]{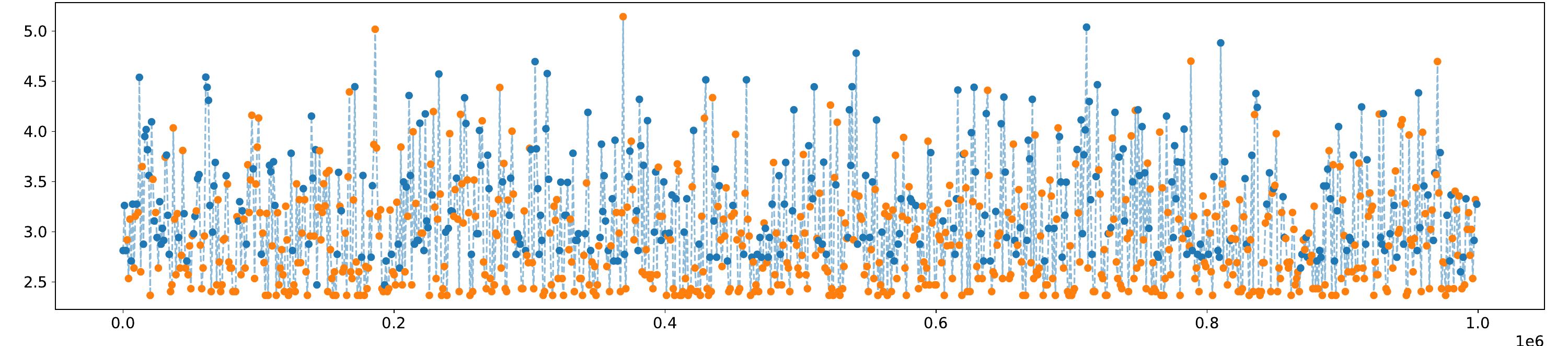}
		\caption{Trace plot of the CCD1 expected RF distance coloured with the two clustering.}
	\end{subfigure}
	\caption{MDS embedding and spectral clustering of two independent chains for DS3. The two identified clusters/modes correspond well with the MDS coordinates. Trace plots show no apparent mixing issues between the modes, noting that the blue cluster has a larger average RNNI distance.}
	\label{fig:nomodes:mds}
\end{figure}

\begin{figure}[htb]
	\centering
	\begin{subfigure}{0.48\textwidth}
		\centering
		\includegraphics[width=\linewidth, trim={0 0 0 0.7cm}, clip]{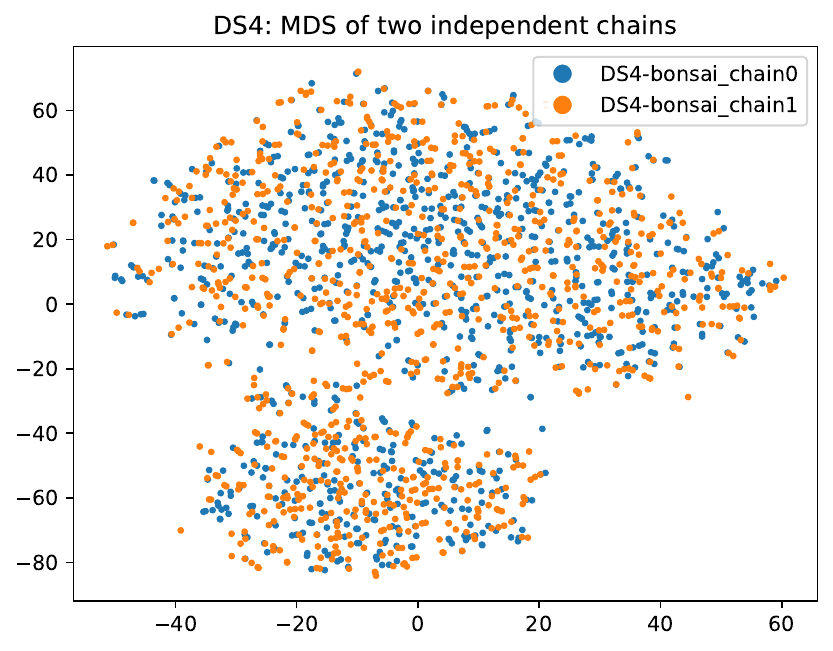}
		\caption{MDS plot with colours for two independent chains}
	\end{subfigure}\hfill
	\begin{subfigure}{0.48\textwidth}
		\centering
		\includegraphics[width=\linewidth, trim={0 0 0 0.7cm}, clip]{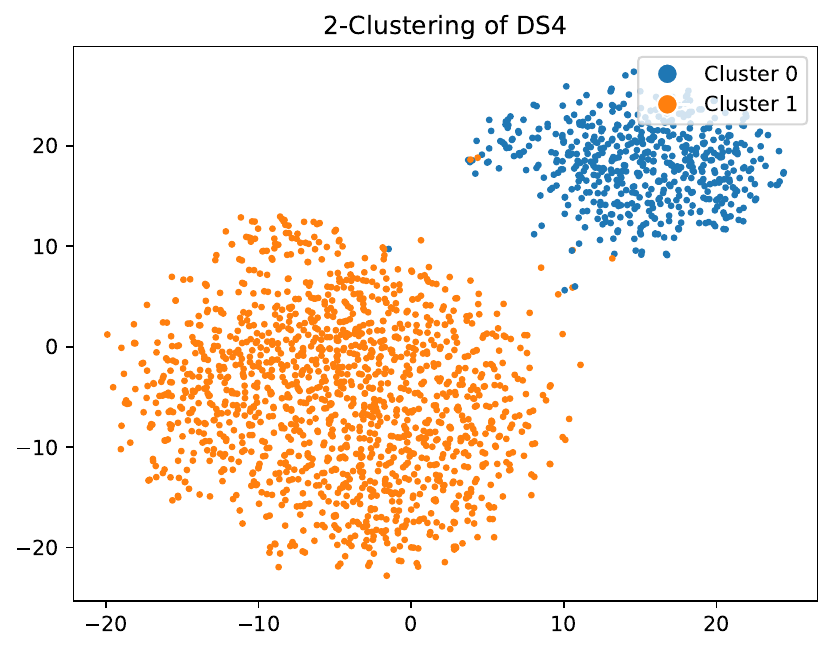}
		\caption{MDS plot with 2 clustering identified on RNNI distances.}
	\end{subfigure}\hfill
	\begin{subfigure}{0.98\textwidth}
		\centering
		\includegraphics[width=\linewidth, trim={0 13.5cm 0 76.9cm}, clip]{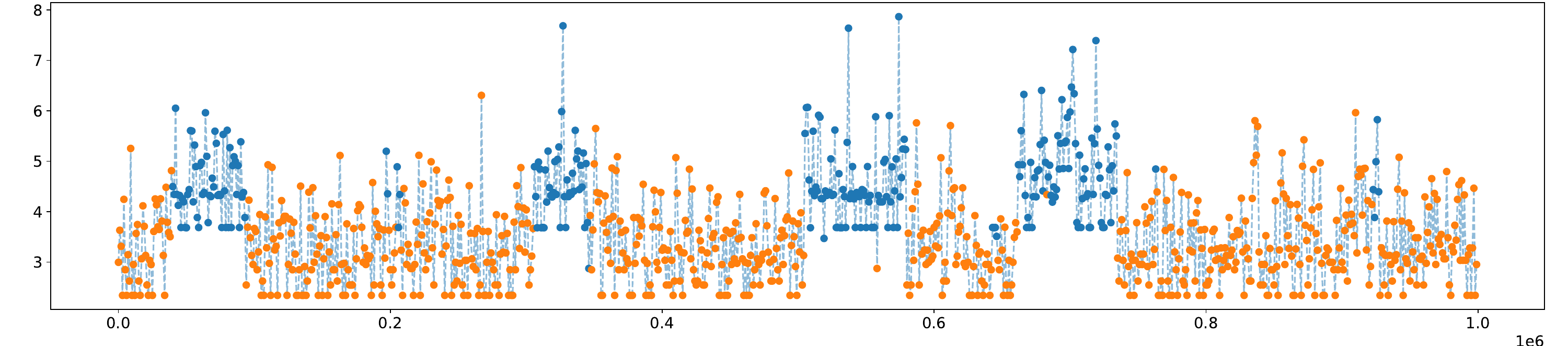}
		\caption{Trace plot of the CCD1 expected RF distance coloured with the two clustering.}
	\end{subfigure}
	\caption{MDS embedding and spectral clustering of two independent chains for DS4. The clustering aligns with the MDS coordinates. Trace plots reveal poor mixing between modes, explaining the robustness problems observed across ESS estimators.}
	\label{fig:multimodes:mds}
\end{figure}

Multimodality was assessed by visually inspecting the MDS plots, as summarised in~\cref{tbl:ds}.
Poor mixing between modes, where chains remain in one mode for an extended period before transitioning to another, can lead to lower ESS estimates.
Conversely, when multiple modes are present but the chain mixes well across them (see~\cref{fig:nomodes:mds}), ESS estimates tend to be more consistent and robust to chain fragmentation.
These mixing issues are particularly evident in DS4 (see~\cref{fig:multimodes:mds}); as well as DS1 (~\cref{fig:mds:ds1,sfig:jumpdistance:ds1}), DS6 (~\cref{fig:mds:ds6,sfig:jumpdistance:ds6}), and DS10 (~\cref{fig:mds:ds10,sfig:jumpdistance:ds10}); and are reflected in both the robustness and accuracy analyses.
In the robustness evaluation, the dataset itself appears to drive the observed instability in ESS estimates across chain fragmentation.
In the accuracy assessment, poor mixing contributes to increase variability in the estimates for DS4 relative to DS3 (see~\cref{fig:acc_results_xy:rel}), although the overall bias of the estimators is less affected.
The MDS visualizations provide an intuitive explanation for the observed differences in ESS behaviour across datasets.
By contrast, the ten simulated Yule datasets are all unimodal (see Supplementary Material~\autoref{asec:yule:mds}).

\section{Discussion} 
\label{sec:discussion}
We evaluated both new and existing estimators for the effective sample size of tree topologies in MCMC samples, using a testing framework that combined simulated chains with empirical datasets.
For the simulations, we integrated ideas from previous approaches with the versatility of CCDs: chains were simulated by sampling from a CCD fitted to the posterior tree distribution, providing access to the true ESS (or at least a lower bound).
To control the autocorrelation time, we extended these chains by connecting consecutive independent samples through random trees drawn from shortest RNNI paths between them.

Overall, we did not observe extreme differences in accuracy between most estimators, with the notable exception of the newly introduced probabilistic method (\edcf), which substantially overestimated ESS in most settings.
Methods that map trees to numerical traces and subsequently apply the initial positive sequence estimator (as implemented in Tracer) yielded similar results across datasets.
Consistent with the findings of Fabreti and Höhna~\cite{fabreti22} for the initial positive sequence estimator, these methods tended to slightly underestimate the ESS.
The two methods that average over multiple traces, \cladeInd{} and \medianRF{}, displayed lower variance and more stable accuracy, and are therefore our recommended estimators.

While \frechess{} performed comparably well in terms of stability and robustness, it requires the computation of pairwise distances between all trees, which becomes prohibitively expensive for repeated computations and long chains (see \autoref{sec:runtime}).
In practice, we imposed time limits in our experiments.
Similar limitations apply to \approximateESS{} and \edcf{}.
Although downsampling could in principle mitigate these computational costs, our preliminary experiments indicated that estimator behaviour was not stable under downsampling.
Note that this differs from our stability experiment, where we thinned already-oversampled chains \cref{fig:stability}.
Here we are referring to downsampling a regular, non-oversampled chain to reduce computational runtime, which results in highly variable ESS estimates.
We therefore recommend the simpler and faster estimators, which achieve comparable performance without substantial computational overhead.

The motivation behind \edcf was to leverage a probabilistic model of clade observations for ESS estimation.
By construction, these estimates are independent of both tree order and distance metrics.
This approach was recently applied to estimate the sampling error on clade counts from two independent MCMC chains, under the assumption of a fixed ESS \cite{douglas2025evolution}.
Here we extended this strategy by estimating the ESS directly from observed clade frequencies, noting that clade frequencies have previously been used for convergence assessment.
Although initial results were encouraging, the method lacks the accuracy and robustness of existing approaches, which likely results from the incorrect assumption of clade count independence.
While improvements may be found in the further tuning of the model and parameters, or with a Gaussian approximation of the differences in clade frequencies, the strong behaviour of existing and other novel methods, combined with the computational demands of this estimator, suggests that the practical utility of such refinements is limited.

Our new trace-based methods, which map trees to numerical sequences via CCD log probabilities (\logP{}), distances to the CCD MAP tree (\mapRF{}), or expected RF distances with respect to the CCD (\expRF{}), perform competitively with \medianRF{}, albeit with higher variance.
While not our primary recommendation for final ESS estimation, these methods offer distinct conceptual advantages.
For instance, \logP{} is independent of any explicit tree distance metric.
Using the CCD MAP tree provides a fixed and reproducible reference point, valuable for tree-distance trace plots ~\cite{lanfear16estimating} and for ensuring comparability across analyses.
Similarly, the expected RF distance does not require fixing a specific focal tree to compute a trace.
Overall, our practical recommendation for tree-based ESS estimation is to use \cladeInd{} by Fabreti and Höhna~\cite{fabreti22} or \medianRF{} (implemented as \texttt{pseudoESS}) by Lanfear \etal~\cite{lanfear16estimating}.

We also introduced a new algorithm for efficiently computing the expected RF distance of a tree to a CCD.
This quantity represents the expected distance between a given tree and a randomly drawn tree from the CCD distribution.
We expect this algorithm to be useful beyond its role in the \expRF{} estimator as it provides a principled way of measuring distances between individual trees and entire tree distributions.
Extending this approach to other tree distance measures represents a promising direction for future work.

Furthermore, we briefly discuss the role of ESS for convergence assessment in practical phylogenetic analyses.
Estimating tree ESS across multiple chains provides a valuable diagnostic of convergence: a high combined ESS suggests that chains are sampling from similar distributions, and therefore likely approximating the target stationary distribution.
Conversely, a low combined ESS indicates that chains may not yet have explored sufficiently similar regions of treespace, signalling a lack of convergence.
While this strategy has been applied previously~\cite{berling24automated}, it is important to emphasize that a high ESS alone does not guarantee convergence to the correct target distribution.
Independent checks, such as visual inspection of parameter traces and posterior distributions or the use of diagnostics like the Gelman–Rubin diagnostic, remain essential.
Ultimately, the ESS should be interpreted as a measure of how many effectively independent samples have been obtained, and does not provide a proof of convergence in any~case.

Some tree-based ESS estimators can be computed online during the MCMC run and are therefore candidates for inclusion in automated convergence testing frameworks, such as the ASM package.
This is straightforward for \cladeInd{}, and it is also feasible to maintain a CCD during MCMC in order to compute log-probability or RF-based traces online.

Finally, we observed that the combination of poor mixing and multimodality can lead to unreliable ESS estimation (as demonstrated by DS1, 4, 6, and 10).
When multiple well-separated modes are present, for example as indicated by MDS combined with clustering or by visual inspection of topology trace plots, we recommend considering ESS estimates both within individual modes and for transitions between modes.
High ESS values for each component provide stronger evidence of adequate mixing than a single global ESS estimate, which may be misleading when transitions between modes are rare.
We emphasize that methods such as MDS and clustering provide only coarse and potentially unreliable views of multimodality in treespace, with results that can depend strongly on the underlying choice of tree distance metric.
Developing more principled and scalable approaches for detecting and characterizing multimodal posterior tree distributions remains an important direction for future work.
Once such problems are detected, several strategies can help address them, for example guided tree proposals~\cite{hoehna12guided}, Metropolis-coupled MCMC~\cite{altekar2004parallel,mueller20adaptive}, or the handling of outlier samples discussed by Gao et al.~\cite{gao2026biological}.

\section{Conclusion and Outlook} 
\label{sec:conclusion}

We have extended the toolkit for phylogenetic MCMC diagnostics by introducing CCD-based ESS estimators and a probabilistic approach based on clade frequencies, alongside a systematic comparison with existing methods.
Our practical recommendation is to use \cladeInd{} (Fabreti and Höhna, 2022) or \medianRF{} (implemented as \texttt{pseudoESS}; Lanfear et al., 2022) for reliable and efficient ESS estimation of tree topologies, while our new expected RF distance algorithm provides additional flexibility for trace computation.

More broadly, our results highlight that ESS estimation for phylogenetic trees is fundamentally challenging due to the discrete and non-euclidean nature of treespace.
Distinguishing between insufficient sampling within a single mode and poor mixing across multiple modes therefore requires careful interpretation of ESS values across chains and within chain fragments.
Methods computable during MCMC runs enable integration into automated convergence frameworks such as the ASM package~\cite{berling24automated}, supporting real-time diagnostics.

Our evaluation framework of combining simulated chains with known ESS bounds and robustness testing on real datasets, offers a general template for future method development.
Looking forward, benchmarking efforts would benefit from expanding beyond the standard DS1--11 datasets~\cite{lakner2008efficiency}, which we and many others in the field have relied on extensively since 2008~\cite{whidden2015quantifying,meyer2021adaptive,magee23,berling24automated}.
While these datasets have enabled foundational comparisons, increasingly diverse applications (e.g., viral phylodynamics, the mulispecies coalescent) remain underrepresented in treespace method development and comparisons.

Finally, our simulated chains introduced controlled autocorrelation structures into i.i.d. samples, inspired by approaches in~\cite{fabreti22, magee23}.
A systematic comparison of these construction methods against empirical MCMC traces would provide valuable insight into how faithfully such simulations capture real sampling behaviour.

As phylogenetic analyses continue to grow in scale and complexity, robust ESS estimation for tree topologies remains fundamental to ensuring the validity of posterior inference.

\section*{Availability}\label{sec:availability}

All methods and software developed in this paper are publicly available across several repositories and packages.
The trace-based tree ESS estimators can be calculated with the \texttt{TreeStat2} package at \url{github.com/alexeid/TreeStat2} and the resulting log files can be viewed in \texttt{Tracer}~\cite{rambaut2018tracer}.

For RNNI-based computations, we utilized the \texttt{tetres} Python package at \url{github.com/bioDS/tetres}.
The RNNI chain simulation from CCD samples and the evaluations presented in this paper are available at \url{github.com/CompEvol/CCD-Research/tree/main/treeESSEstimation}.

The algorithm to compute the expected RF distance of a tree to a CCD is available in the \texttt{CCD} package at \url{github.com/CompEvol/CCD}.
Sampling trees from a CCD can be done with the \texttt{CCDSampler} application within the \texttt{CCD} package and is executable via the BEAST2 applauncher.

Our implementations for the \edcf methods, \frechess, and \approximateESS are hosted at \url{https://github.com/CompEvol/TreeESSExtra}.
Closely related tools that we reimplemented include those from Lanfear \etal~\cite{lanfear16estimating} which provide tree ESS estimators (\medianRF, \approximateESS) and methods for trace and jump distance plots in the R package \texttt{RWTY} (R We There Yet?) at \url{github.com/danlwarren/RWTY/}.
Magee \etal~\cite{magee23} also implement \frechess (among other estimators) in their R package \texttt{treess} at \url{github.com/afmagee/treess/}.

\section*{Acknowledgements} 
JK, JD, and AJD were partially supported by the Beyond Prediction Data Science Research Programme (MBIE grant UOAX1932).
LB was partially supported by the Federal Government of Canada’s Canada 150 Research Chair program.

This research was enabled in part by the BC DRI Group and the Digital Research Alliance of Canada through the \texttt{fir} high performance computing cluster.

\section*{Conflict of Interest Disclosure} 
The authors declare they have no conflict of interest relating to the content of this article.

\pdfbookmark[1]{References}{References}

\bibliographystyle{plainurl}
\bibliography{sources.bib}

\begin{thebibliography}{10}

\bibitem{allen01}
Benjamin~L. Allen and Mike Steel.
\newblock {Subtree Transfer Operations and Their Induced Metrics on
  Evolutionary Trees}.
\newblock {\em Annals of Combinatorics}, 5(1):1--15, 2001.
\newblock \href {https://doi.org/10.1007/s00026-001-8006-8}
  {\path{doi:10.1007/s00026-001-8006-8}}.

\bibitem{altekar2004parallel}
Gautam Altekar, Sandhya Dwarkadas, John~P. Huelsenbeck, and Fredrik Ronquist.
\newblock Parallel metropolis coupled markov chain monte carlo for bayesian
  phylogenetic inference.
\newblock {\em Bioinformatics}, 20(3):407--415, 2004.
\newblock \href {https://doi.org/10.1093/bioinformatics/btg427}
  {\path{doi:10.1093/bioinformatics/btg427}}.

\bibitem{hipstr25}
Guy Baele, Luiz~M Carvalho, Marius Brusselmans, Gytis Dudas, Xiang Ji, John~T
  McCrone, Philippe Lemey, Marc~A Suchard, and Andrew Rambaut.
\newblock {HIPSTR: highest independent posterior subtree reconstruction in
  TreeAnnotator X}.
\newblock {\em Bioinformatics}, 41(10):btaf488, 2025.
\newblock \href {https://doi.org/10.1093/bioinformatics/btaf488}
  {\path{doi:10.1093/bioinformatics/btaf488}}.

\bibitem{berling24automated}
Lars Berling, Remco Bouckaert, and Alex Gavryushkin.
\newblock An automated convergence diagnostic for phylogenetic {MCMC} analyses.
\newblock {\em IEEE/ACM Transactions on Computational Biology and
  Bioinformatics}, pages 1--13, 2024.
\newblock \href {https://doi.org/10.1109/TCBB.2024.3457875}
  {\path{doi:10.1109/TCBB.2024.3457875}}.

\bibitem{berling2025accurate}
Lars Berling, Jonathan Klawitter, Remco Bouckaert, Dong Xie, Alex Gavryushkin,
  and Alexei~J Drummond.
\newblock Accurate bayesian phylogenetic point estimation using a tree
  distribution parameterized by clade probabilities.
\newblock {\em PLOS Computational Biology}, 21(2):e1012789, 2025.
\newblock \href {https://doi.org/10.1371/journal.pcbi.1012789}
  {\path{doi:10.1371/journal.pcbi.1012789}}.

\bibitem{bordewich05}
Magnus Bordewich and Charles Semple.
\newblock {On the Computational Complexity of the Rooted Subtree Prune and
  Regraft Distance}.
\newblock {\em Annals of Combinatorics}, 8(4):409--423, 2005.
\newblock \href {https://doi.org/10.1007/s00026-004-0229-z}
  {\path{doi:10.1007/s00026-004-0229-z}}.

\bibitem{bouckaert2019beast}
Remco Bouckaert, Timothy~G Vaughan, Jo{\"e}lle Barido-Sottani, Sebasti{\'a}n
  Duch{\^e}ne, Mathieu Fourment, Alexandra Gavryushkina, Joseph Heled, Graham
  Jones, Denise Kühnert, Nicola De~Maio, et~al.
\newblock {BEAST} 2.5: An advanced software platform for {Bayesian}
  evolutionary analysis.
\newblock {\em PLoS Computational Biology}, 15(4):e1006650, 2019.
\newblock \href {https://doi.org/10.1371/journal.pcbi.1006650}
  {\path{doi:10.1371/journal.pcbi.1006650}}.

\bibitem{brusselmans2024importance}
Marius Brusselmans, Luiz~Max Carvalho, Samuel~L Hong, Jiansi Gao, Frederick~A
  Matsen~{IV}, Andrew Rambaut, Philippe Lemey, Marc~A Suchard, Gytis Dudas, and
  Guy Baele.
\newblock {On the importance of assessing topological convergence in Bayesian
  phylogenetic inference}.
\newblock {\em Virus Evolution}, 10(1):veae081, 2024.

\bibitem{collienne2021computing}
Lena Collienne and Alex Gavryushkin.
\newblock Computing nearest neighbour interchange distances between ranked
  phylogenetic trees.
\newblock {\em Journal of Mathematical Biology}, 82(1):8, 2021.
\newblock \href {https://doi.org/10.1007/s00285-021-01567-5}
  {\path{doi:10.1007/s00285-021-01567-5}}.

\bibitem{douglas2025evolution}
Jordan Douglas, Remco Bouckaert, Simon~C Harris, Charles~W Carter~Jr, and
  Peter~R Wills.
\newblock Evolution is coupled with branching across many granularities of
  life.
\newblock {\em Proceedings of the Royal Society B: Biological Sciences},
  292(2047), 2025.
\newblock \href {https://doi.org/10.1098/rspb.2025.0182}
  {\path{doi:10.1098/rspb.2025.0182}}.

\bibitem{douglas2022starbeast3}
Jordan Douglas, Cinthy~L Jim{\'e}nez-Silva, and Remco Bouckaert.
\newblock {StarBeast3: adaptive parallelized Bayesian inference under the
  multispecies coalescent}.
\newblock {\em Systematic Biology}, 71(4):901--916, 2022.
\newblock \href {https://doi.org/10.1093/sysbio/syac010}
  {\path{doi:10.1093/sysbio/syac010}}.

\bibitem{drummond23linguaphylo}
Alexei~J Drummond, Kylie Chen, F{\'a}bio~K Mendes, and Dong Xie.
\newblock {LinguaPhylo:} a probabilistic model specification language for
  reproducible phylogenetic analyses.
\newblock {\em PLOS Computational Biology}, 19(7):e1011226, 2023.
\newblock \href {https://doi.org/10.1371/journal.pcbi.1011226}
  {\path{doi:10.1371/journal.pcbi.1011226}}.

\bibitem{MCMCSE21}
James~M. Flegal, John Hughes, Dootika Vats, Ning Dai, Kushagra Gupta, and
  Uttiya Maji.
\newblock {\em mcmcse: Monte Carlo Standard Errors for MCMC}.
\newblock Riverside, CA, and Kanpur, India, 2021.
\newblock R package version 1.5-0.

\bibitem{flegal10}
James~M. Flegal and Galin~L. Jones.
\newblock Batch means and spectral variance estimators in {Markov chain Monte
  Carlo}.
\newblock {\em The Annals of Statistics}, 38(2):1034--1070, 2010.
\newblock \href {https://doi.org/10.1214/09-AOS735}
  {\path{doi:10.1214/09-AOS735}}.

\bibitem{gao2026biological}
Jiansi Gao, Marius Brusselmans, Luiz~M Carvalho, Marc~A Suchard, Guy Baele, and
  Frederick~A Matsen~{IV}.
\newblock Biological causes and impacts of rugged tree landscapes in
  phylodynamic inference.
\newblock {\em Proceedings of the National Academy of Sciences},
  123(2):e2510938123, 2026.

\bibitem{DS2-garey96}
James~R. Garey, Thomas~J. Near, Michael~R. Nonnemacher, and Steven~A. Nadler.
\newblock {Molecular evidence for Acanthocephala as a subtaxon of Rotifera}.
\newblock {\em Journal of Molecular Evolution}, 43(3):287--292, 1996.
\newblock \href {https://doi.org/10.1007/BF02338837}
  {\path{doi:10.1007/BF02338837}}.

\bibitem{gelman95bayesian}
Andrew Gelman, John~B Carlin, Hal~S Stern, and Donald~B Rubin.
\newblock {\em Bayesian Data Analysis}.
\newblock Chapman and Hall/CRC, 1995.

\bibitem{geyer11}
Charles~J Geyer.
\newblock {Introduction to Markov Chain Monte Carlo}.
\newblock In Steve Brooks, Andrew Gelman, Galin Jones, and Xiao-Li Meng,
  editors, {\em Handbook of Markov Chain Monte Carlo}, pages 3--48, 2011.

\bibitem{fabreti22}
Luiza Guimar{\~a}es~Fabreti and Sebastian H{\"o}hna.
\newblock Convergence assessment for {{Bayesian}} phylogenetic analysis using
  {{MCMC}} simulation.
\newblock {\em Methods in Ecology and Evolution}, 13(1):77--90, 2022.
\newblock \href {https://doi.org/10.1111/2041-210X.13727}
  {\path{doi:10.1111/2041-210X.13727}}.

\bibitem{hamilton94}
James~D. Hamilton.
\newblock {\em TIME SERIES ANALYSIS}.
\newblock Princeton University Press, 1994.

\bibitem{harrington2021properties}
Sean~M Harrington, Van Wishingrad, and Robert~C Thomson.
\newblock Properties of markov chain monte carlo performance across many
  empirical alignments.
\newblock {\em Molecular Biology and Evolution}, 38(4):1627--1640, 2021.

\bibitem{hasegawa1985dating}
Masami Hasegawa, Hirohisa Kishino, and Taka-aki Yano.
\newblock Dating of the human-ape splitting by a molecular clock of
  mitochondrial {DNA}.
\newblock {\em Journal of molecular evolution}, 22:160--174, 1985.
\newblock \href {https://doi.org/10.1007/BF02101694}
  {\path{doi:10.1007/BF02101694}}.

\bibitem{DS1-hedges90}
S~B Hedges, K~D Moberg, and L~R Maxson.
\newblock {Tetrapod phylogeny inferred from 18S and 28S ribosomal RNA sequences
  and a review of the evidence for amniote relationships}.
\newblock {\em Molecular Biology and Evolution}, 7(6):607--633, 1990.
\newblock \href {https://doi.org/10.1093/oxfordjournals.molbev.a040628}
  {\path{doi:10.1093/oxfordjournals.molbev.a040628}}.

\bibitem{heidelberger81spectral}
Philip Heidelberger and Peter~D. Welch.
\newblock A spectral method for confidence interval generation and run length
  control in simulations.
\newblock {\em Communications of the ACM}, 24(4):233--–245, 1981.
\newblock \href {https://doi.org/10.1145/358598.358630}
  {\path{doi:10.1145/358598.358630}}.

\bibitem{hein90}
Jotun Hein.
\newblock Reconstructing evolution of sequences subject to recombination using
  parsimony.
\newblock {\em Mathematical Biosciences}, 98(2):185--200, 1990.
\newblock \href {https://doi.org/10.1016/0025-5564(90)90123-G}
  {\path{doi:10.1016/0025-5564(90)90123-G}}.

\bibitem{DS4-henk03}
Daniel~A. Henk, Alex Weir, and Meredith Blackwell.
\newblock Laboulbeniopsis termitarius, an ectoparasite of termites newly
  recognized as a member of the {Laboulbeniomycetes}.
\newblock {\em Mycologia}, 95(4):561--564, 2003.
\newblock \href {https://doi.org/10.1080/15572536.2004.11833059}
  {\path{doi:10.1080/15572536.2004.11833059}}.

\bibitem{hoehna12guided}
Sebastian Höhna and Alexei~J. Drummond.
\newblock Guided tree topology proposals for {B}ayesian phylogenetic inference.
\newblock {\em Systematic Biology}, 61(1):1--11, 2012.
\newblock \href {https://doi.org/10.1093/sysbio/syr074}
  {\path{doi:10.1093/sysbio/syr074}}.

\bibitem{hoehna16revbayes}
Sebastian Höhna, Michael~J. Landis, Tracy~A. Heath, Bastien Boussau, Nicolas
  Lartillot, Brian~R. Moore, John~P. Huelsenbeck, and Fredrik Ronquist.
\newblock {RevBayes: Bayesian} phylogenetic inference using graphical models
  and an interactive model-specification language.
\newblock {\em Systematic Biology}, 65(4):726--736, 2016.
\newblock \href {https://doi.org/10.1093/sysbio/syw021}
  {\path{doi:10.1093/sysbio/syw021}}.

\bibitem{DS9-ingram04}
Amanda~L. Ingram and Jeff~J. Doyle.
\newblock Is eragrostis (poaceae) monophyletic? insights from nuclear and
  plastid sequence data.
\newblock {\em Systematic Botany}, 29(3):545--552, 2004.
\newblock \href {https://doi.org/doi:10.1600/0363644041744392}
  {\path{doi:doi:10.1600/0363644041744392}}.

\bibitem{berger14}
M.~J.~Bayarri James~Berger and L.~R. Pericchi.
\newblock The effective sample size.
\newblock {\em Econometric Reviews}, 33(1-4):197--217, 2014.
\newblock \href {https://doi.org/10.1080/07474938.2013.807157}
  {\path{doi:10.1080/07474938.2013.807157}}.

\bibitem{jun23}
Seong-Hwan Jun, Hassan Nasif, Chris Jennings-Shaffer, David~H. Rich, Anna
  Kooperberg, Mathieu Fourment, Cheng Zhang, Marc~A. Suchard, and Frederick~A.
  Matsen.
\newblock A topology-marginal composite likelihood via a generalized
  phylogenetic pruning algorithm.
\newblock {\em Algorithms for Molecular Biology}, 18(10), 2023.
\newblock \href {https://doi.org/10.1186/s13015-023-00235-1}
  {\path{doi:10.1186/s13015-023-00235-1}}.

\bibitem{klawitter24rogue}
Jonathan Klawitter, Remco Bouckaert, and Alexei~J. Drummond.
\newblock Skeletons in the forest: Using entropy-based rogue detection on
  {B}ayesian phylogenetic tree distributions.
\newblock {\em bioRxiv}, 2024.
\newblock \href {https://doi.org/10.1101/2024.09.25.615070}
  {\path{doi:10.1101/2024.09.25.615070}}.

\bibitem{klawitter25}
Jonathan Klawitter and Alexei~J. Drummond.
\newblock Credible sets of phylogenetic tree topology distributions, 2025.
\newblock \href {https://arxiv.org/abs/2505.14532} {\path{arXiv:2505.14532}},
  \href {https://doi.org/10.48550/arXiv.2505.14532}
  {\path{doi:10.48550/arXiv.2505.14532}}.

\bibitem{DS11-kroken00}
Scott Kroken and John~W. Taylor.
\newblock Phylogenetic species, reproductive mode, and specificity of the green
  alga trebouxia forming lichens with the fungal genus letharia.
\newblock {\em The Bryologist}, 103(4):645--660, 2000.

\bibitem{lakner2008efficiency}
Clemens Lakner, Paul Van Der~Mark, John~P Huelsenbeck, Bret Larget, and Fredrik
  Ronquist.
\newblock Efficiency of {Markov chain Monte Carlo} tree proposals in {B}ayesian
  phylogenetics.
\newblock {\em Systematic Biology}, 57(1):86--103, 2008.
\newblock \href {https://doi.org/10.1080/10635150801886156}
  {\path{doi:10.1080/10635150801886156}}.

\bibitem{lanfear16estimating}
Robert Lanfear, Xia Hua, and Dan~L. Warren.
\newblock {Estimating the Effective Sample Size of Tree Topologies from
  Bayesian Phylogenetic Analyses}.
\newblock {\em Genome Biology and Evolution}, 8(8):2319--2332, 2016.
\newblock \href {https://doi.org/10.1093/gbe/evw171}
  {\path{doi:10.1093/gbe/evw171}}.

\bibitem{larget13estimation}
Bret Larget.
\newblock The estimation of tree posterior probabilities using conditional
  clade probability distributions.
\newblock {\em Systematic Biology}, 62(4):501--511, 2013.
\newblock \href {https://doi.org/10.1093/sysbio/syt014}
  {\path{doi:10.1093/sysbio/syt014}}.

\bibitem{lewis16estimating}
Paul~O. Lewis, Ming-Hui Chen, Lynn Kuo, Louise~A. Lewis, Karolina Fu{\v
  c}{\'\i}kov{\'a}, Suman Neupane, Yu-Bo Wang, and Daoyuan Shi.
\newblock Estimating {B}ayesian phylogenetic information content.
\newblock {\em Systematic Biology}, 65(6):1009--1023, 2016.
\newblock \href {https://doi.org/10.1093/sysbio/syw042}
  {\path{doi:10.1093/sysbio/syw042}}.

\bibitem{magee23}
Andrew Magee, Michael Karcher, Frederick~A Matsen~IV, and Volodymyr~M Minin.
\newblock How trustworthy is your tree? {Bayesian} phylogenetic effective
  sample size through the lens of {Monte Carlo} error.
\newblock {\em Bayesian Analysis}, 1(1):1--29, 2023.
\newblock \href {https://doi.org/10.1214/22-BA1339}
  {\path{doi:10.1214/22-BA1339}}.

\bibitem{meyer2021adaptive}
Xavier Meyer.
\newblock {Adaptive tree proposals for Bayesian phylogenetic inference}.
\newblock {\em Systematic Biology}, 70(5):1015--1032, 2021.

\bibitem{mueller20adaptive}
Nicola~F. Müller and Remco~R. Bouckaert.
\newblock Adaptive metropolis-coupled mcmc for beast 2.
\newblock {\em PeerJ}, 8:e9473, 2020.
\newblock \href {https://doi.org/10.7717/peerj.9473}
  {\path{doi:10.7717/peerj.9473}}.

\bibitem{sklearn}
F.~Pedregosa, G.~Varoquaux, A.~Gramfort, V.~Michel, B.~Thirion, O.~Grisel,
  M.~Blondel, P.~Prettenhofer, R.~Weiss, V.~Dubourg, J.~Vanderplas, A.~Passos,
  D.~Cournapeau, M.~Brucher, M.~Perrot, and E.~Duchesnay.
\newblock Scikit-learn: Machine learning in {P}ython.
\newblock {\em Journal of Machine Learning Research}, 12:2825--2830, 2011.

\bibitem{coda}
Martyn Plummer, Nicky Best, Kate Cowles, and Karen Vines.
\newblock {CODA: Convergence Diagnosis and Output Analysis for MCMC}.
\newblock {\em R News}, 6(1):7--11, 2006.

\bibitem{rambaut2018tracer}
Andrew Rambaut, Alexei~J Drummond, Dong Xie, Guy Baele, and Marc~A Suchard.
\newblock Posterior {{Summarization}} in {{Bayesian Phylogenetics Using
  Tracer}} 1.7.
\newblock {\em Systematic Biology}, 67(5):901--904, 2018.
\newblock \href {https://doi.org/10.1093/sysbio/syy032}
  {\path{doi:10.1093/sysbio/syy032}}.

\bibitem{ronquist12mrbayes}
Fredrik Ronquist, Maxim Teslenko, Paul van~der Mark, Daniel~L. Ayres, Aaron
  Darling, Sebastian Höhna, Bret Larget, Liang Liu, Marc~A. Suchard, and
  John~P. Huelsenbeck.
\newblock {MrBayes 3.2:} efficient {B}ayesian phylogenetic inference and model
  choice across a large model space.
\newblock {\em Systematic Biology}, 61(3):539--542, 2012.
\newblock \href {https://doi.org/10.1093/sysbio/sys029}
  {\path{doi:10.1093/sysbio/sys029}}.

\bibitem{DS8-rossman01}
Amy~Y. Rossman, John~M. McKemy, Rebecca~A. Pardo-Schultheiss, and Hans-Josef
  Schroers.
\newblock {Molecular studies of the Bionectriaceae using large subunit rDNA
  sequences}.
\newblock {\em Mycologia}, 93(1):100--110, 2001.
\newblock \href {https://doi.org/10.1080/00275514.2001.12061283}
  {\path{doi:10.1080/00275514.2001.12061283}}.

\bibitem{steel93distributions}
Mike~A. Steel and David Penny.
\newblock Distributions of tree comparison metrics—some new results.
\newblock {\em Systematic Biology}, 42(2):126--141, 1993.
\newblock \href {https://doi.org/10.1093/sysbio/42.2.126}
  {\path{doi:10.1093/sysbio/42.2.126}}.

\bibitem{suchard2018bayesian}
Marc~A Suchard, Philippe Lemey, Guy Baele, Daniel~L Ayres, Alexei~J Drummond,
  and Andrew Rambaut.
\newblock Bayesian phylogenetic and phylodynamic data integration using {BEAST}
  1.10.
\newblock {\em Virus Evolution}, 4(1):vey016, 2018.
\newblock \href {https://doi.org/10.1093/ve/vey016}
  {\path{doi:10.1093/ve/vey016}}.

\bibitem{DS10-suh99}
Sung-Oui Suh and Meredith Blackwell.
\newblock {Molecular phylogeny of the cleistothecial fungi placed in
  Cephalothecaceae and Pseudeurotiaceae}.
\newblock {\em Mycologia}, 91(5):836--848, 1999.
\newblock \href {https://doi.org/10.1080/00275514.1999.12061089}
  {\path{doi:10.1080/00275514.1999.12061089}}.

\bibitem{szollosi13efficient}
Gergely~J. Szöllősi, Wojciech Rosikiewicz, Bastien Boussau, Eric Tannier, and
  Vincent Daubin.
\newblock Efficient exploration of the space of reconciled gene trees.
\newblock {\em Systematic Biology}, 62(6):901--912, 2013.
\newblock \href {https://doi.org/10.1093/sysbio/syt054}
  {\path{doi:10.1093/sysbio/syt054}}.

\bibitem{straatsma86}
H.J.C.~Berendsen T.P.~Straatsma and A.J. Stam.
\newblock Estimation of statistical errors in molecular simulation
  calculations.
\newblock {\em Molecular Physics}, 57(1):89--95, 1986.
\newblock \href {https://doi.org/10.1080/00268978600100071}
  {\path{doi:10.1080/00268978600100071}}.

\bibitem{van2008visualizing}
Laurens Van~der Maaten and Geoffrey Hinton.
\newblock Visualizing data using t-sne.
\newblock {\em Journal of Machine Learning Research}, 9(11), 2008.

\bibitem{vats2021revisiting}
Dootika Vats and Christina Knudson.
\newblock Revisiting the {{Gelman}}\textendash{{Rubin Diagnostic}}.
\newblock {\em Statistical Science}, 36(4):518--529, 2021.
\newblock \href {https://doi.org/10.1214/20-STS812}
  {\path{doi:10.1214/20-STS812}}.

\bibitem{wasserman2000bayesian}
Larry Wasserman.
\newblock Bayesian model selection and model averaging.
\newblock {\em Journal of mathematical psychology}, 44(1):92--107, 2000.
\newblock \href {https://doi.org/10.1006/jmps.1999.1278}
  {\path{doi:10.1006/jmps.1999.1278}}.

\bibitem{whidden2015quantifying}
Chris Whidden and Frederick~A Matsen~IV.
\newblock {Quantifying MCMC exploration of phylogenetic tree space}.
\newblock {\em Systematic biology}, 64(3):472--491, 2015.

\bibitem{yang2013searching}
Ziheng Yang and Carlos~E Rodr{\'\i}guez.
\newblock Searching for efficient markov chain monte carlo proposal kernels.
\newblock {\em Proceedings of the National Academy of Sciences},
  110(48):19307--19312, 2013.
\newblock \href {https://doi.org/10.1073/pnas.1311790110}
  {\path{doi:10.1073/pnas.1311790110}}.

\bibitem{DS3-yang03}
Ziheng Yang and Anne~D. Yoder.
\newblock Comparison of likelihood and {Bayesian} methods for estimating
  divergence times using multiple gene loci and calibration points, with
  application to a radiation of cute-looking mouse lemur species.
\newblock {\em Systematic Biology}, 52(5):705--716, 2003.
\newblock \href {https://doi.org/10.1080/10635150390235557}
  {\path{doi:10.1080/10635150390235557}}.

\bibitem{DS7-yoder04}
Anne~D. Yoder and Ziheng Yang.
\newblock Divergence dates for {Malagasy} lemurs estimated from multiple gene
  loci: geological and evolutionary context.
\newblock {\em Molecular Ecology}, 13(4):757--773, 2004.
\newblock \href {https://doi.org/10.1046/j.1365-294X.2004.02106.x}
  {\path{doi:10.1046/j.1365-294X.2004.02106.x}}.

\bibitem{zhang18generalizing}
Cheng Zhang and Frederick~A Matsen~IV.
\newblock Generalizing tree probability estimation via {Bayesian} networks.
\newblock In S.~Bengio, H.~Wallach, H.~Larochelle, K.~Grauman, N.~Cesa-Bianchi,
  and R.~Garnett, editors, {\em Advances in Neural Information Processing
  Systems}, volume~31, pages 1444--1453, 2018.

\bibitem{zhang18variational}
Cheng Zhang and Frederick~A Matsen~IV.
\newblock Variational {Bayesian} phylogenetic inference.
\newblock In {\em International Conference on Learning Representations (ICLR)},
  2018.

\bibitem{DS6-zhang01}
Ning Zhang and Meredith Blackwell.
\newblock {Molecular phylogeny of dogwood anthracnose fungus (Discula
  destructiva) and the Diaporthales}.
\newblock {\em Mycologia}, 93(2):355--365, 2001.
\newblock \href {https://doi.org/10.1080/00275514.2001.12063167}
  {\path{doi:10.1080/00275514.2001.12063167}}.

\end{thebibliography}

\newpage
\appendix
\pdfbookmark[1]{Supplementary Material}{Appendix}
\section*{Supplementary Material}
We provide additional results on the autocorrelation signatures in~\cref{asec:autocorr:signature}, on estimator accuracy in~\cref{asec:accuracy}, on their stability in~\cref{asec:stab}, and on their robustness in~\cref{asec:robustness} as well as on the modality analysis (\cref{asec:mds:cluster}) and trace plots (\cref{asec:traces}).

\section{Autocorrelation Signature} \label{asec:autocorr:signature}  

In addition to the autocorrelation signature plot for DS4 with a target ACT of~25 shown in \cref{fig:signature}, we include the corresponding autocorrelation signature plots for all datasets DS1--11 in
\Crefrange{fig:signature:ds1}{fig:signature:ds11}.

\signatureGrid{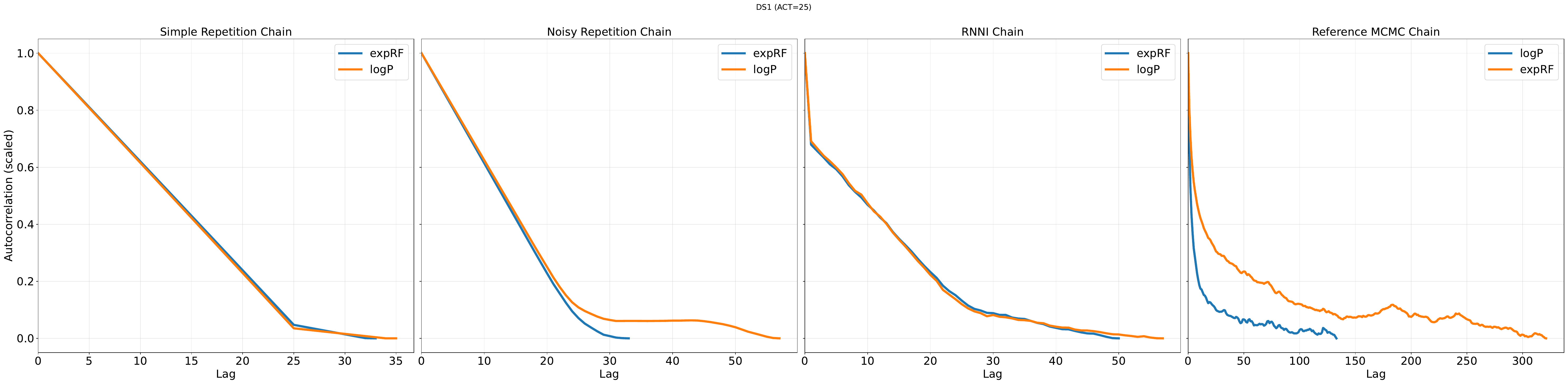}
\signatureGrid{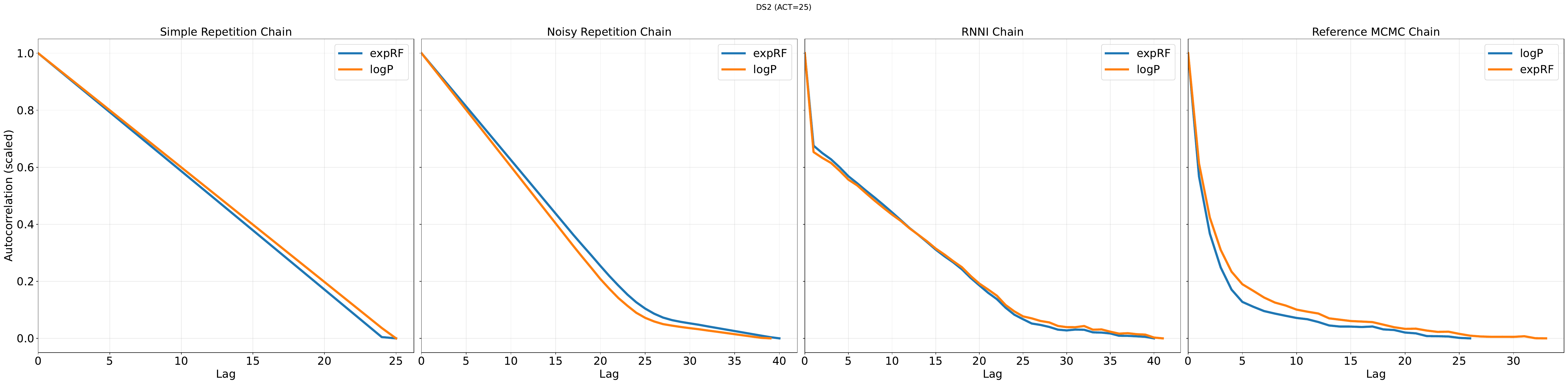}
\signatureGrid{ds3}
\signatureGrid{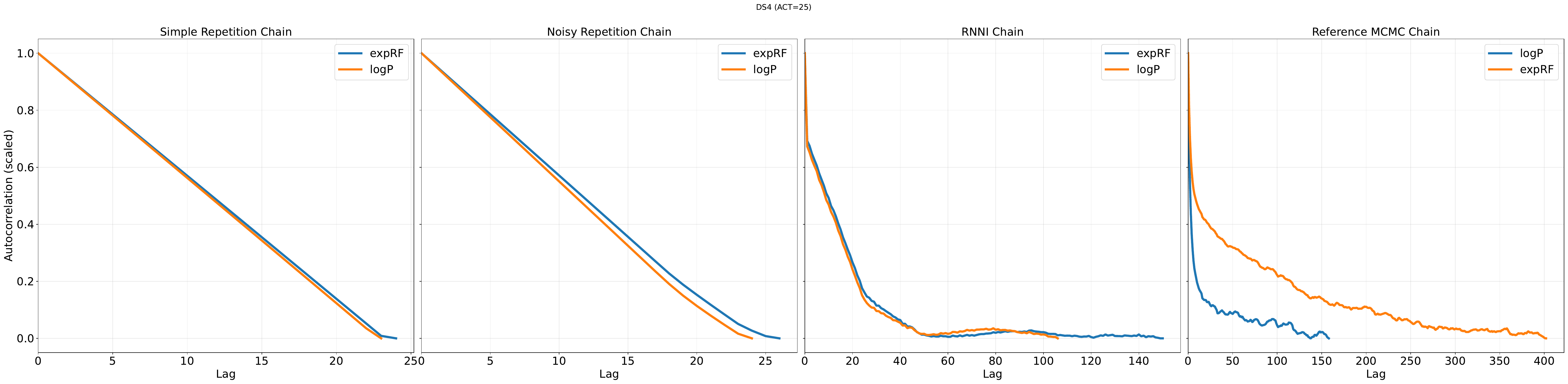}
\signatureGrid{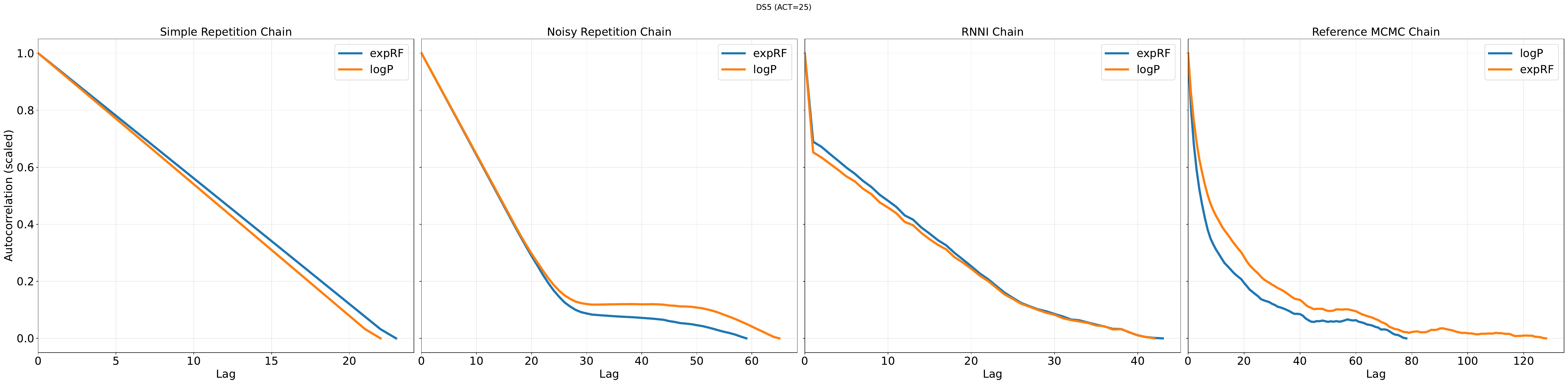}
\signatureGrid{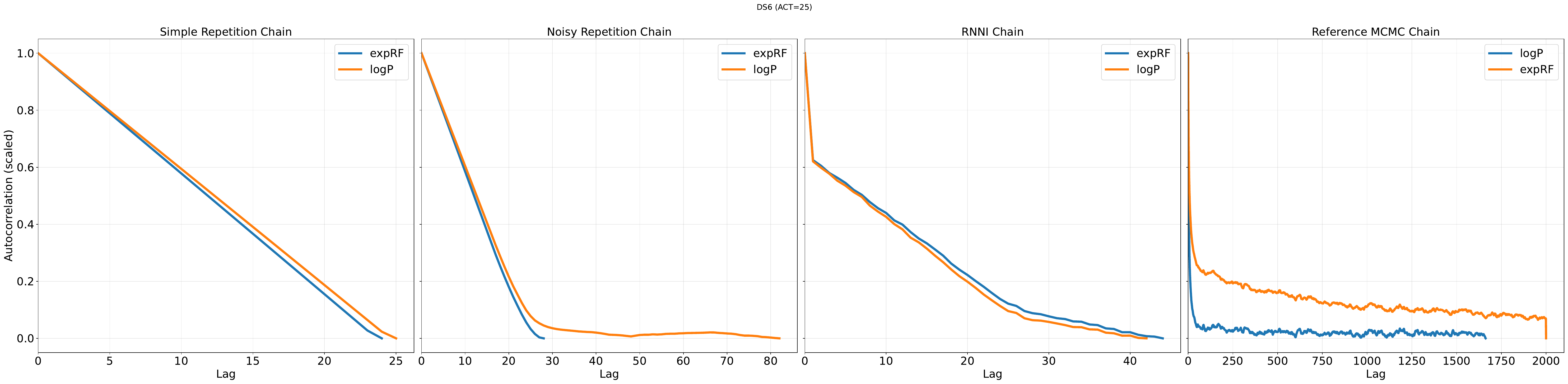}
\signatureGrid{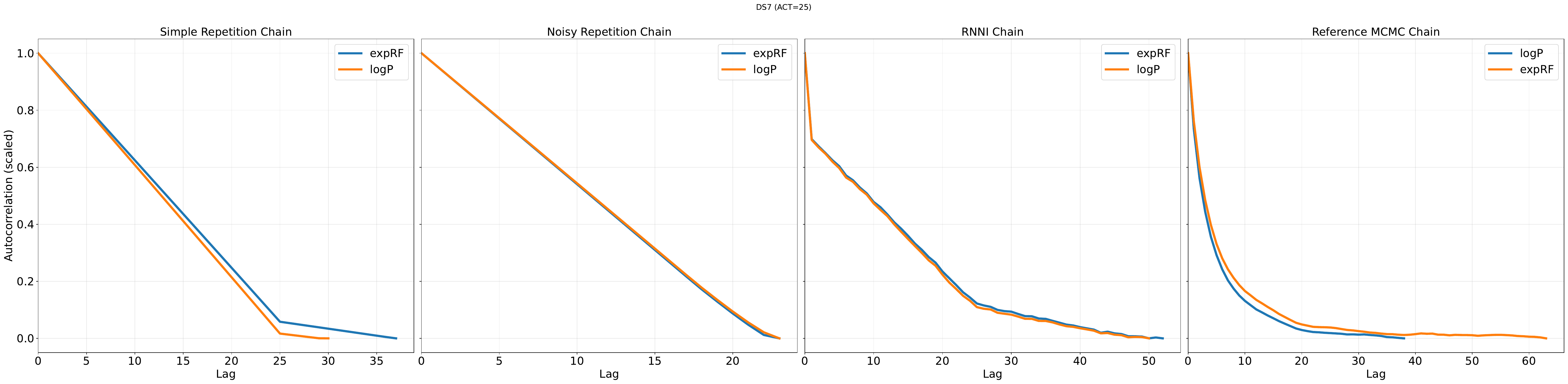}
\signatureGrid{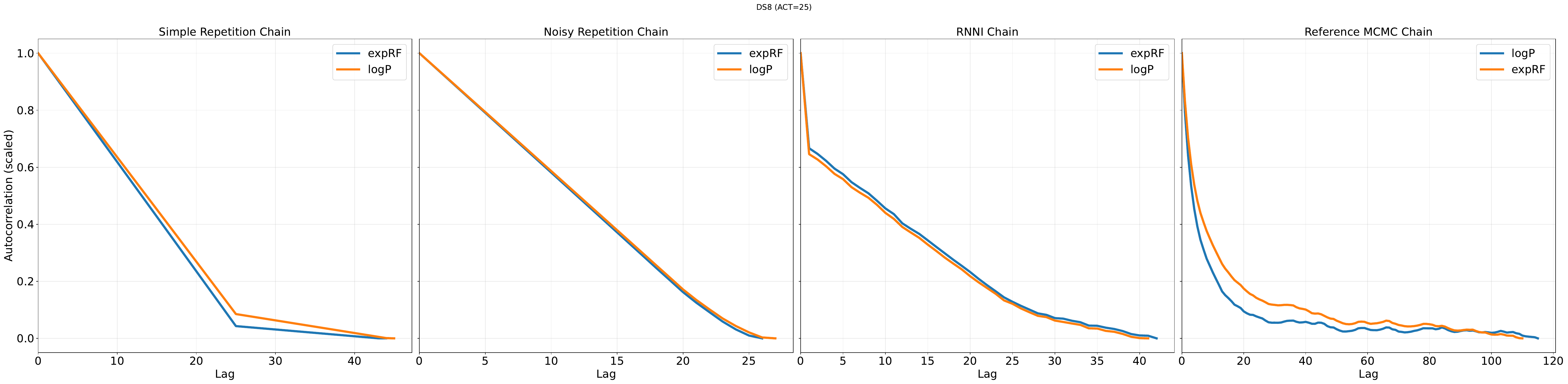}
\signatureGrid{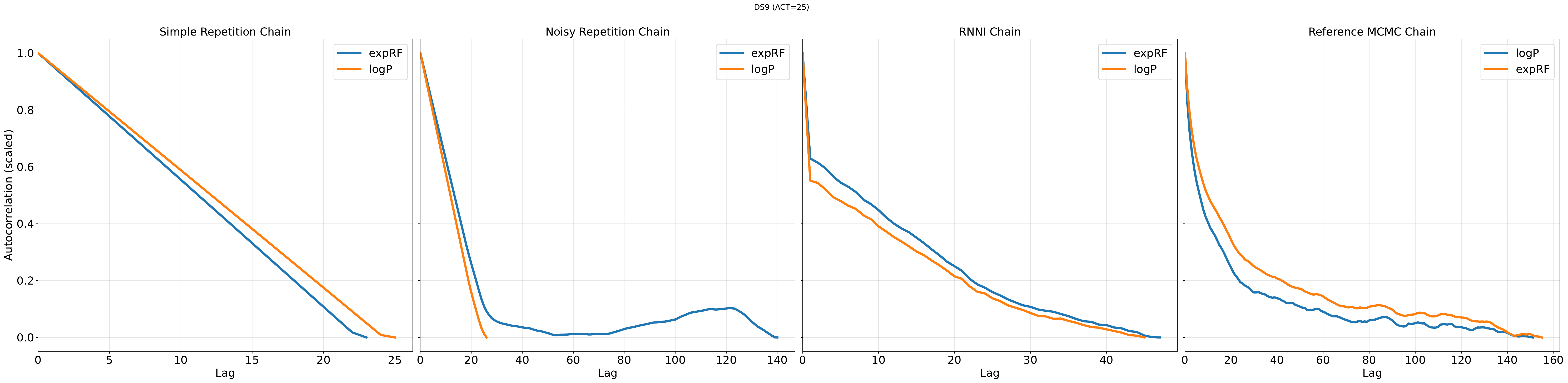}
\signatureGrid{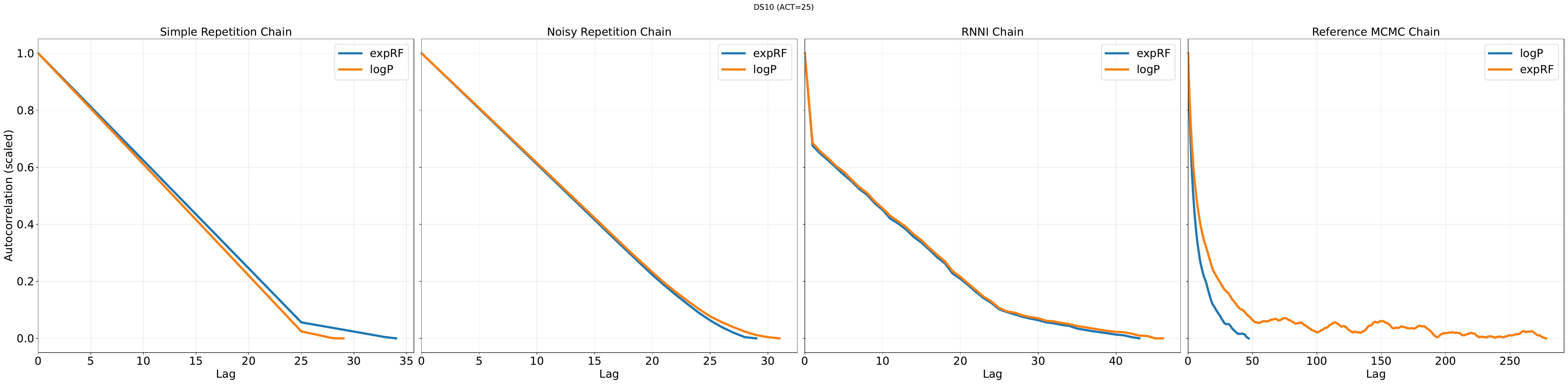}
\signatureGrid{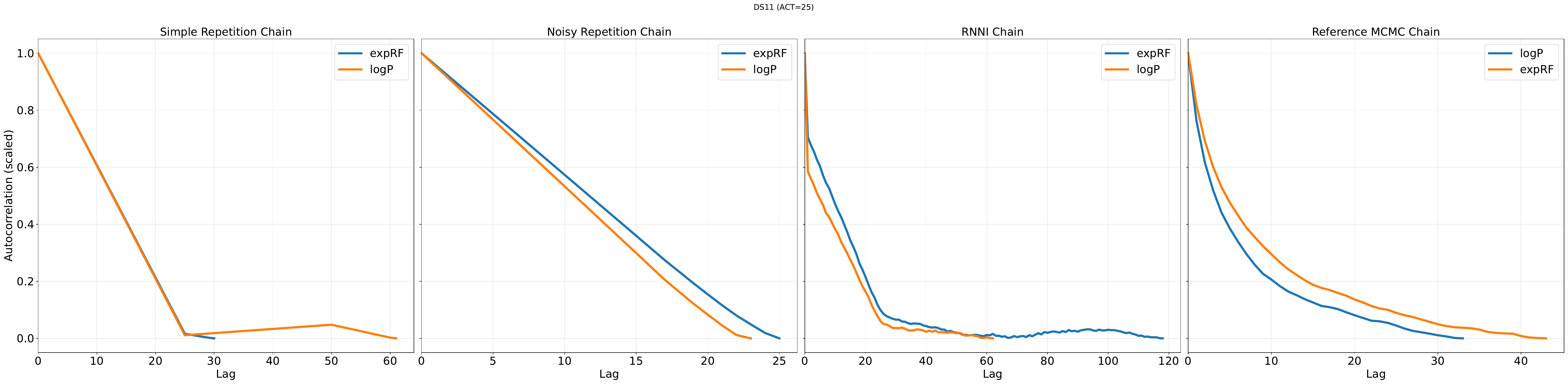}

\FloatBarrier
\newpage
\section{Accuracy} \label{asec:accuracy}  
Here we provide additional plots on the accuracy experiments in \cref{sec:accuracy}.

\subsection{Relative Accuracy}
In addition to the plots for DS3 and DS4 in \cref{fig:acc_results_xy:rel}, we show here the relative accuracy plots (including the EDSF estimators) for DS1--11 and a target ACT of 5 in
\Crefrange{sfig:accuracy:rel:DS1}{sfig:accuracy:rel:DS11}.

\fullAccuracyRel{1}
\fullAccuracyRel{2}
\fullAccuracyRel{3}
\fullAccuracyRel{4}
\fullAccuracyRel{5}
\fullAccuracyRel{6}
\fullAccuracyRel{7}
\fullAccuracyRel{8}
\fullAccuracyRel{9}
\fullAccuracyRel{10}
\fullAccuracyRel{11}

\FloatBarrier
\subsection{Accuracy Summaries}  
\begin{figure}[ht]
	\centering
	\includegraphics[width=\textwidth]{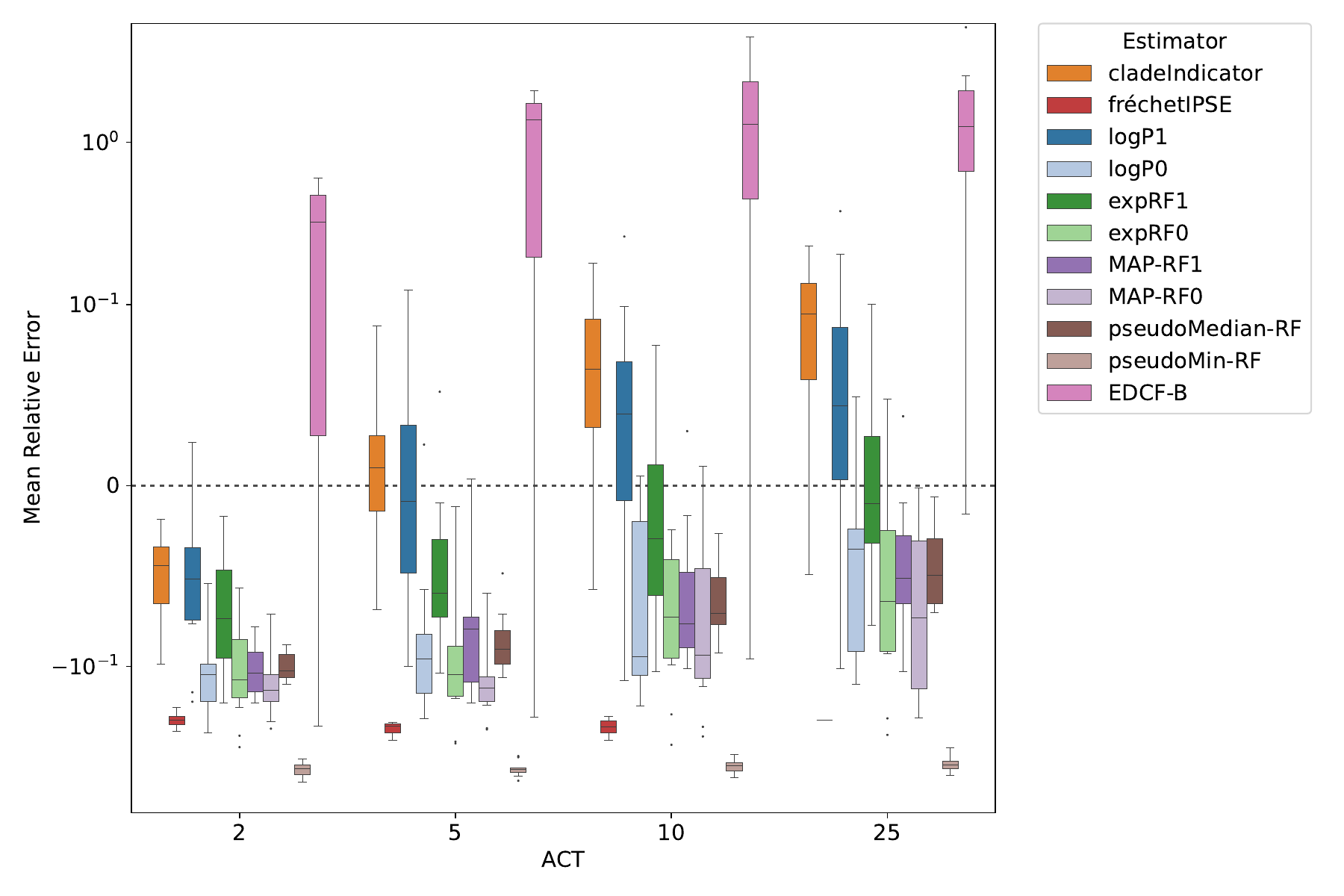}
	\caption{Summary of estimator accuracy on simulated RNNI chains for the 11 real datasets.}
	\label{sfig:accuracy:summary:ds}
\end{figure}

\begin{figure}[!ht]
	\centering
	\includegraphics[width=\textwidth]{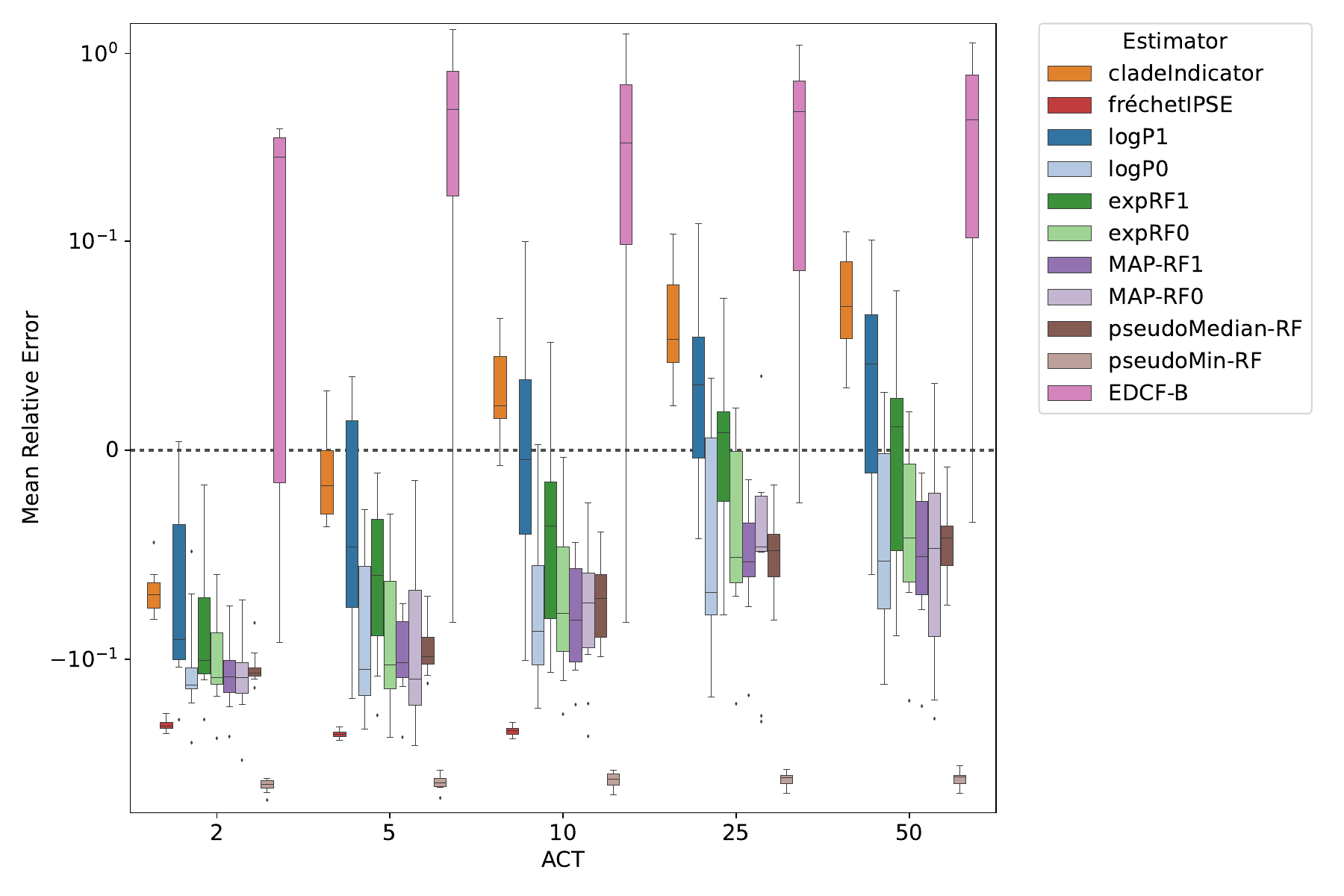}
	\caption{Summary of estimator accuracy on simulated RNNI chains for the 10 Yule simulations.}
	\label{sfig:accuracy:summary:yule}
\end{figure}

In \cref{sfig:accuracy:summary:ds,sfig:accuracy:summary:yule} we provide summary plots of the estimator accuracy across different target ACTs and DS1--11 as well as the Yule datasets.
Note that some estimators are only shown for small ACTs due to their computational cost.

\FloatBarrier
\newpage
\section{Stability} \label{asec:stab}  
Here we provide additional plots on the stability experiment in \cref{sec:stability},
where we started with oversampled versions of DS1--11 and then sequentially thinned out the tree sets.
The plots shown in \Crefrange{sfig:stab:ds1}{sfig:stab:ds11}
also include additional tree ESS estimators.

\stabilitySingle{1}
\stabilitySingle{2}
\stabilitySingle{3}
\stabilitySingle{4}
\stabilitySingle{5}
\stabilitySingle{6}
\stabilitySingle{7}
\stabilitySingle{8}
\stabilitySingle{9}
\stabilitySingle{10}
\stabilitySingle{11}

\FloatBarrier
\newpage
\section{Robustness} \label{asec:robustness}  
In \Crefrange{sfig:robustness:ds1}{sfig:robustness:ds11}, we provide additional plots on the robustness experiments in \cref{sec:robustness} that include more tree ESS estimators.

\robustnessAppendix{1}
\robustnessAppendix{2}
\robustnessAppendix{3}
\robustnessAppendix{4}
\robustnessAppendix{5}
\robustnessAppendix{6}
\robustnessAppendix{7}
\robustnessAppendix{8}
\robustnessAppendix{9}
\robustnessAppendix{10}
\robustnessAppendix{11}

\FloatBarrier
\subsection{Summarizing Robustness} \label{asec:summary:robustness}
In \cref{stab:robustness:summary:10k:run1:ds,stab:robustness:summary:10k:run1:yule}, we provide summary tables of the robustness experiments that contain all the estimators for the DS1--11 and Yule datasets.

\begin{table}[!ht]
	\centering
	\caption{Robustness of ESS estimates across 11 datasets (10,000 trees), with lower values indicating more robust performance.}
	\label{stab:robustness:summary:10k:run1:ds}
	\resizebox{1\textwidth}{!}{
		\begin{tabular}{lccccccccccc}
        \toprule
         & 1 & 2 & 3 & 4 & 5 & 6 & 7 & 8 & 9 & 10 & 11 \\
        \midrule
        cladeIndicator & $\colorbox{red!25}{0.18} \pm \colorbox{red!25}{0.12}$ & $\colorbox{green!25}{0.02} \pm \colorbox{green!25}{0.02}$ & $\colorbox{green!25}{0.02} \pm \colorbox{green!25}{0.01}$ & $\colorbox{red!25}{0.10} \pm \colorbox{orange!25}{0.06}$ & $\colorbox{orange!25}{0.09} \pm \colorbox{orange!25}{0.06}$ & $\colorbox{orange!25}{0.09} \pm \colorbox{orange!25}{0.06}$ & $\colorbox{green!25}{0.01} \pm \colorbox{green!25}{0.01}$ & $\colorbox{orange!25}{0.08} \pm \colorbox{green!25}{0.05}$ & $\colorbox{orange!25}{0.08} \pm \colorbox{green!25}{0.05}$ & $\colorbox{red!25}{0.10} \pm \colorbox{orange!25}{0.06}$ & $\colorbox{green!25}{0.02} \pm \colorbox{green!25}{0.02}$ \\
        fréchetIPSE & $\colorbox{red!25}{0.27} \pm \colorbox{red!25}{0.19}$ & $\colorbox{green!25}{0.03} \pm \colorbox{green!25}{0.02}$ & $\colorbox{green!25}{0.02} \pm \colorbox{green!25}{0.02}$ & $\colorbox{red!25}{0.40} \pm \colorbox{red!25}{0.27}$ & $\colorbox{orange!25}{0.09} \pm \colorbox{orange!25}{0.06}$ & $\colorbox{red!25}{0.31} \pm \colorbox{red!25}{0.28}$ & $\colorbox{orange!25}{0.05} \pm \colorbox{green!25}{0.04}$ & $\colorbox{orange!25}{0.05} \pm \colorbox{green!25}{0.04}$ & $\colorbox{orange!25}{0.09} \pm \colorbox{orange!25}{0.06}$ & $\colorbox{red!25}{0.29} \pm \colorbox{red!25}{0.19}$ & $\colorbox{orange!25}{0.05} \pm \colorbox{green!25}{0.03}$ \\
        logP1 & $\colorbox{red!25}{0.21} \pm \colorbox{red!25}{0.16}$ & $\colorbox{orange!25}{0.06} \pm \colorbox{green!25}{0.04}$ & $\colorbox{green!25}{0.04} \pm \colorbox{green!25}{0.03}$ & $\colorbox{red!25}{0.26} \pm \colorbox{red!25}{0.18}$ & $\colorbox{red!25}{0.14} \pm \colorbox{orange!25}{0.09}$ & $\colorbox{red!25}{0.18} \pm \colorbox{red!25}{0.13}$ & $\colorbox{green!25}{0.05} \pm \colorbox{green!25}{0.04}$ & $\colorbox{red!25}{0.19} \pm \colorbox{red!25}{0.12}$ & $\colorbox{red!25}{0.19} \pm \colorbox{red!25}{0.13}$ & $\colorbox{red!25}{0.15} \pm \colorbox{red!25}{0.11}$ & $\colorbox{red!25}{0.17} \pm \colorbox{red!25}{0.12}$ \\
        logP0 & $\colorbox{red!25}{0.18} \pm \colorbox{red!25}{0.12}$ & $\colorbox{orange!25}{0.07} \pm \colorbox{green!25}{0.04}$ & $\colorbox{green!25}{0.04} \pm \colorbox{green!25}{0.03}$ & $\colorbox{red!25}{0.42} \pm \colorbox{red!25}{0.28}$ & $\colorbox{red!25}{0.18} \pm \colorbox{red!25}{0.11}$ & $\colorbox{red!25}{0.30} \pm \colorbox{red!25}{0.27}$ & $\colorbox{green!25}{0.03} \pm \colorbox{green!25}{0.02}$ & $\colorbox{red!25}{0.22} \pm \colorbox{red!25}{0.14}$ & $\colorbox{red!25}{0.26} \pm \colorbox{red!25}{0.17}$ & $\colorbox{red!25}{0.21} \pm \colorbox{red!25}{0.14}$ & $\colorbox{red!25}{0.25} \pm \colorbox{red!25}{0.16}$ \\
        expRF1 & $\colorbox{red!25}{0.19} \pm \colorbox{red!25}{0.13}$ & $\colorbox{orange!25}{0.06} \pm \colorbox{green!25}{0.04}$ & $\colorbox{green!25}{0.02} \pm \colorbox{green!25}{0.01}$ & $\colorbox{red!25}{0.43} \pm \colorbox{red!25}{0.29}$ & $\colorbox{red!25}{0.16} \pm \colorbox{red!25}{0.11}$ & $\colorbox{red!25}{0.30} \pm \colorbox{red!25}{0.29}$ & $\colorbox{green!25}{0.04} \pm \colorbox{green!25}{0.03}$ & $\colorbox{red!25}{0.20} \pm \colorbox{red!25}{0.13}$ & $\colorbox{red!25}{0.23} \pm \colorbox{red!25}{0.15}$ & $\colorbox{red!25}{0.26} \pm \colorbox{red!25}{0.17}$ & $\colorbox{red!25}{0.27} \pm \colorbox{red!25}{0.20}$ \\
        expRF0 & $\colorbox{red!25}{0.18} \pm \colorbox{red!25}{0.12}$ & $\colorbox{orange!25}{0.06} \pm \colorbox{green!25}{0.04}$ & $\colorbox{green!25}{0.02} \pm \colorbox{green!25}{0.01}$ & $\colorbox{red!25}{0.43} \pm \colorbox{red!25}{0.29}$ & $\colorbox{red!25}{0.14} \pm \colorbox{orange!25}{0.10}$ & $\colorbox{red!25}{0.31} \pm \colorbox{red!25}{0.27}$ & $\colorbox{green!25}{0.03} \pm \colorbox{green!25}{0.02}$ & $\colorbox{red!25}{0.18} \pm \colorbox{red!25}{0.12}$ & $\colorbox{red!25}{0.24} \pm \colorbox{red!25}{0.16}$ & $\colorbox{red!25}{0.27} \pm \colorbox{red!25}{0.19}$ & $\colorbox{red!25}{0.20} \pm \colorbox{red!25}{0.14}$ \\
        MAP-RF1 & $\colorbox{red!25}{0.18} \pm \colorbox{red!25}{0.12}$ & $\colorbox{orange!25}{0.06} \pm \colorbox{green!25}{0.04}$ & $\colorbox{green!25}{0.03} \pm \colorbox{green!25}{0.02}$ & $\colorbox{red!25}{0.41} \pm \colorbox{red!25}{0.28}$ & $\colorbox{red!25}{0.12} \pm \colorbox{orange!25}{0.08}$ & $\colorbox{red!25}{0.35} \pm \colorbox{red!25}{0.31}$ & $\colorbox{green!25}{0.05} \pm \colorbox{green!25}{0.03}$ & $\colorbox{red!25}{0.12} \pm \colorbox{orange!25}{0.09}$ & $\colorbox{red!25}{0.16} \pm \colorbox{red!25}{0.11}$ & $\colorbox{red!25}{0.33} \pm \colorbox{red!25}{0.21}$ & $\colorbox{red!25}{0.18} \pm \colorbox{red!25}{0.13}$ \\
        MAP-RF0 & $\colorbox{red!25}{0.33} \pm \colorbox{red!25}{0.20}$ & $\colorbox{orange!25}{0.06} \pm \colorbox{green!25}{0.04}$ & $\colorbox{green!25}{0.03} \pm \colorbox{green!25}{0.02}$ & $\colorbox{red!25}{0.41} \pm \colorbox{red!25}{0.28}$ & $\colorbox{red!25}{0.13} \pm \colorbox{orange!25}{0.09}$ & $\colorbox{red!25}{0.34} \pm \colorbox{red!25}{0.30}$ & $\colorbox{green!25}{0.05} \pm \colorbox{green!25}{0.03}$ & $\colorbox{red!25}{0.15} \pm \colorbox{red!25}{0.11}$ & $\colorbox{red!25}{0.19} \pm \colorbox{red!25}{0.12}$ & $\colorbox{red!25}{0.29} \pm \colorbox{red!25}{0.20}$ & $\colorbox{red!25}{0.12} \pm \colorbox{red!25}{0.14}$ \\
        pseudoMedian-RF & $\colorbox{red!25}{0.28} \pm \colorbox{red!25}{0.20}$ & $\colorbox{green!25}{0.03} \pm \colorbox{green!25}{0.02}$ & $\colorbox{green!25}{0.04} \pm \colorbox{green!25}{0.03}$ & $\colorbox{red!25}{0.41} \pm \colorbox{red!25}{0.28}$ & $\colorbox{orange!25}{0.08} \pm \colorbox{green!25}{0.05}$ & $\colorbox{red!25}{0.33} \pm \colorbox{red!25}{0.30}$ & $\colorbox{green!25}{0.04} \pm \colorbox{green!25}{0.03}$ & $\colorbox{orange!25}{0.07} \pm \colorbox{green!25}{0.04}$ & $\colorbox{red!25}{0.11} \pm \colorbox{orange!25}{0.07}$ & $\colorbox{red!25}{0.32} \pm \colorbox{red!25}{0.21}$ & $\colorbox{green!25}{0.03} \pm \colorbox{green!25}{0.02}$ \\
        pseudoMin-RF & $\colorbox{red!25}{0.28} \pm \colorbox{red!25}{0.23}$ & $\colorbox{red!25}{0.10} \pm \colorbox{orange!25}{0.08}$ & $\colorbox{red!25}{0.12} \pm \colorbox{orange!25}{0.08}$ & $\colorbox{red!25}{0.39} \pm \colorbox{red!25}{0.28}$ & $\colorbox{orange!25}{0.08} \pm \colorbox{orange!25}{0.06}$ & $\colorbox{red!25}{0.27} \pm \colorbox{red!25}{0.31}$ & $\colorbox{red!25}{0.13} \pm \colorbox{orange!25}{0.09}$ & $\colorbox{red!25}{0.10} \pm \colorbox{orange!25}{0.08}$ & $\colorbox{orange!25}{0.10} \pm \colorbox{orange!25}{0.08}$ & $\colorbox{red!25}{0.27} \pm \colorbox{red!25}{0.20}$ & $\colorbox{orange!25}{0.09} \pm \colorbox{orange!25}{0.10}$ \\
        EDCF-B & $\colorbox{red!25}{0.24} \pm \colorbox{red!25}{0.24}$ & $\colorbox{green!25}{0.03} \pm \colorbox{green!25}{0.02}$ & $\colorbox{green!25}{0.01} \pm \colorbox{green!25}{0.00}$ & $\colorbox{red!25}{0.18} \pm \colorbox{red!25}{0.14}$ & $\colorbox{orange!25}{0.07} \pm \colorbox{orange!25}{0.06}$ & $\colorbox{red!25}{0.26} \pm \colorbox{red!25}{0.24}$ & $\colorbox{red!25}{0.22} \pm \colorbox{red!25}{0.15}$ & $\colorbox{red!25}{0.15} \pm \colorbox{red!25}{0.11}$ & $\colorbox{green!25}{0.03} \pm \colorbox{green!25}{0.02}$ & $\colorbox{red!25}{0.28} \pm \colorbox{red!25}{0.18}$ & $\colorbox{orange!25}{0.05} \pm \colorbox{green!25}{0.03}$ \\
        EDCF-ML & $\colorbox{red!25}{0.22} \pm \colorbox{red!25}{0.29}$ & $\colorbox{red!25}{0.15} \pm \colorbox{red!25}{0.11}$ & $\colorbox{red!25}{0.15} \pm \colorbox{red!25}{0.11}$ & $\colorbox{red!25}{0.20} \pm \colorbox{red!25}{0.14}$ & $\colorbox{red!25}{0.12} \pm \colorbox{orange!25}{0.09}$ & $\colorbox{red!25}{0.24} \pm \colorbox{red!25}{0.26}$ & $\colorbox{red!25}{0.23} \pm \colorbox{red!25}{0.20}$ & $\colorbox{red!25}{0.16} \pm \colorbox{red!25}{0.13}$ & $\colorbox{orange!25}{0.10} \pm \colorbox{orange!25}{0.07}$ & $\colorbox{red!25}{0.27} \pm \colorbox{red!25}{0.18}$ & $\colorbox{orange!25}{0.06} \pm \colorbox{orange!25}{0.06}$ \\
        \bottomrule
\end{tabular}
	}
\end{table}

\begin{table}[!th]
	\centering
	\caption{Robustness of ESS estimates across 10 Yule datasets (10,000 trees), with lower values indicating more stable performance.}
	\label{stab:robustness:summary:10k:run1:yule}
	\resizebox{1\textwidth}{!}{
		\begin{tabular}{lcccccccccc}
        \toprule
         & 1 & 2 & 3 & 4 & 5 & 6 & 7 & 8 & 9 & 10 \\
        \midrule
        cladeIndicator & $\colorbox{green!25}{0.00} \pm \colorbox{green!25}{0.00}$ & $\colorbox{green!25}{0.01} \pm \colorbox{green!25}{0.00}$ & $\colorbox{green!25}{0.01} \pm \colorbox{green!25}{0.01}$ & $\colorbox{green!25}{0.00} \pm \colorbox{green!25}{0.00}$ & $\colorbox{green!25}{0.00} \pm \colorbox{green!25}{0.00}$ & $\colorbox{green!25}{0.01} \pm \colorbox{green!25}{0.01}$ & $\colorbox{green!25}{0.01} \pm \colorbox{green!25}{0.00}$ & $\colorbox{green!25}{0.00} \pm \colorbox{green!25}{0.00}$ & $\colorbox{green!25}{0.00} \pm \colorbox{green!25}{0.00}$ & $\colorbox{green!25}{0.01} \pm \colorbox{green!25}{0.00}$ \\
        fréchetIPSE & $\colorbox{green!25}{0.00} \pm \colorbox{green!25}{0.00}$ & $\colorbox{green!25}{0.00} \pm \colorbox{green!25}{0.00}$ & $\colorbox{green!25}{0.00} \pm \colorbox{green!25}{0.00}$ & $\colorbox{green!25}{0.00} \pm \colorbox{green!25}{0.00}$ & $\colorbox{green!25}{0.02} \pm \colorbox{green!25}{0.01}$ & $\colorbox{green!25}{0.00} \pm \colorbox{green!25}{0.00}$ & $\colorbox{green!25}{0.00} \pm \colorbox{green!25}{0.00}$ & $\colorbox{green!25}{0.00} \pm \colorbox{green!25}{0.00}$ & $\colorbox{green!25}{0.00} \pm \colorbox{green!25}{0.00}$ & $\colorbox{green!25}{0.01} \pm \colorbox{green!25}{0.00}$ \\
        logP1 & $\colorbox{green!25}{0.04} \pm \colorbox{green!25}{0.03}$ & $\colorbox{green!25}{0.02} \pm \colorbox{green!25}{0.01}$ & $\colorbox{green!25}{0.02} \pm \colorbox{green!25}{0.01}$ & $\colorbox{green!25}{0.01} \pm \colorbox{green!25}{0.01}$ & $\colorbox{green!25}{0.02} \pm \colorbox{green!25}{0.02}$ & $\colorbox{green!25}{0.04} \pm \colorbox{green!25}{0.03}$ & $\colorbox{green!25}{0.02} \pm \colorbox{green!25}{0.01}$ & $\colorbox{green!25}{0.01} \pm \colorbox{green!25}{0.01}$ & $\colorbox{green!25}{0.04} \pm \colorbox{green!25}{0.02}$ & $\colorbox{green!25}{0.01} \pm \colorbox{green!25}{0.01}$ \\
        logP0 & $\colorbox{green!25}{0.02} \pm \colorbox{green!25}{0.02}$ & $\colorbox{green!25}{0.01} \pm \colorbox{green!25}{0.01}$ & $\colorbox{green!25}{0.02} \pm \colorbox{green!25}{0.01}$ & $\colorbox{green!25}{0.01} \pm \colorbox{green!25}{0.01}$ & $\colorbox{green!25}{0.03} \pm \colorbox{green!25}{0.02}$ & $\colorbox{green!25}{0.02} \pm \colorbox{green!25}{0.01}$ & $\colorbox{green!25}{0.03} \pm \colorbox{green!25}{0.02}$ & $\colorbox{green!25}{0.01} \pm \colorbox{green!25}{0.01}$ & $\colorbox{green!25}{0.02} \pm \colorbox{green!25}{0.02}$ & $\colorbox{green!25}{0.01} \pm \colorbox{green!25}{0.01}$ \\
        expRF1 & $\colorbox{green!25}{0.02} \pm \colorbox{green!25}{0.01}$ & $\colorbox{green!25}{0.01} \pm \colorbox{green!25}{0.01}$ & $\colorbox{green!25}{0.02} \pm \colorbox{green!25}{0.01}$ & $\colorbox{green!25}{0.01} \pm \colorbox{green!25}{0.01}$ & $\colorbox{green!25}{0.02} \pm \colorbox{green!25}{0.02}$ & $\colorbox{green!25}{0.04} \pm \colorbox{green!25}{0.04}$ & $\colorbox{green!25}{0.03} \pm \colorbox{green!25}{0.02}$ & $\colorbox{green!25}{0.01} \pm \colorbox{green!25}{0.00}$ & $\colorbox{green!25}{0.02} \pm \colorbox{green!25}{0.02}$ & $\colorbox{green!25}{0.01} \pm \colorbox{green!25}{0.01}$ \\
        expRF0 & $\colorbox{green!25}{0.02} \pm \colorbox{green!25}{0.01}$ & $\colorbox{green!25}{0.01} \pm \colorbox{green!25}{0.01}$ & $\colorbox{green!25}{0.01} \pm \colorbox{green!25}{0.01}$ & $\colorbox{green!25}{0.01} \pm \colorbox{green!25}{0.01}$ & $\colorbox{green!25}{0.03} \pm \colorbox{green!25}{0.02}$ & $\colorbox{orange!25}{0.06} \pm \colorbox{green!25}{0.05}$ & $\colorbox{green!25}{0.02} \pm \colorbox{green!25}{0.02}$ & $\colorbox{green!25}{0.01} \pm \colorbox{green!25}{0.01}$ & $\colorbox{green!25}{0.01} \pm \colorbox{green!25}{0.01}$ & $\colorbox{green!25}{0.01} \pm \colorbox{green!25}{0.01}$ \\
        MAP-RF1 & $\colorbox{green!25}{0.02} \pm \colorbox{green!25}{0.01}$ & $\colorbox{green!25}{0.01} \pm \colorbox{green!25}{0.01}$ & $\colorbox{green!25}{0.02} \pm \colorbox{green!25}{0.02}$ & $\colorbox{green!25}{0.04} \pm \colorbox{green!25}{0.02}$ & $\colorbox{green!25}{0.03} \pm \colorbox{green!25}{0.02}$ & $\colorbox{green!25}{0.02} \pm \colorbox{green!25}{0.02}$ & $\colorbox{green!25}{0.04} \pm \colorbox{green!25}{0.03}$ & $\colorbox{green!25}{0.02} \pm \colorbox{green!25}{0.02}$ & $\colorbox{green!25}{0.02} \pm \colorbox{green!25}{0.01}$ & $\colorbox{green!25}{0.02} \pm \colorbox{green!25}{0.01}$ \\
        MAP-RF0 & $\colorbox{green!25}{0.03} \pm \colorbox{green!25}{0.02}$ & $\colorbox{green!25}{0.03} \pm \colorbox{green!25}{0.02}$ & $\colorbox{green!25}{0.02} \pm \colorbox{green!25}{0.02}$ & $\colorbox{green!25}{0.03} \pm \colorbox{green!25}{0.02}$ & $\colorbox{green!25}{0.04} \pm \colorbox{green!25}{0.03}$ & $\colorbox{green!25}{0.04} \pm \colorbox{green!25}{0.03}$ & $\colorbox{green!25}{0.04} \pm \colorbox{green!25}{0.03}$ & $\colorbox{green!25}{0.01} \pm \colorbox{green!25}{0.01}$ & $\colorbox{green!25}{0.02} \pm \colorbox{green!25}{0.02}$ & $\colorbox{green!25}{0.02} \pm \colorbox{green!25}{0.01}$ \\
        pseudoMedian-RF & $\colorbox{green!25}{0.01} \pm \colorbox{green!25}{0.01}$ & $\colorbox{green!25}{0.01} \pm \colorbox{green!25}{0.01}$ & $\colorbox{green!25}{0.01} \pm \colorbox{green!25}{0.01}$ & $\colorbox{green!25}{0.01} \pm \colorbox{green!25}{0.01}$ & $\colorbox{green!25}{0.01} \pm \colorbox{green!25}{0.01}$ & $\colorbox{green!25}{0.01} \pm \colorbox{green!25}{0.01}$ & $\colorbox{green!25}{0.01} \pm \colorbox{green!25}{0.00}$ & $\colorbox{green!25}{0.01} \pm \colorbox{green!25}{0.01}$ & $\colorbox{green!25}{0.01} \pm \colorbox{green!25}{0.01}$ & $\colorbox{green!25}{0.01} \pm \colorbox{green!25}{0.01}$ \\
        pseudoMin-RF & $\colorbox{red!25}{0.10} \pm \colorbox{orange!25}{0.07}$ & $\colorbox{red!25}{0.11} \pm \colorbox{orange!25}{0.08}$ & $\colorbox{orange!25}{0.10} \pm \colorbox{orange!25}{0.08}$ & $\colorbox{red!25}{0.10} \pm \colorbox{orange!25}{0.06}$ & $\colorbox{orange!25}{0.07} \pm \colorbox{green!25}{0.05}$ & $\colorbox{orange!25}{0.08} \pm \colorbox{green!25}{0.05}$ & $\colorbox{orange!25}{0.09} \pm \colorbox{orange!25}{0.07}$ & $\colorbox{orange!25}{0.10} \pm \colorbox{orange!25}{0.07}$ & $\colorbox{orange!25}{0.10} \pm \colorbox{orange!25}{0.08}$ & $\colorbox{red!25}{0.11} \pm \colorbox{orange!25}{0.07}$ \\
        EDCF-B & $\colorbox{green!25}{0.01} \pm \colorbox{green!25}{0.01}$ & $\colorbox{orange!25}{0.06} \pm \colorbox{orange!25}{0.06}$ & $\colorbox{green!25}{0.03} \pm \colorbox{green!25}{0.04}$ & $\colorbox{orange!25}{0.09} \pm \colorbox{red!25}{0.16}$ & $\colorbox{green!25}{0.02} \pm \colorbox{green!25}{0.02}$ & $\colorbox{green!25}{0.03} \pm \colorbox{green!25}{0.03}$ & $\colorbox{green!25}{0.00} \pm \colorbox{green!25}{0.00}$ & $\colorbox{red!25}{0.11} \pm \colorbox{red!25}{0.13}$ & $\colorbox{green!25}{0.00} \pm \colorbox{green!25}{0.00}$ & $\colorbox{red!25}{0.10} \pm \colorbox{orange!25}{0.07}$ \\
        EDCF-ML & $\colorbox{green!25}{0.00} \pm \colorbox{green!25}{0.00}$ & $\colorbox{green!25}{0.03} \pm \colorbox{green!25}{0.03}$ & $\colorbox{green!25}{0.04} \pm \colorbox{green!25}{0.04}$ & $\colorbox{green!25}{0.03} \pm \colorbox{green!25}{0.03}$ & $\colorbox{orange!25}{0.06} \pm \colorbox{green!25}{0.05}$ & $\colorbox{red!25}{0.11} \pm \colorbox{red!25}{0.11}$ & $\colorbox{orange!25}{0.07} \pm \colorbox{orange!25}{0.07}$ & $\colorbox{green!25}{0.04} \pm \colorbox{green!25}{0.03}$ & $\colorbox{green!25}{0.00} \pm \colorbox{green!25}{0.01}$ & $\colorbox{red!25}{0.11} \pm \colorbox{red!25}{0.12}$ \\
        \bottomrule
\end{tabular}

	}
\end{table}

\FloatBarrier
\newpage
\section{MDS and Clustering} \label{asec:mds:cluster}  

For both multidimensional scaling and clustering of posterior tree sets, we use methods implemented in scikit-learn~\cite{sklearn}.

To obtain low-dimensional embeddings, we apply t-distributed Stochastic Neighbour Embedding (t-SNE) to pairwise tree distance matrices.
This method constructs a two-dimensional representation that preserves local neighbourhood structure, making it suitable for visualizing clustering and potential multimodality in treespace.
As with all t-SNE embeddings, global geometric relationships should not be over-interpreted.

To cluster posterior tree sets, we employ spectral clustering.
Distances are first converted into affinities using a heat kernel, after which clustering is performed on a low-dimensional embedding of the resulting graph Laplacian.
This approach allows us to identify groups of similar trees based on their pairwise tree distances.

\subsection{Figures Yule}
\label{asec:yule:mds}

In \Crefrange{fig:mds:yule1}{fig:mds:yule10},we provide MDS plots of two independent chains for the Yule simulations.

\mdsFigureYule{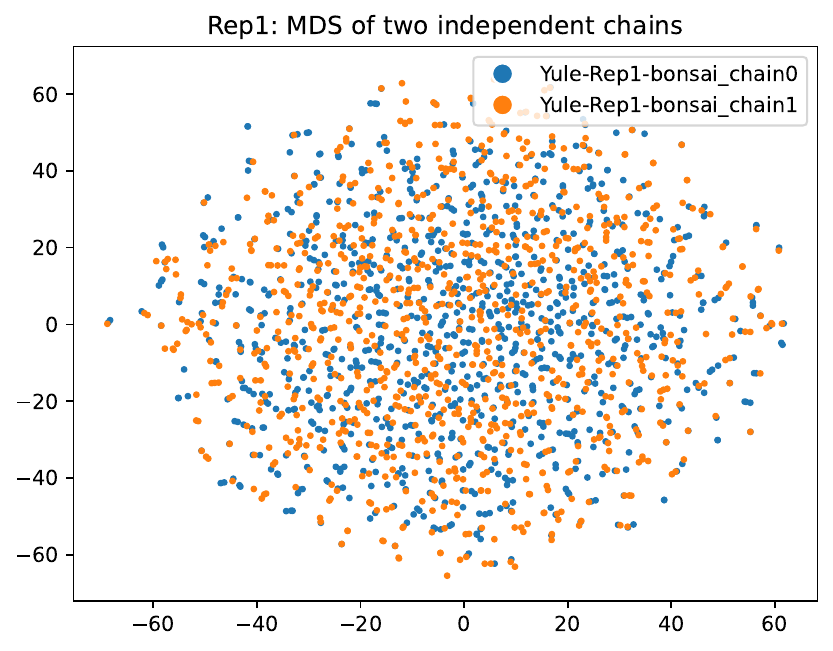}
\mdsFigureYule{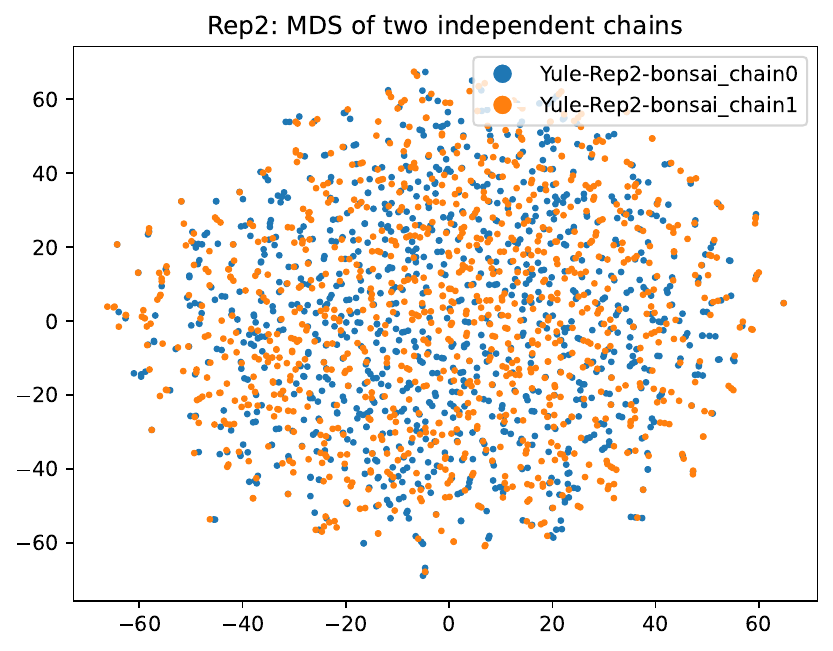}
\mdsFigureYule{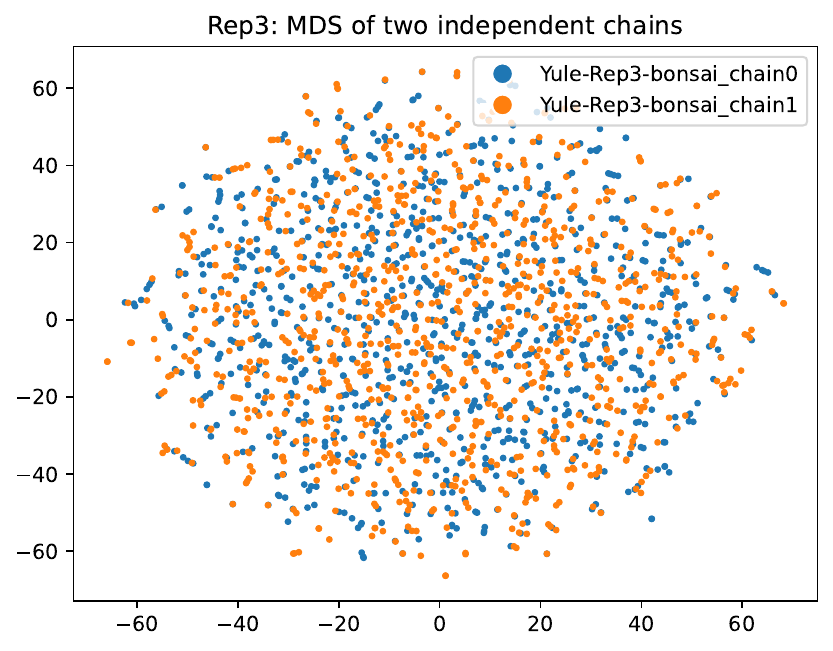}
\mdsFigureYule{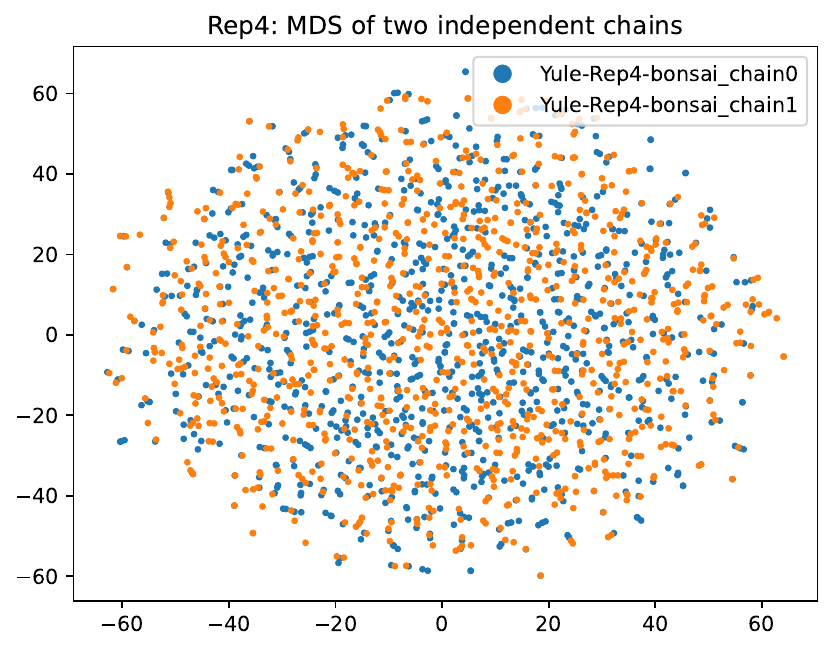}
\mdsFigureYule{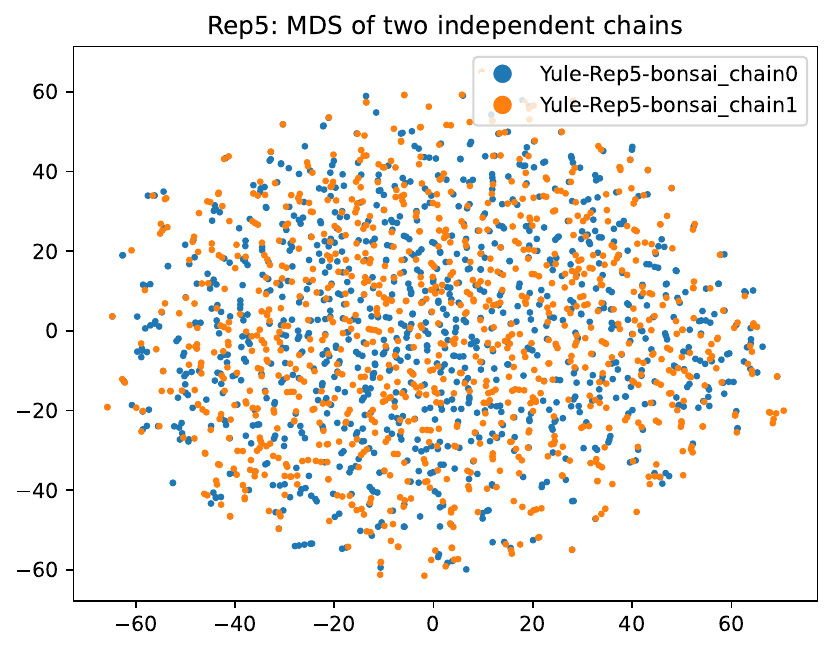}
\mdsFigureYule{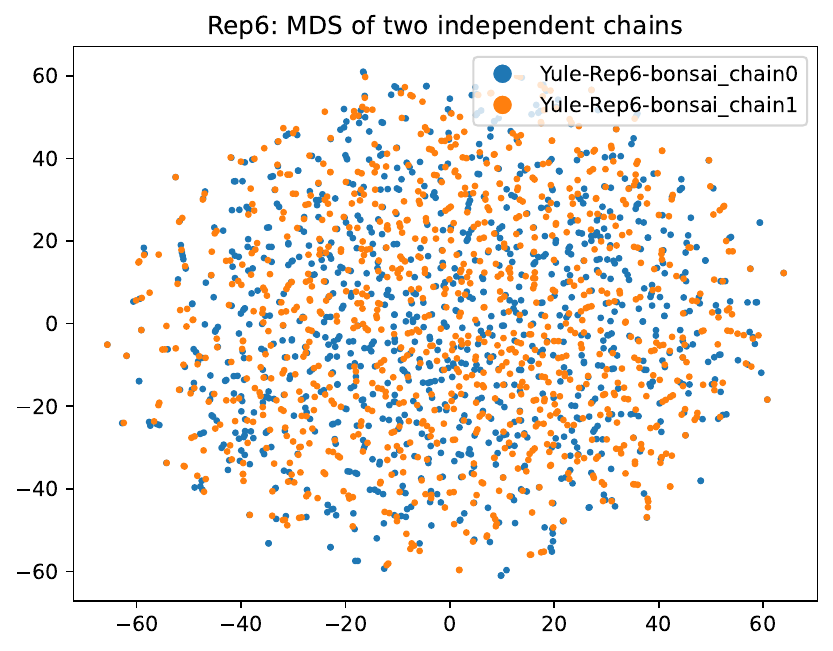}
\mdsFigureYule{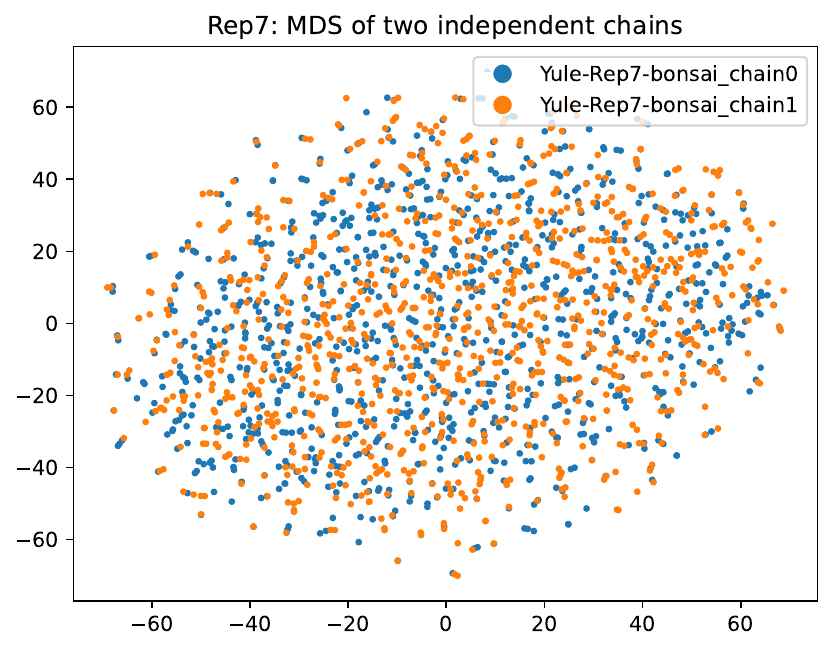}
\mdsFigureYule{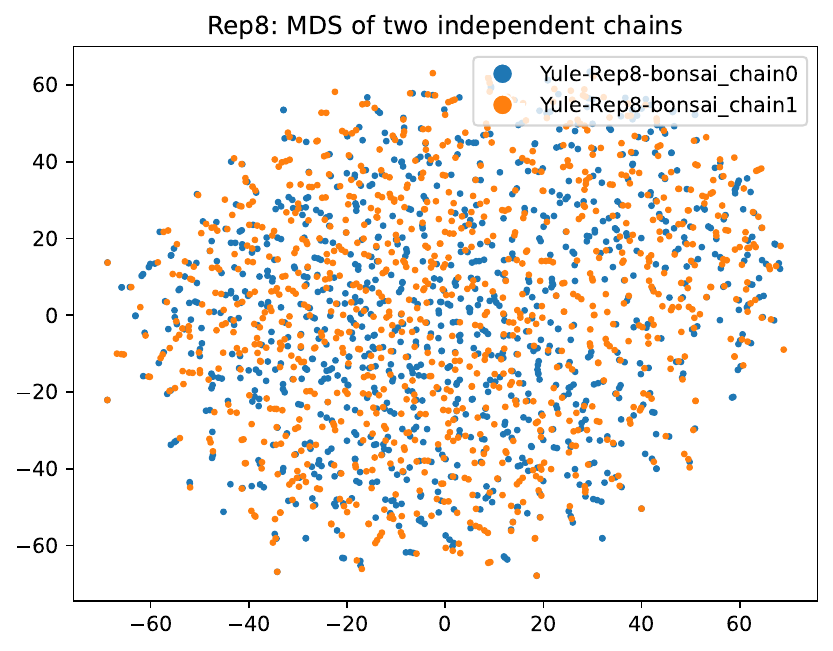}
\mdsFigureYule{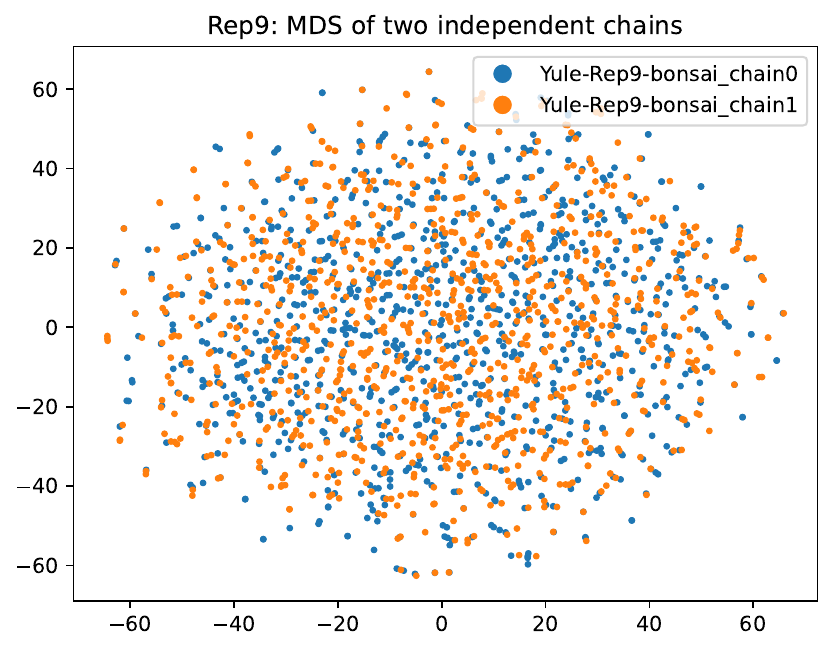}
\mdsFigureYule{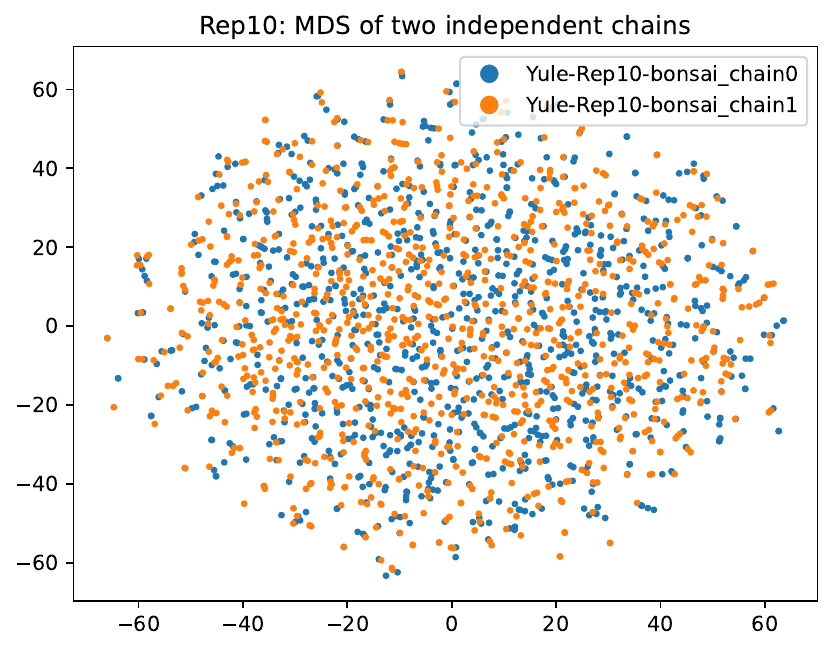}

\FloatBarrier

\subsection{Figures DS1--11}

In \Crefrange{fig:mds:ds1}{fig:mds:ds11}, we provide MDS plots and different clustering visualizations for DS1--11.

\mdsFigure{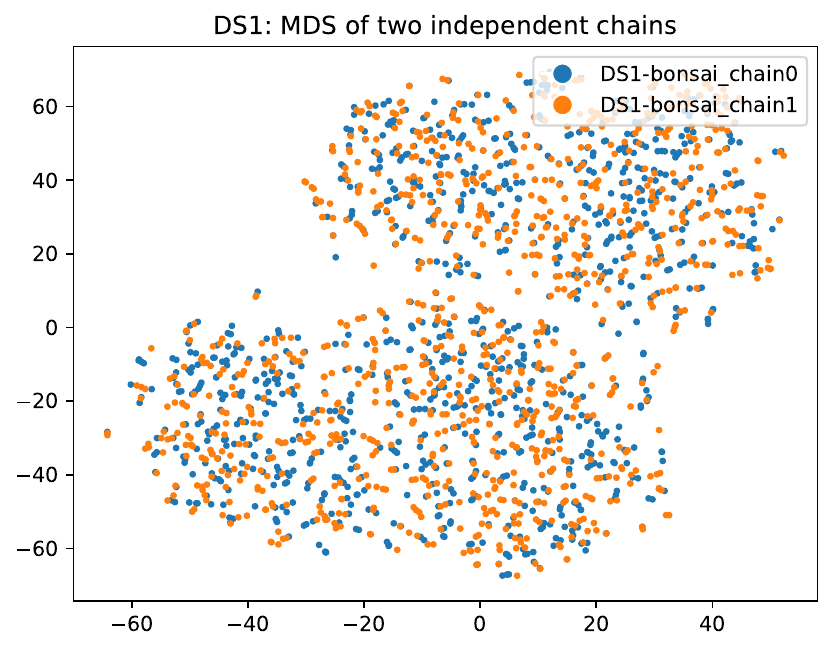}[2][3]
\mdsFigure{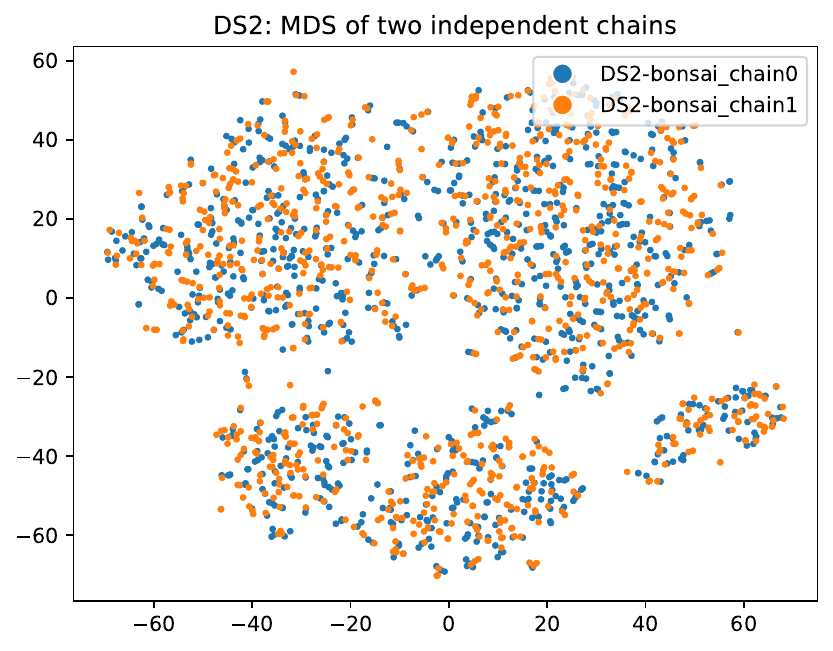}[3][4]
\mdsFigure{3}[2]
\mdsFigure{4}[2][3]
\mdsFigure{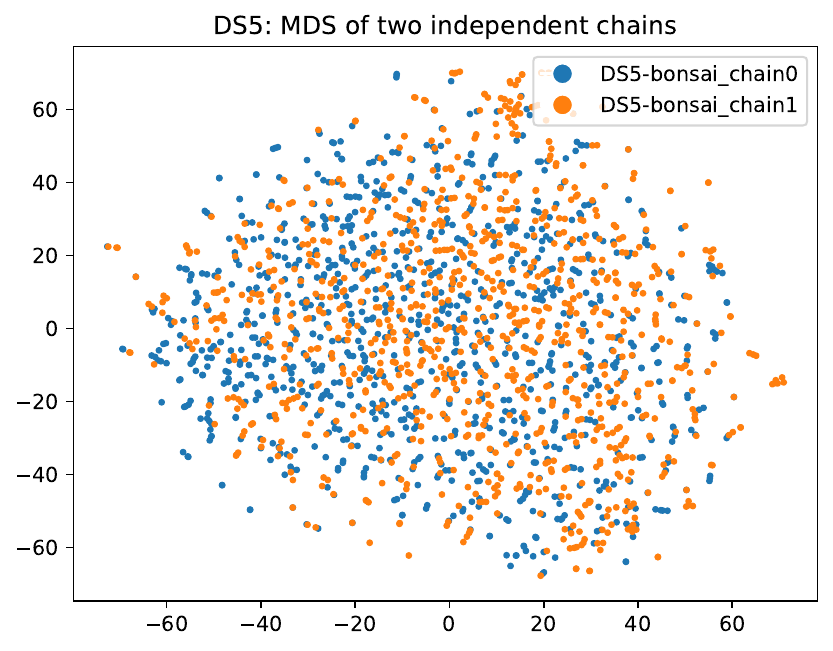}[3]
\mdsFigure{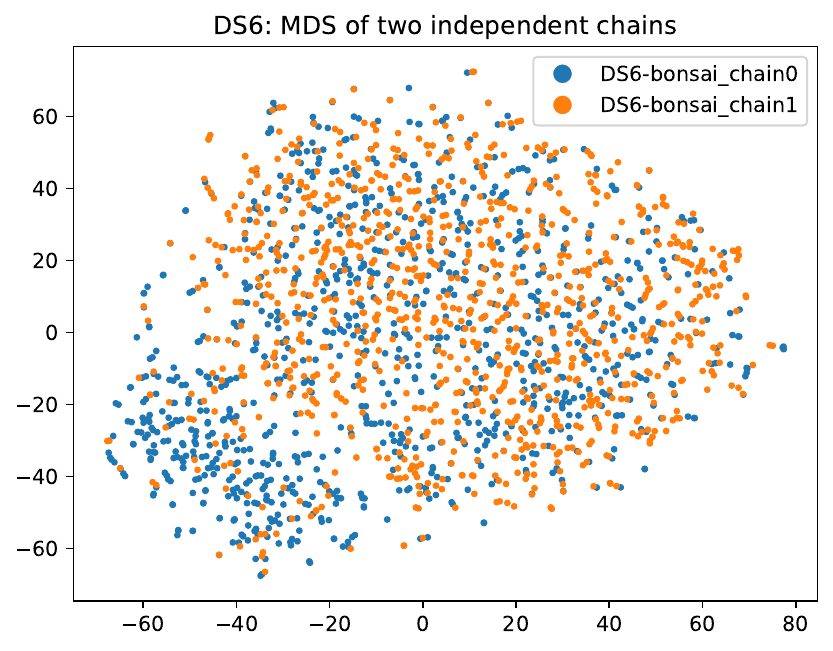}[2][3]
\mdsFigure{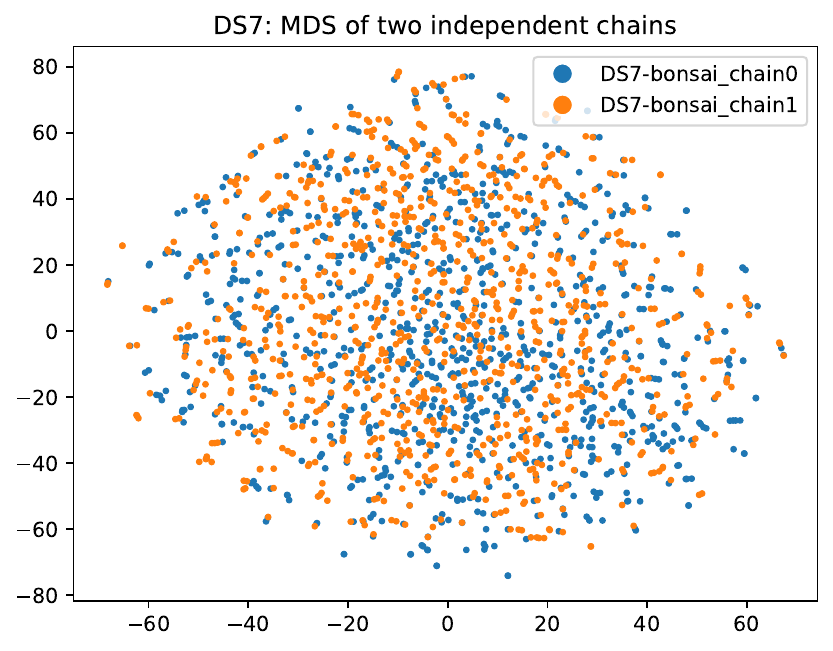}
\mdsFigure{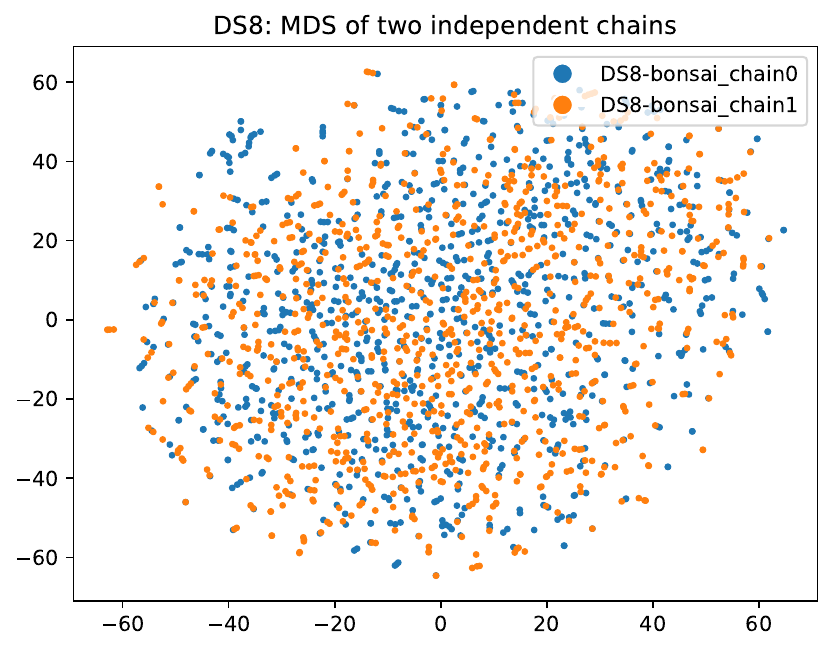}
\mdsFigure{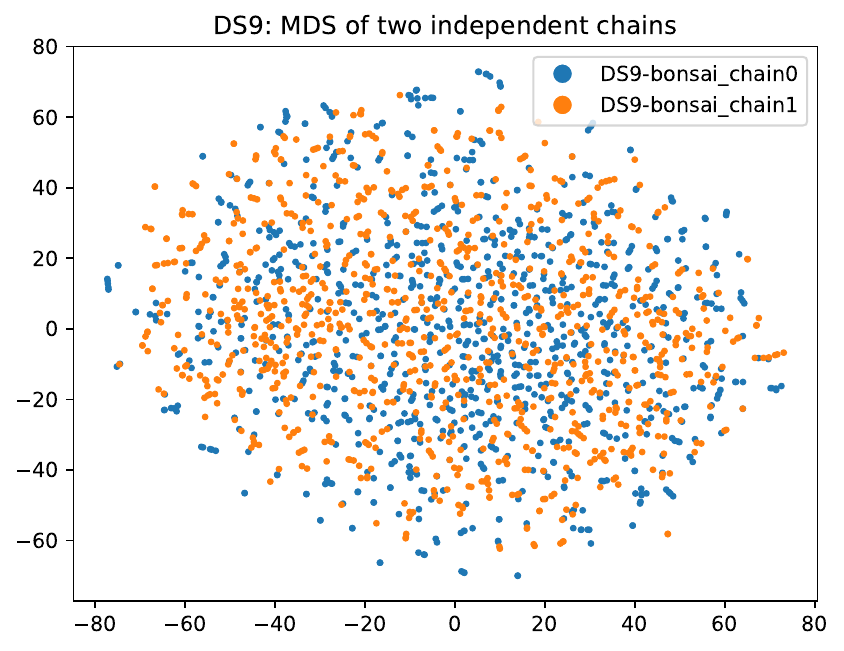}[2]
\mdsFigure{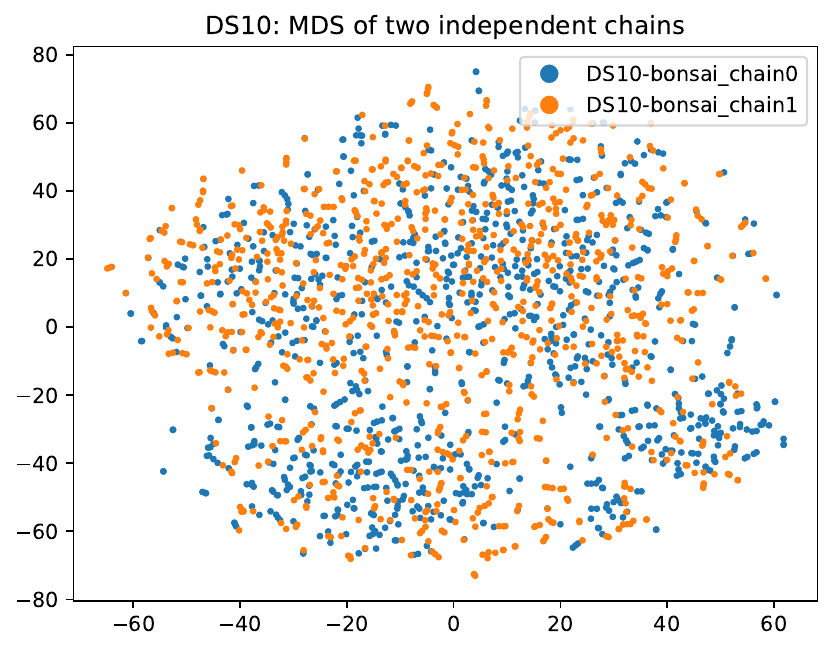}[2][3]
\mdsFigure{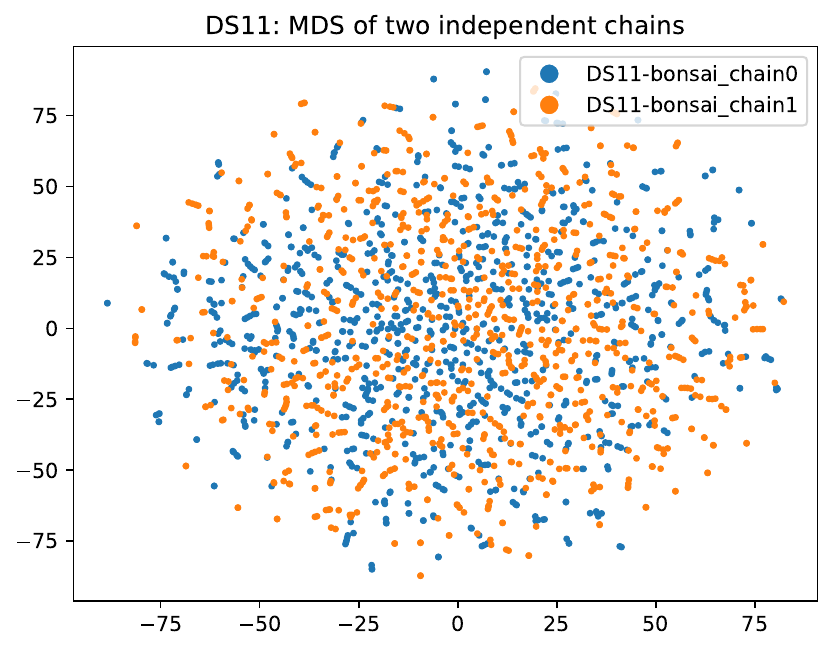}[2][4]

\FloatBarrier
\newpage
\section{Trace Plots} \label{asec:traces}  
In \Crefrange{sfig:jumpdistance:ds1}{sfig:jumpdistance:ds11}, we provide trace plots for DS1--11 and various traces.

\jumpDistance{1}{2}
\jumpDistance{2}{4}
\jumpDistance{3}{2}
\jumpDistance{4}{2}
\jumpDistance{5}{1}
\jumpDistance{6}{2}
\jumpDistance{7}{1}
\jumpDistance{8}{1}
\jumpDistance{9}{1}
\jumpDistance{10}{2}
\jumpDistance{11}{1}

\end{document}